\documentclass[a4paper,11pt]{article}
\usepackage[latin1]{inputenc}
\usepackage{indentfirst}
\usepackage[left]{lineno}
\usepackage{amsfonts}
\usepackage{float}

\usepackage{latexsym,amssymb,amsmath}
\usepackage[dvips]{graphics,graphicx,epsfig,color}
\usepackage[lofdepth,lotdepth]{subfig}
\usepackage[lofdepth,lotdepth]{subfig}
\begin{document}
\title{\bf Parametric study of the seismic response of a  hill or mountain}
\author{Armand Wirgin\thanks{LMA, CNRS, UPR 7051, Aix-Marseille Univ, Centrale Marseille, F-13453 Marseille Cedex 13, France, ({\tt wirgin@lma.cnrs-mrs.fr})} }
\date{\today}
\maketitle
\begin{abstract}
The problem of the response of a cylindrical protuberance of rectangular shape to a SH seismic plane wave is studied in parametric manner so as to provide answers to the questions: (i) where and how should one measure this response, (ii) is the normal-incidence response a valid indication of response at other incident angles of the seismic plane wave, iii) is it possible to predict the maximal response without a detailed knowledge of the subsurface composition,  iv) how does the aspect ratio of the protuberance  affect the resonant response, and v) is the resonant wavefield uniformly distributed in the interior of the protuberance?
\end{abstract}
\newline
\newline
Abbreviated title: Earthquakes in  a hill or mountain
\newline
\newline
Corresponding author: Armand Wirgin, \\ e-mail: wirgin@lma.cnrs-mrs.fr
\newpage
\tableofcontents
\newpage
\newpage
\section{Introduction}\label{intro}
This contribution deals with a seismic scattering problem identical to the one studied in our two recent contributions \cite{wi20a,wi20b} so that all the theoretical and numerical details can be found therein and will not be repeated here. However, it is necessary to  state what precisely is the seismic scattering problem  we are here interested in.
\subsection{Statement of the problem for a bilayer protuberance of arbitrary shape}\label{state}
In the first approximation, the earth's surface is considered to be (horizontally-) flat (termed "ground" for short) and to separate the vacuum (above) from a linear, isotropic, homogeneous (LIH) solid (below), so as to be stress-free. In the second approximation the flat ground is locally deformed so as to penetrate into what was formerly the vacuum half space. We now define the {\it protuberance}  (such as a hill or mountain) as the region between the locally-deformed stress-free surface and what was formerly a portion of the flat ground. This protuberance is underlain by the same LIH solid as previously, but the solid material within the protuberance is now assumed to be  only linear and isotropic (i.e., not homogeneous). In fact, we consider the specific  case in which the material within the protuberance is in the form of a horizontal bilayer so as to be able to account for various empirically-observed effects that are thought to be due to inhomogeneity of the protuberance material. Furthermore, we assume that: the protuberance is of infinite extent along one ($z$) of the cartesian ($xyz$) coordinates,  and its stress-free boundary to be of arbitrary shape (in its $xy$ cross-section plane). The underlying problem  of much of what follows  is the prediction of the seismic wave response of this earth model.

The earthquake sources are assumed to be located in the lower half-space and to be infinitely-distant from the ground so that the seismic (pulse-like) solicitation takes the form of a body (plane) wave in the neighborhood of the protuberance. This plane wavefield is assumed to be of the shear-horizontal ($SH$) variety, which means that: only one (i.e., the cartesian coordinate $z$) component of the incident displacement field is non-nil and this field does not depend on $z$.

We  assume, not only that the protuberance boundary does not depend on $z$ but also, that the (often relatively-soft) medium filling the protuberance as well as the (usually relatively-hard) medium below the protuberance are both linear and isotropic. Furthermore the medium of the below-ground half space is assumed to be homogeneous, whereas that of the protuberance to be piecewise homogeneous (however, this heterogeneity is such as to not depend on $z$). It ensues that the scattered and total displacement fields within and outside the protuberance do not depend on $z$. Thus, the problem we are faced with is 2D ($z$ being the ignorable coordinate), and it is sufficient to search for the $z$-component of the scattered displacement field, designated by $u_{z}^{s}(\mathbf{x};\omega)$ in the sagittal (i.e., $x-y$) plane, when $u_{z}^{i}(\mathbf{x};\omega)$ designates the incident displacement field, with $\mathbf{x}=(x,y)$ and $\omega=2\pi f$  the angular frequency, $f$ the frequency. Since we now know that only the $z$ component of the field is non-nil, we drop the index $z$ in the incident, scattered, and total displacement fields. The temporal version of the displacement field is $u_{z}(\mathbf{x};t)=2\Re\int_{0}^{\infty}u_{z}^{i}(\mathbf{x};\omega)\exp(-i\omega t)d\omega$ wherein $t$ is the temporal variable. Since we now know that only the $z$ component of the field is non-nil, we drop the index $z$ in the incident, scattered, and total displacement fields in all that follows.
\begin{figure}[ht]
\begin{center}
\includegraphics[width=0.75\textwidth]{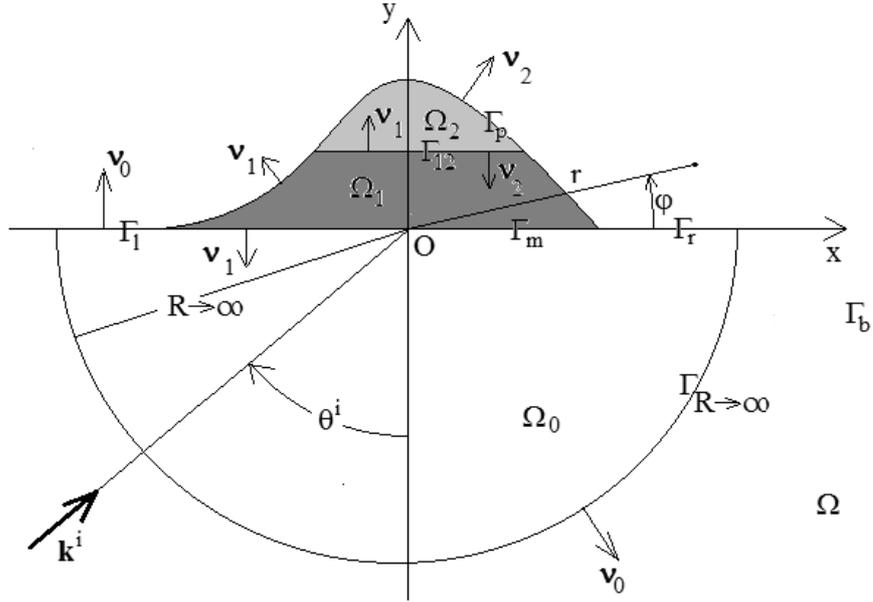}
\caption{Sagittal plane view of the 2D scattering configuration. The protuberance occupies the shaded areas and the medium within it is a horizontal bilayer.}
\label{protuberance}
\end{center}
\end{figure}

Fig. \ref{protuberance} describes the scattering configuration in the sagittal ($xy$) plane. In this figure, in which the stress-free component of the boundary of the protuberance is of arbitrary shape, $\mathbf{k}^{i}=\mathbf{k}^{i}(\theta^{i},\omega)$ is the incident wavevector oriented so that its $z$  component is nil, and $\theta^{i}$ is the angle of incidence.

The portion of the ground outside the protuberance is stress-free but since the protuberance is assumed to be in welded contact with the surrounding below-ground medium, its lower, flat, boundary is the locus of continuous displacement and stress. Thus,  the incident field is able to  penetrate into  the protuberance and then  be scattered outside the protuberance in the remaining lower half space.

The three media (other than the one of the portion of the  space above the protuberance, being occupied by the vacuum, is of no interest since the field cannot penetrate therein)  are $M^{[l]}~;~l=0,1,2$  within which the real shear modulii $\mu^{[l]}~;~l=0,1,2$  and the generally-complex shear body wave velocities are $\beta^{[l]}~;~l=0,1,2$  i.e., $\beta^{[l]}=\beta^{'[l]}+i\beta^{''[l]}$, with $\beta^{'[l]}\ge 0$, $\beta^{''[l]}\le 0$, $\beta^{[l]}=\sqrt{\frac{\mu^{[l]}}{\rho^{[l]}}}$, and  $\rho^{[l]}$ the (generally-complex) mass density. The shear-wave velocity $\beta^{[0]}$ is assumed to be real, i.e., $\beta^{''[0]}=0$.
\subsection{Case of a bilayer protuberance  of rectangular shape}
\begin{figure}[ht]
\begin{center}
\includegraphics[width=0.85\textwidth]{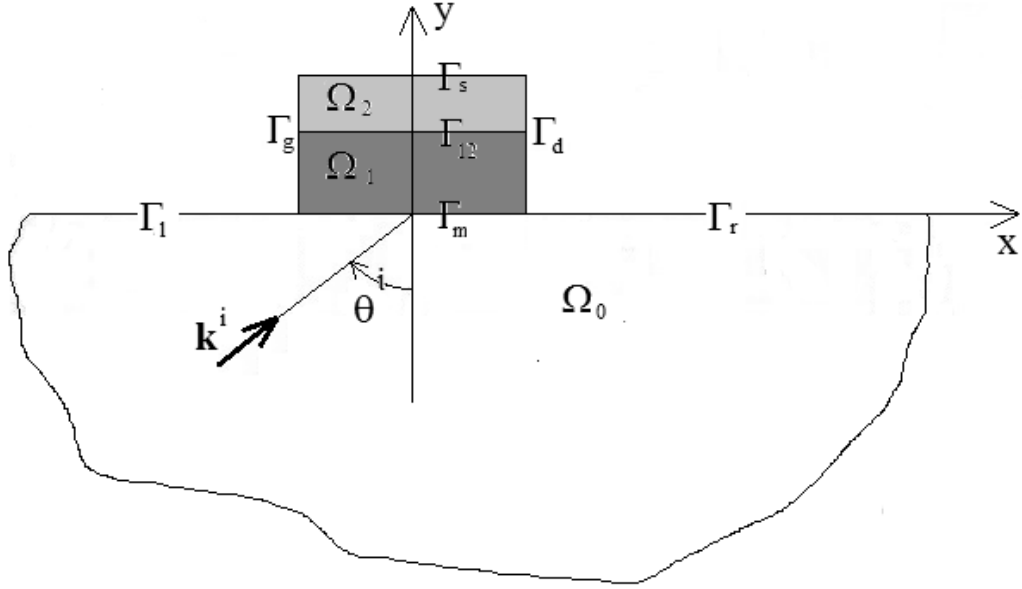}
\caption{Sagittal plane view of the 2D rectangular protuberance scattering configuration. Note that now the boundary $\Gamma_{p}$ of the above-ground feature is composed of three connected portions, $\Gamma_{g}$, $\Gamma_{s}$ and $\Gamma_{d}$.}
\label{hill}
\end{center}
\end{figure}

From now on, the option is to completely solve the forward scattering problem for the configuration depicted in
fig.\ref{hill}. The important feature thereof is the rectangular shape (in the sagittal plane) of the protuberance.

The choice of a protuberance with such simple shape is dictated by the fact that key aspects of its seismic response can be unveiled in a relatively-simple manner, both from the theoretical and numerical angles, the latter  (numerical) feature being  very useful in a parametric study such as the one undertaken in the present contribution.

As previously, the width of the protuberance is $w$, and its other characteristic dimensions are the bottom ($h_{1}$) and top ($h_{2}$) layer thicknesses, with $h=h_{1}+h_{2}$ being the height of the protuberance. What was formerly $\Gamma_{p}$ is now $\Gamma_{g}\cup\Gamma_{s}\cup\Gamma_{d}$, wherein $\Gamma_{g}$ is the leftmost vertical segment of height $h$, $\Gamma_{s}$ is the top segment of width $w$ (located between $x=-w/2$ and $x=w/2$) and $\Gamma_{d}$ is the rightmost vertical segment of height $h$. Everything else is as in fig.\ref{protuberance}.

Of primary interest is the total displacement field $u(\mathbf{x};f)$ at location $\mathbf{x}$ and frequency $f$ as well as  the transfer function $T=u(\mathbf{x};f)/a^{i}(f)$, wherein $a^{i}(f)$ is the spectral amplitude of the incident (seismic) pulse. Since we shall assume that $a^{i}(f)=1$ for all frequencies, $T=u(\mathbf{x};f)$.
\subsection{The contours of the parametric study}\label{param}
Our parametric study is justified by the fact that a protuberance (e.g., dike, hill or mountain) gives rise to a large variety of seismic responses and that it  has not, until now, been possible to identify the principal features of the incident wave, the protuberance, and the underground (to which the protuberance is connected), that condition these responses. In \cite{wi20b} we showed why these, often amplified, responses are largely the result of the coupling of the incident wave to (surface shape) resonances, but, we still do not know how these resonances (i.e., their frequencies of occurrence and amplitudes) depend on the various parameters of the wave-structure interaction.

As seen in the previous section, our scattering problem involves  three incident wave parameters $\theta^{i}$, $a^{i}(f)$, $f$, and eleven configurational (i.e., geometrical and compositional) parameters: $w,~h_{1},~h_{2},~ \mu^{[0]},~\beta^{[0]},~\mu^{[1]},~\beta^{[1]'},~$ $\beta^{[1]''},~\mu^{[2]},~\beta^{[2]'},~\beta^{[2]''}$. If
   we ignore all but $\theta^{i}$ amongst the incident-wave parameters, and, for example,  allow each of the remaining twelve parameters  to take on three different values, one at a time. We will thus be  faced with the difficult task of the computation of the response of $3^{12}$ different scattering configurations.

 Another difficulty is that these computations should ordinarily be done for a whole (quasi-continuous) range of frequencies $f$, and if temporal responses are of interest, for several choices of the spectrum of $a^{i}(f)$ of the pulse-like solicitation.  Due to the fact that we deal here only with the transfer functions (i.e., the frequential response), we  choose $a^{i}(f)=1$ for all  frequencies. For the rather  low-frequencies of interest in seismics, we will show that it is sufficient to determine the displacement field   at up to four resonant frequencies and quite a bit more if we focus on the full transfer function.

 Moreover, our previous contribution \cite{wi20b} showed that the seismic response of each configuration, at a given frequency, tends to be spatially non-uniform, so that this is the cause of another computational difficulty (i.e., the determination of the field at a great number of points in a domain that includes the protuberance and a portion of the underground). But, all in all, these voluminous computations are not out of reach of what modern digital computers can do in a reasonable length of time.

 The remaining, although not less considerable, problem is how to graphically represent, and store in a file of reasonable size (a requisite for publication), this large mass of information (concerning the computed responses).

The way to reduce the computational and representational burdens to more reasonable proportions is first to reduce the number of times each parameter, or groups of parameters, is (are) varied. For this reason, many parametric studies are restricted to one angle of incidence, which is usually chosen to be $\theta^{i}=0^{\circ}$, but, as is seen further on, this does not enable a proper appreciation of the possible magnitude of amplified response. It is also tempting to skip the computation of the internal fields (i.e., concentrate one's attention on the three-point transfer functions), but, as will be demonstrated hereafter, this can be quite risky for the prediction of the overall and/or maximal (spatially-speaking) resonant response of the protuberance. Nevertheless, due to the volume of the computations, and that of the storage space required to graphically-represent the solutions,  the usual practice is to content oneself with three-point (or even less)  transfer functions. and perhaps a few examples of internal fields for a selection of the  combinations of the various configurations at a few  frequencies.

 Our parametric study, which obviates many of the aforementioned restrictions, will enable us to show that the most important parameters are the incident angle and  material damping ratio, for a given geometry and composition (i.e., the shear modulus and real part of the wavespeed) of the protuberance and underground. The resonance frequencies, as well as the response at these frequencies, turn out also to be  sensitive to changes in the  shape (i.e., aspect ratio) of the protuberance, composition of the protuberance and composition of the underground.
\section{Variation of the incident angle}
\begin{figure}[ht]
\begin{center}
\includegraphics[width=0.6\textwidth]{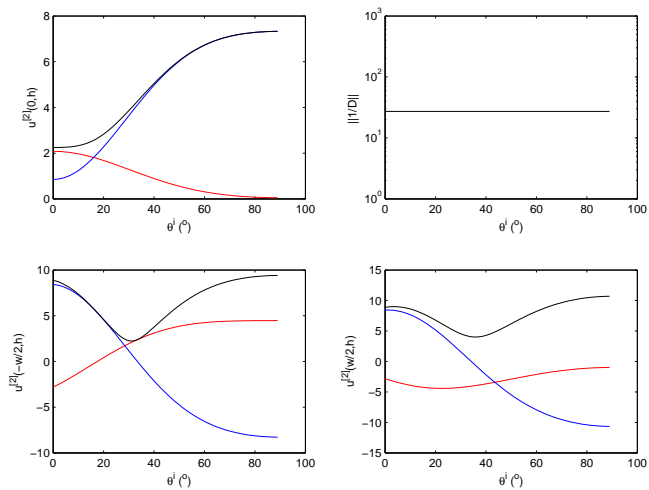}
\caption{3-point transfer functions  versus incident angle. Note that $1/\|D(f^{R}\|$ does not depend on the incident angle.}
\label{ia-010}
\end{center}
\end{figure}

\begin{figure}[ht]
\begin{center}
\includegraphics[width=0.6\textwidth]{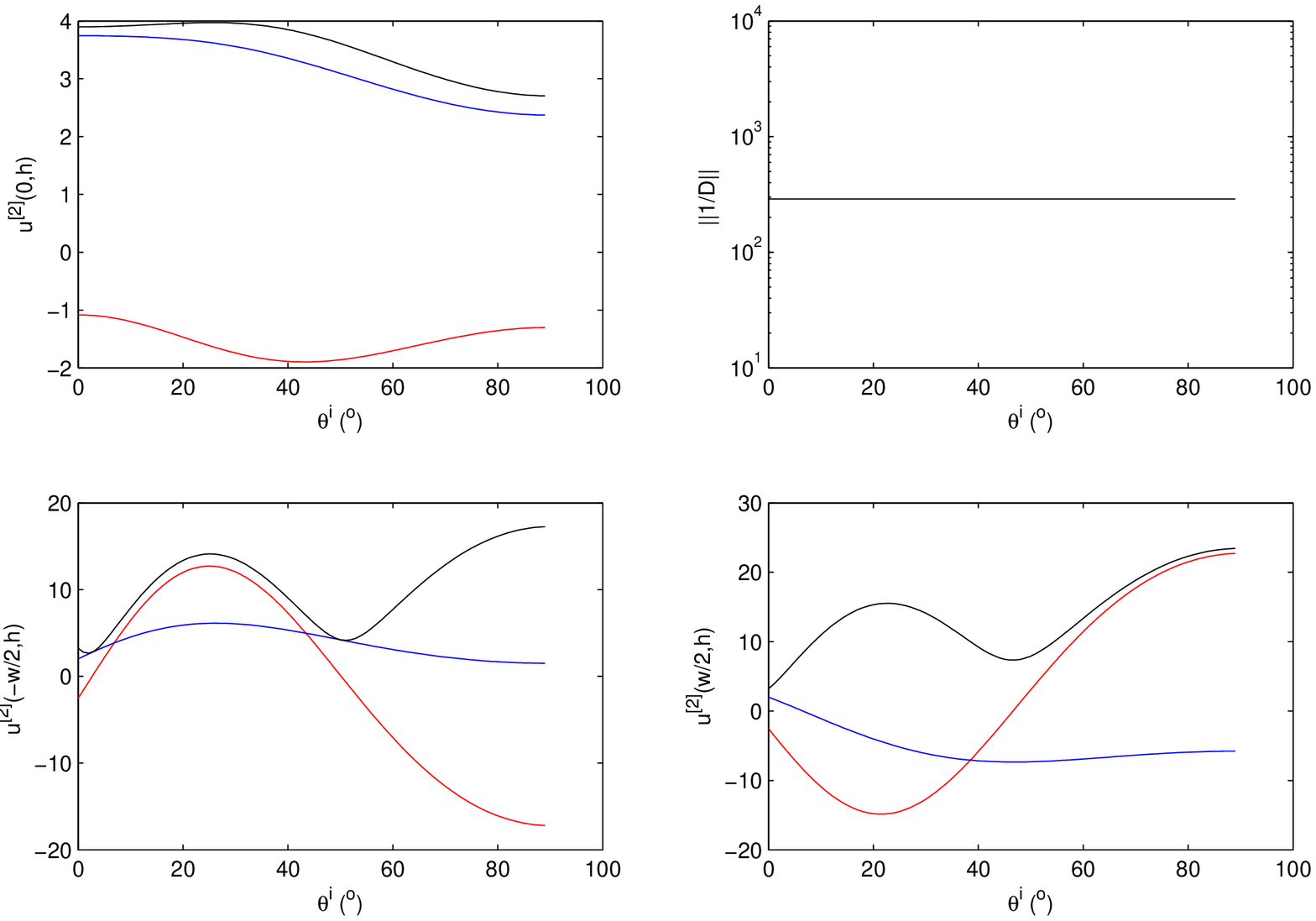}
\caption{3-point transfer functions  versus incident angle.}
\label{ia-020}
\end{center}
\end{figure}

\begin{figure}[h]
\begin{center}
\includegraphics[width=0.6\textwidth]{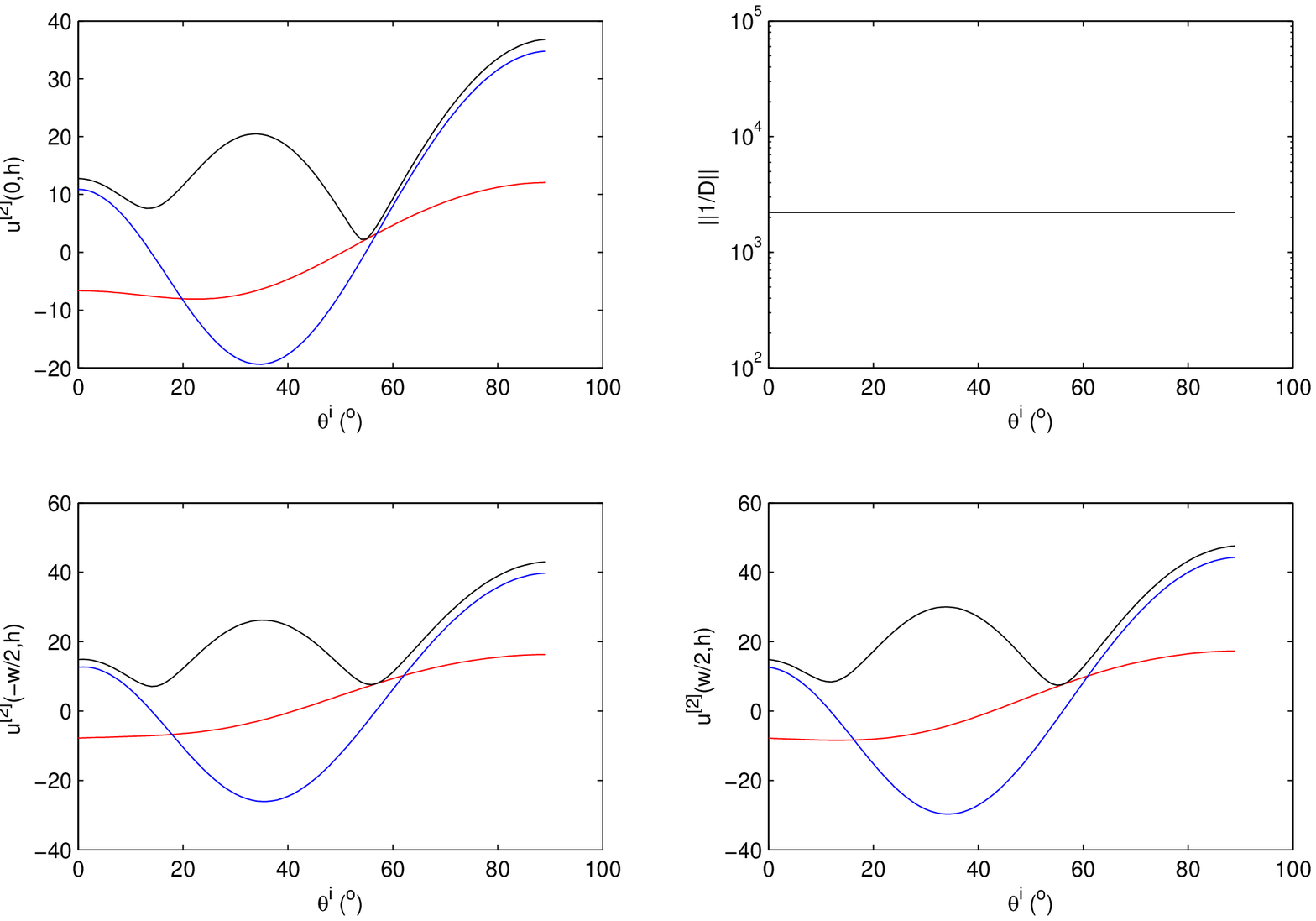}
\caption{3-point transfer functions  versus incident angle.}
\label{ia-030}
\end{center}
\end{figure}

\begin{figure}[h]
\begin{center}
\includegraphics[width=0.6\textwidth]{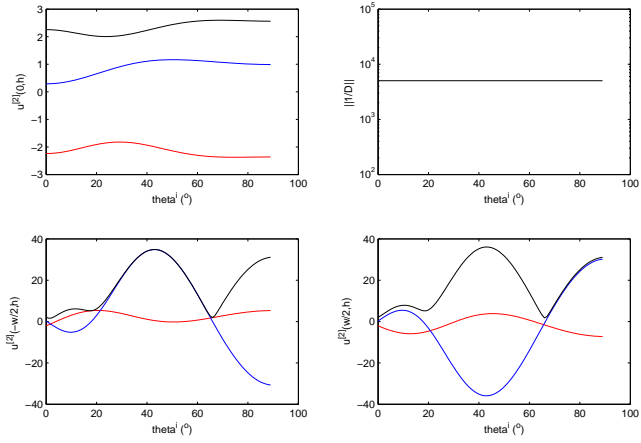}
\caption{3-point transfer functions  versus incident angle.}
\label{ia-040}
\end{center}
\end{figure}

In the totality of the set of figs. \ref{ia-010}-\ref{ia-040} we assume: $w=1000~m$,
 $h_{1}=h_{2}=75~m$, $\mu^{[0]}=6.85~GPa$, $\beta^{[0]}=1629.4~ms^{-1}$,$\mu^{[1]}=\mu^{[1]}=2~GPa$, $\beta^{[1]'}=\beta^{[2]'}=1000~ms^{-1}$. Furthermore,  $\beta^{[1]''}=\beta^{[2]''}$  equals $0~ms^{-1}$ or $-20~ms^{-1}$ (in graphs not shown here).

 In each of these figures:  the upper left-hand, lower left-hand and lower right-hand panels depict $T(0,h;f^{R})$, $T(-w/2,h;f^{R})$ and $T(w/2,h;f^{R})$ as a function of $\theta^{i}$, whereas the upper right-hand panel depicts $\|1/D(f^{R})\|$ as a function of $\theta^{i}$, $f_{R}$ being the indicated resonance frequency and $D(f)$ the determinant of the linear matrix equation by which we compute the seismic response (as indicated in \cite{wi20b}, the positions  of the minima of $D(f)$ define the frequencies of resonance). Red, blue and black curves relate to the real part, the imaginary part and the absolute value of a complex function.
\subsection{First resonant frequency $f^{R}=1.608~Hz$}
\clearpage
\newpage
\subsection{Second resonant frequency $f^{R}=1.975~Hz$}
%
\subsection{Third resonant frequency $f^{R}=2.42~Hz$}
\clearpage
\newpage
\subsection{Fourth resonant frequency $f^{R}=2.874~Hz$}
%
\subsection{Discussion of the results for variations of the incident angle}
Our results show that the variation of the 3-point displacement response with incident angle seems to follow five prototypical patterns when there is no damping in the configuration:
(i) e.g., the pattern, characterized by a monotonic decrease of response with increasing $|\theta^{i}|$, at point $(0,h)$ and resonant frequency $1.975~Hz$;
(ii) e.g., the pattern, characterized by a near-constant response with increasing $|\theta^{i}|$,  at point $(0,h)$ and resonant frequency $2.814~Hz$;
(iii)  e.g., the pattern, characterized by a decreasing, followed by increasing response with increasing $|\theta^{i}|$,  at point $(0, h)$ and resonant frequency $f=1.608$;
(iv) e.g., the pattern, characterized by increasing, then decreasing, followed by increasing response with increasing $|\theta^{i}|$,  at $(\pm w/2,h)$ and resonant frequency $1.975~Hz$;
(v) e.g., the pattern, characterized by decreasing, then increasing, followed  by  decreasing, and ending up as increasing response with increasing $|\theta^{i}|$, at all three points and resonant frequency $f=2.42~Hz$.

Also, the figures not shown here, relative to the case in which damping is present, indicate that the response patterns are essentially smoothed-out versions of the respective undamped response patterns, but the damping essentially reduces (compared to the undamped case)  the intensity of the response for all angles of incidence at the three locations of the protuberance.

Finally, and perhaps most important, is the fact that the field tends to increase with $|\theta^{i}|$ beyond a certain angle of incidence (usually $\ge 60^{\circ}$) except in the above-defined cases (i) and (ii).  For this reason, the majority of the following results will apply to large angles of incidence.
\section{Variation of the material damping within the protuberance}
In the totality of the set of figures (figs. \ref{at10}-\ref{at20}) we assume:
$\theta^{i}=80^{\circ}$, $h_{1}=h_{2}=75~m$, $750~m$, $\mu^{[0]}=6.85~GPa$, $\beta^{[0]}=1629.4~ms^{-1}$, $\mu^{[1]}=\mu^{[1]}=2~GPa$, $\beta^{[1]'}=\beta^{[2]'}=1000~ms^{-1}$.

We analyze the 3-point transfer functions for five damping coefficients: $\beta^{[1]''}=\beta^{[2]''}=0~ms^{-1}$,  $\beta^{[1]''}=\beta^{[2]''}=-15~ms^{-1}$, $\beta^{[1]''}=\beta^{[2]''}=-25~ms^{-1}$, and $\beta^{[1]''}=\beta^{[2]''}=-50~ms^{-1}$.

In these figures; the (1,2) panel is relative to $1/\|D(f)\|$ versus $f$, the other three (1,1), (2,1) and (2,2) panels, to the transfer functions $T(0,h;f)$, $T(-w/2,h;f)$ and $T(w/2,h;f)$ respectively.
\subsection{Variation of $M$ for $\beta^{[1]''}=\beta^{[2]''}=0$}
\begin{figure}[ht]
  \begin{center}
    \subfloat[]{
      \includegraphics[width=0.5\textwidth]{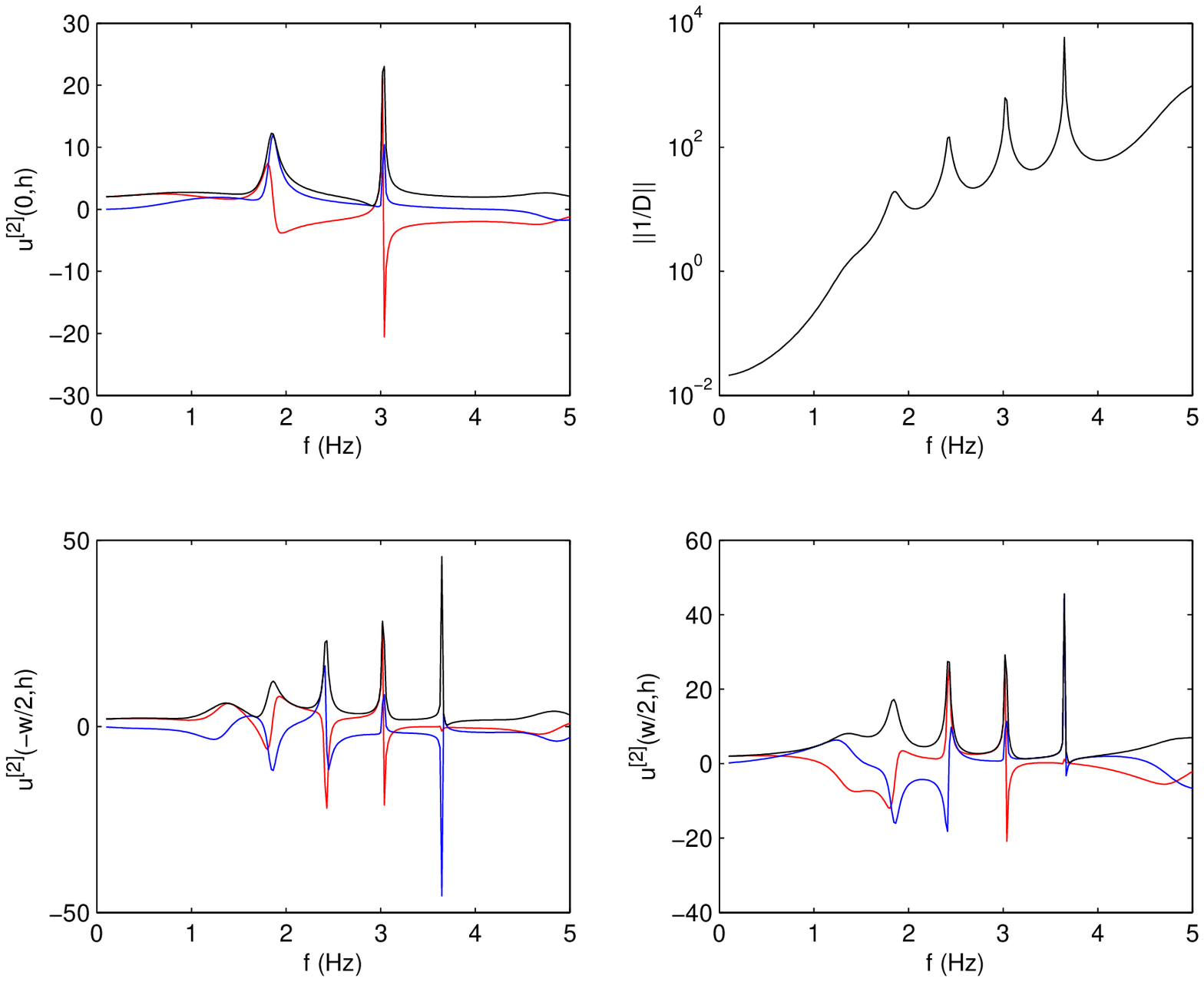}
      \label{a}
                         }
    \subfloat[]{
      \includegraphics[width=0.505\textwidth]{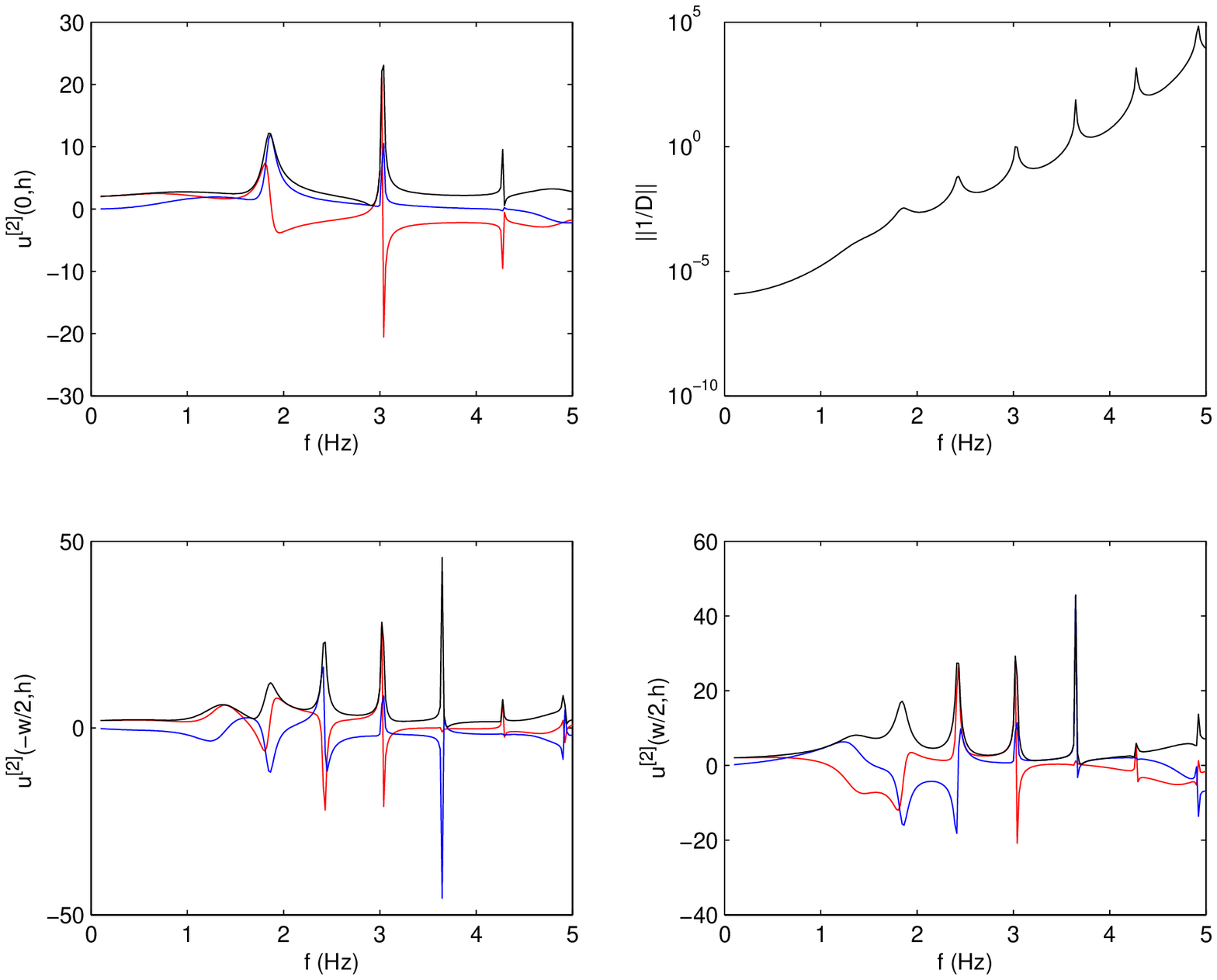}
      \label{b}
                         }

    \caption{
(a): $\beta^{[1]''}=\beta^{[2]''}=0~ms^{-1}$, $M=5$.\\
(b): $\beta^{[1]''}=\beta^{[2]''}=0~ms^{-1}$, $M=8$.
}
    \label{at10}
  \end{center}
\end{figure}

The question we address in fig. \ref{at10} is how large must the approximation order $M$ of the aforementioned matrix equation (see \cite{wi20b}) be for the seismic response to be correctly predicted in the absence of material damping, this being the most severe case (compared to the cases in the presence of material damping) for the prediction of resonant response.
\clearpage
\newpage
The (1,2) panels in parts (a) and (b) of this figure, in which the frequency spans $]0~Hz,5~Hz]$, show that;\\
(1) for $M=5$ (left-hand figure) only five resonances make their appearance, whereas for $M=8$ (right-hand figure) there appear seven resonances;\\
(2) the positions of the first five resonances are the same in the two figures;\\\\
whereas the other three panels in these two figures show that\\
(3) the heights of the 3-point transfer functions $T$ at the first two resonant frequencies in the (1,1) panels), and at the first five resonant frequencies in the (2,1) and (2,2) panels, are the same whereas the behaviors of  all three transfer function $T$ for frequencies beyond the sixth resonance frequencies are different (notably by the fact that there appears no resonant response in this frequency range for the $M=5$ computed solution).

This means that to correctly predict the resonant seismic response of the protuberance one must choose $M$ as large as possible and all the larger the wider is the frequency range in which one attempts to predict this response.

\subsection{Variation of $\beta^{[1]''}=\beta^{[2]''}\le 0$}
\begin{figure}[ht]
  \begin{center}
    \subfloat[]{
      \includegraphics[width=0.225\textwidth]{recthill2layer_13-110919-1854.eps}
      \label{a}
                         }
    \subfloat[]{
      \includegraphics[width=0.225\textwidth]{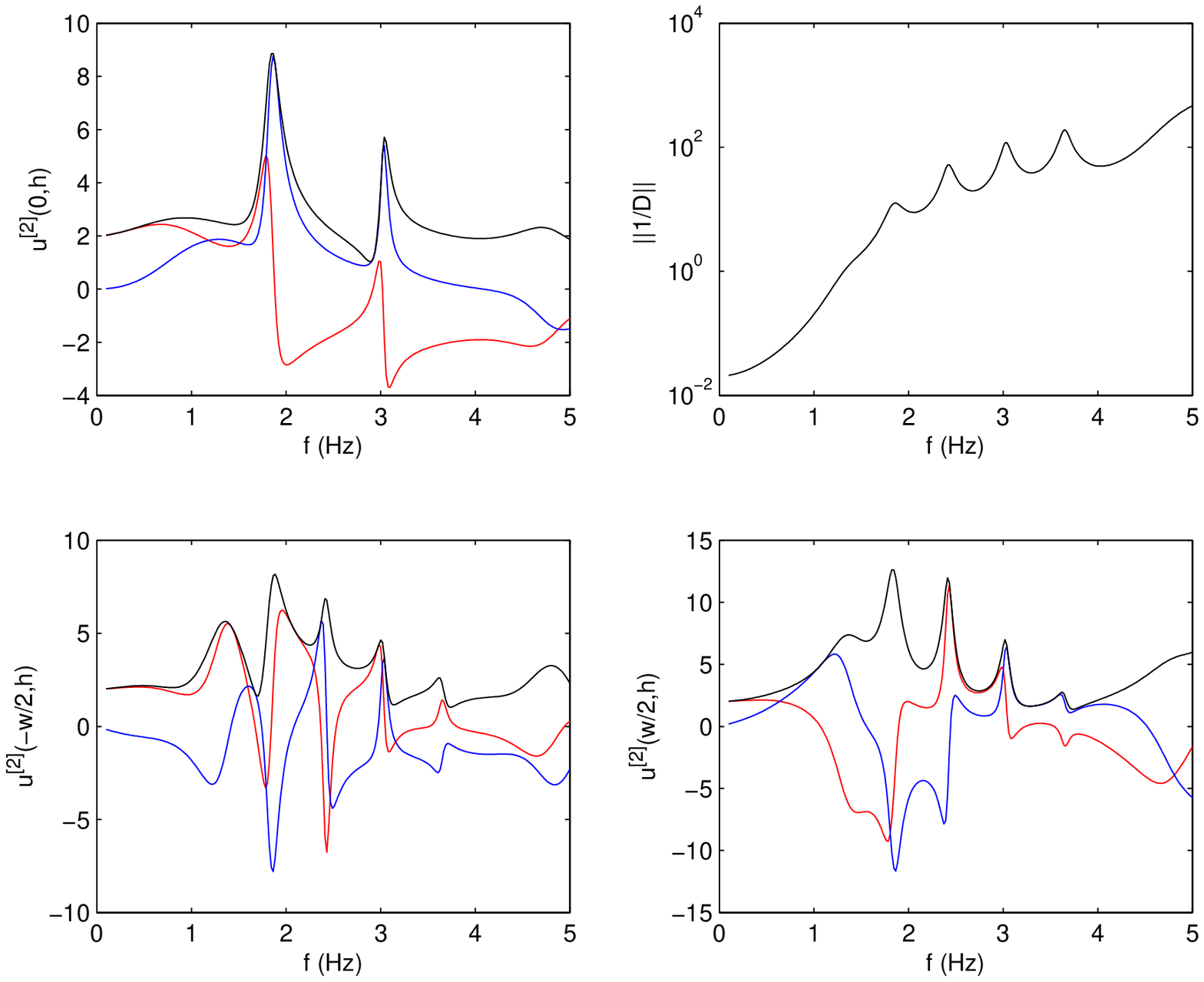}
      \label{b}
                         }
    \subfloat[]{
      \includegraphics[width=0.243\textwidth]{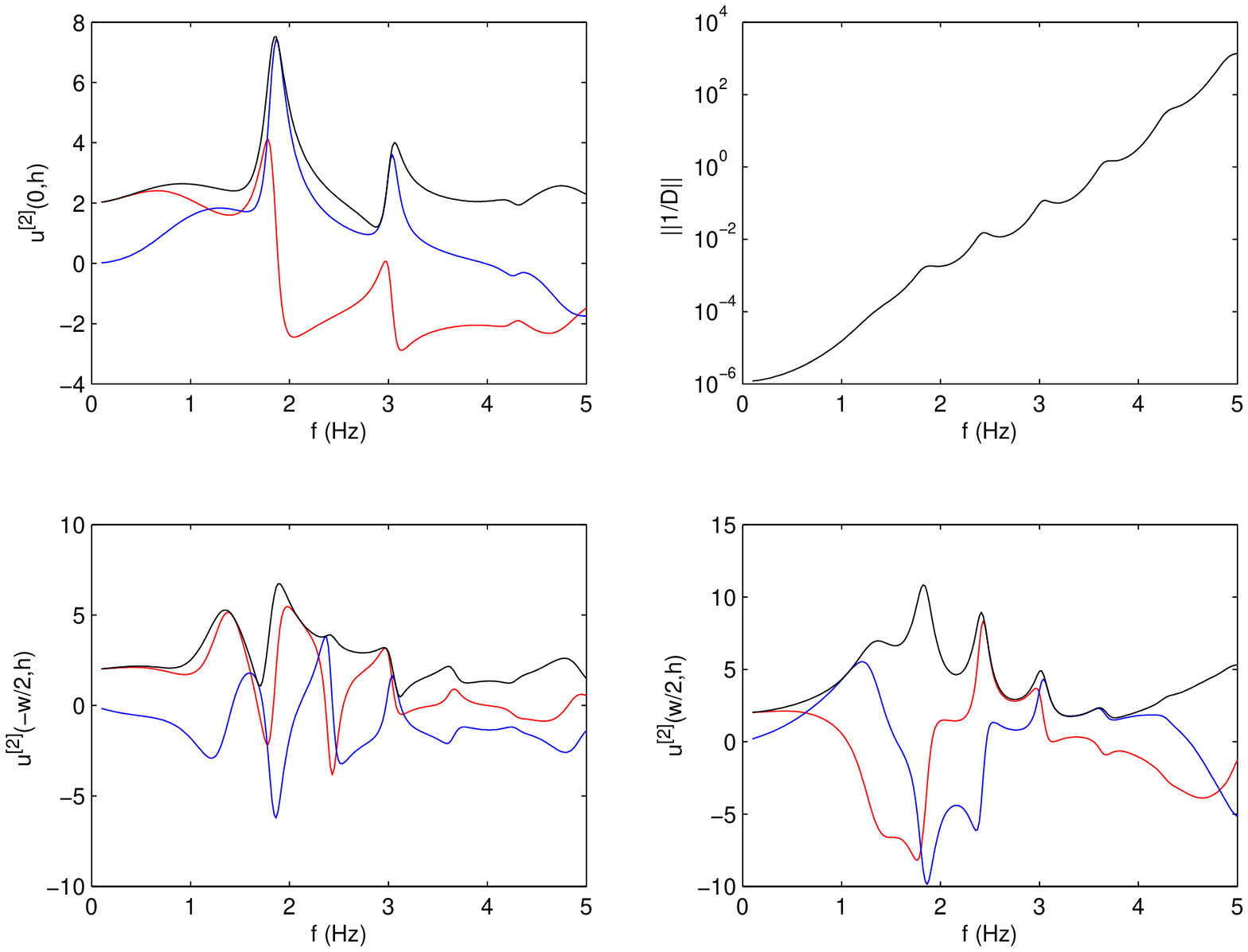}
      \label{c}
                         }
    \subfloat[]{
      \includegraphics[width=0.227\textwidth]{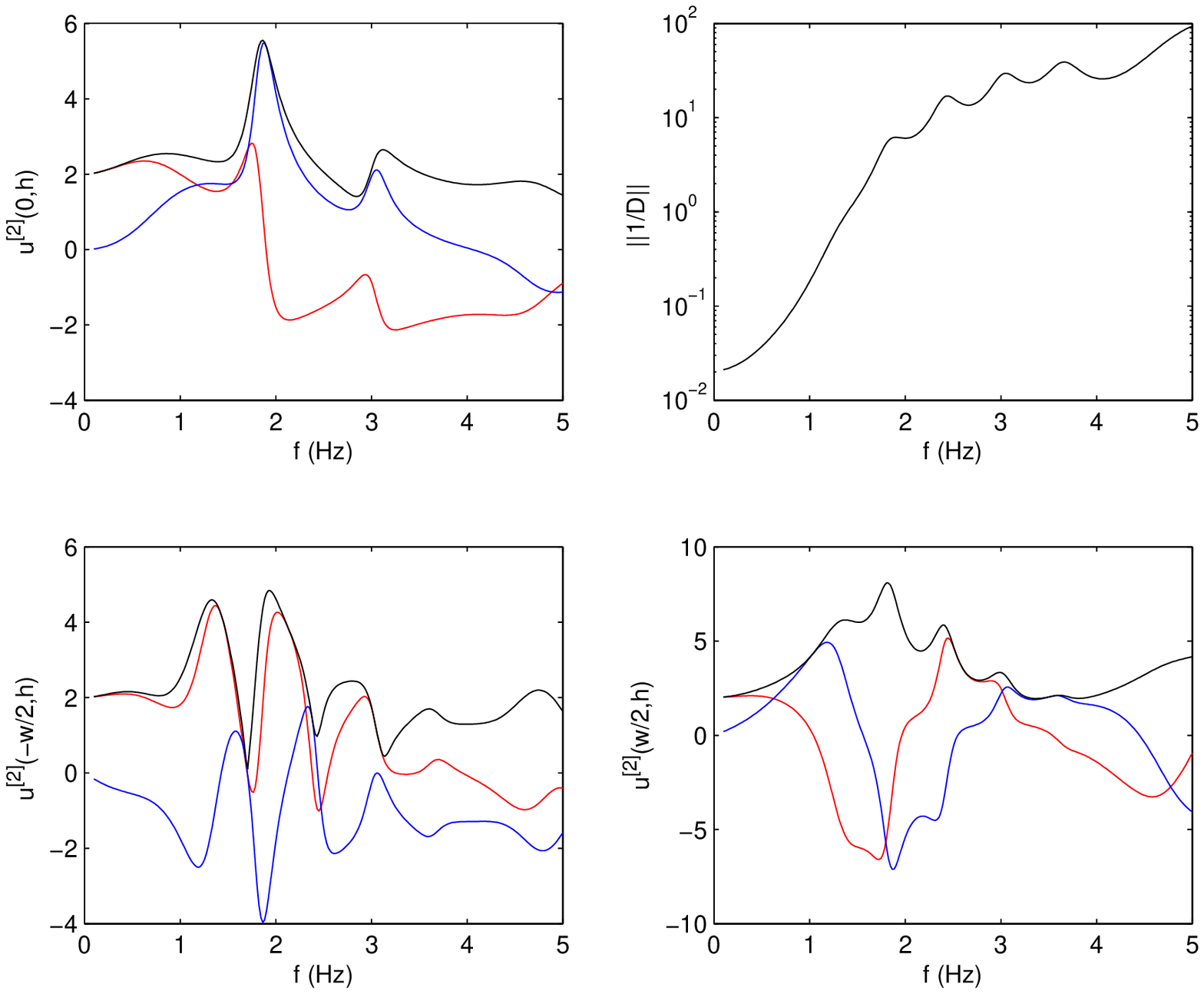}
      \label{d}
                         }
    \caption{
(a): $\beta^{[1]''}=\beta^{[2]''}=0~ms^{-1}$. The maximal ordinates in the (1,1),(1,2),(2,1),(2,2) panels are $30,10^{5},50,60$  respectively.\\
(b): $\beta^{[1]''}=\beta^{[2]''}=-15~ms^{-1}$. The maximal ordinates in the (1,1),(1,2),(2,1),(2,2) panels are $10,10^{4},10,15$  respectively.\\
(c): $\beta^{[1]''}=\beta^{[2]''}=-25~ms^{-1}$. The maximal ordinates in the (1,1),(1,2),(2,1),(2,2) panels are $8,10^{4},10,15$  respectively.\\
(d): $\beta^{[1]''}=\beta^{[2]''}=-50~ms^{-1}$. The maximal ordinates in the (1,1),(1,2),(2,1),(2,2) panels are $6,10^{2},6,10$  respectively.
}
    \label{at20}
  \end{center}
\end{figure}

In fig. \ref{at20} we compare the resonant response at the three locations of the top segment of the protuberance for four values of material damping.
\\
This figure shows clearly that:\\
(i) the position of the resonant frequencies hardly changes with the (relatively-small) amount of material damping in the protuberance, which means that they can (and should) be inferred from the undamped response (for which the resonance peaks are the sharpest);\\
(ii) the heights of the resonant response peaks are diminished considerably by the introduction of material damping, all the more so than this damping is larger and the resonance order is larger (i.e., the first resonance peaks in the 3-point transfer functions are the ones that are the least affected by material damping, but these peaks are usually lower than those of the higher-order  resonance peaks).
\section{Variation of the real part of the velocity within  the protuberance}
\begin{figure}[ht]
  \begin{center}
    \subfloat[]{
      \includegraphics[width=0.5\textwidth]{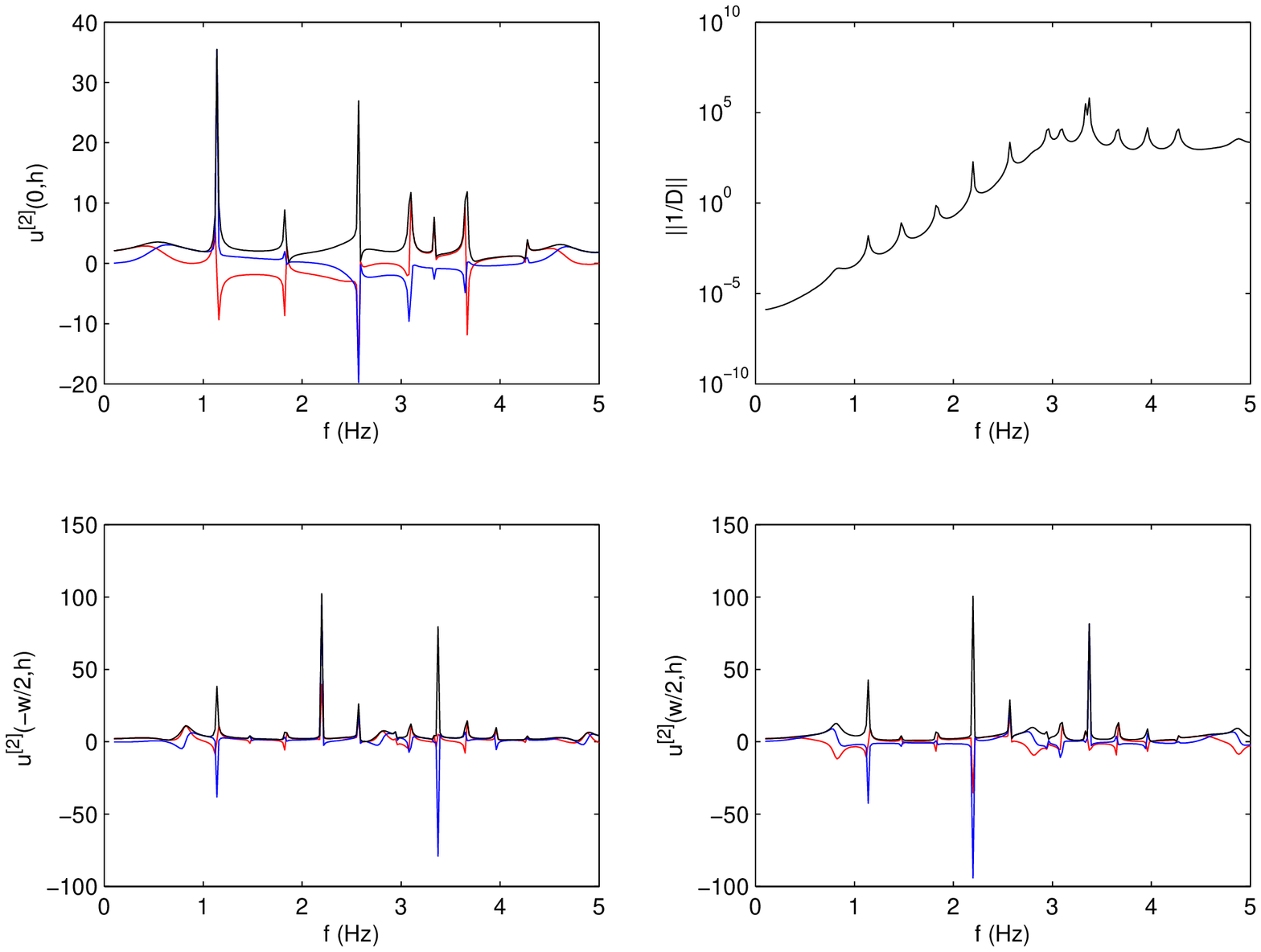}
      \label{a}
                         }
    \subfloat[]{
      \includegraphics[width=0.5\textwidth]{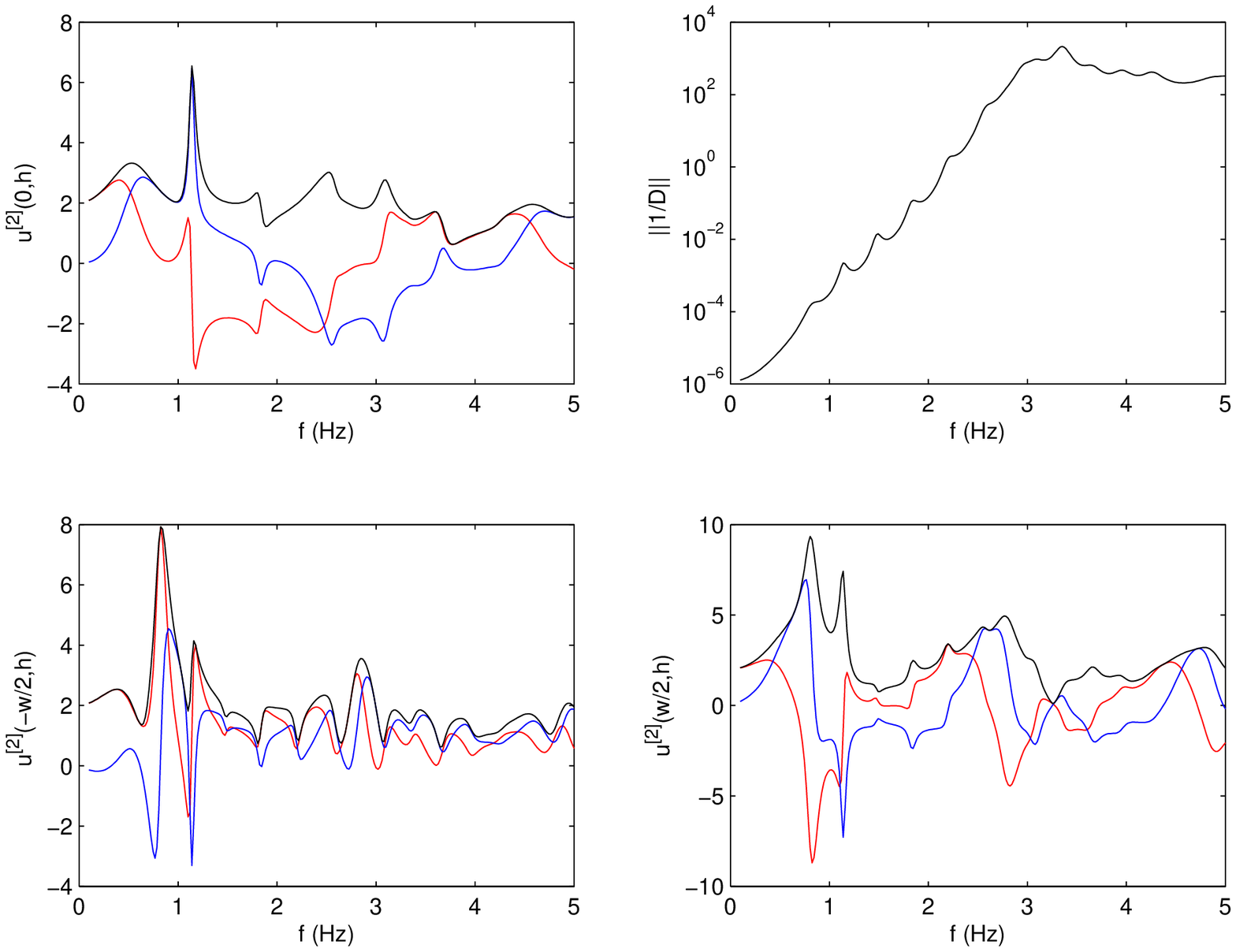}
      \label{b}
                         }

    \caption{
(a): $\beta^{[1]''}=\beta^{[2]''}=0~ms^{-1}$.\\
(b): $\beta^{[1]''}=\beta^{[2]''}=-15~ms^{-1}$.
}
    \label{be-010}
  \end{center}
\end{figure}

In the totality of the  set of figs. \ref{be-010}-\ref{be-030} we assume:
 $h_{1}=h_{2}=75~m$, $w=750~m$, $\mu^{[0]}=6.85~GPa$, $\beta^{[0]}=1629.4~ms^{-1}$, $\mu^{[1]}=\mu^{[1]}=2~GPa$, $\theta^{i}=80^{\circ}$. Furthermore,  $\frac{\beta^{[1]''}}{\beta^{[1]'}}=\frac{\beta^{[2]'}}{\beta^{[2]''}}$ either equals $0$ or $=0.025$.

 In each of these figures:  the upper left-hand, lower left-hand and lower right-hand panels depict $T(0,h;f)$, $T(-w/2,h;f)$ and $T(w/2,h;f)$ as a function of $f$, whereas the upper right panel depicts $\|1/D(f_{R})\|$ as a function of $f$.  Red, blue and black curves relate to the real part, the imaginary part and the absolute value of a complex function.
\subsection{Case $\beta^{[1]'}=\beta^{[2]'}=600~ms^{-1}$}
\clearpage
\newpage
\subsection{Case $\beta^{[1]'}=\beta^{[2]'}=800~ms^{-1}$}
\begin{figure}[h]
  \begin{center}
    \subfloat[]{
      \includegraphics[width=0.5\textwidth]{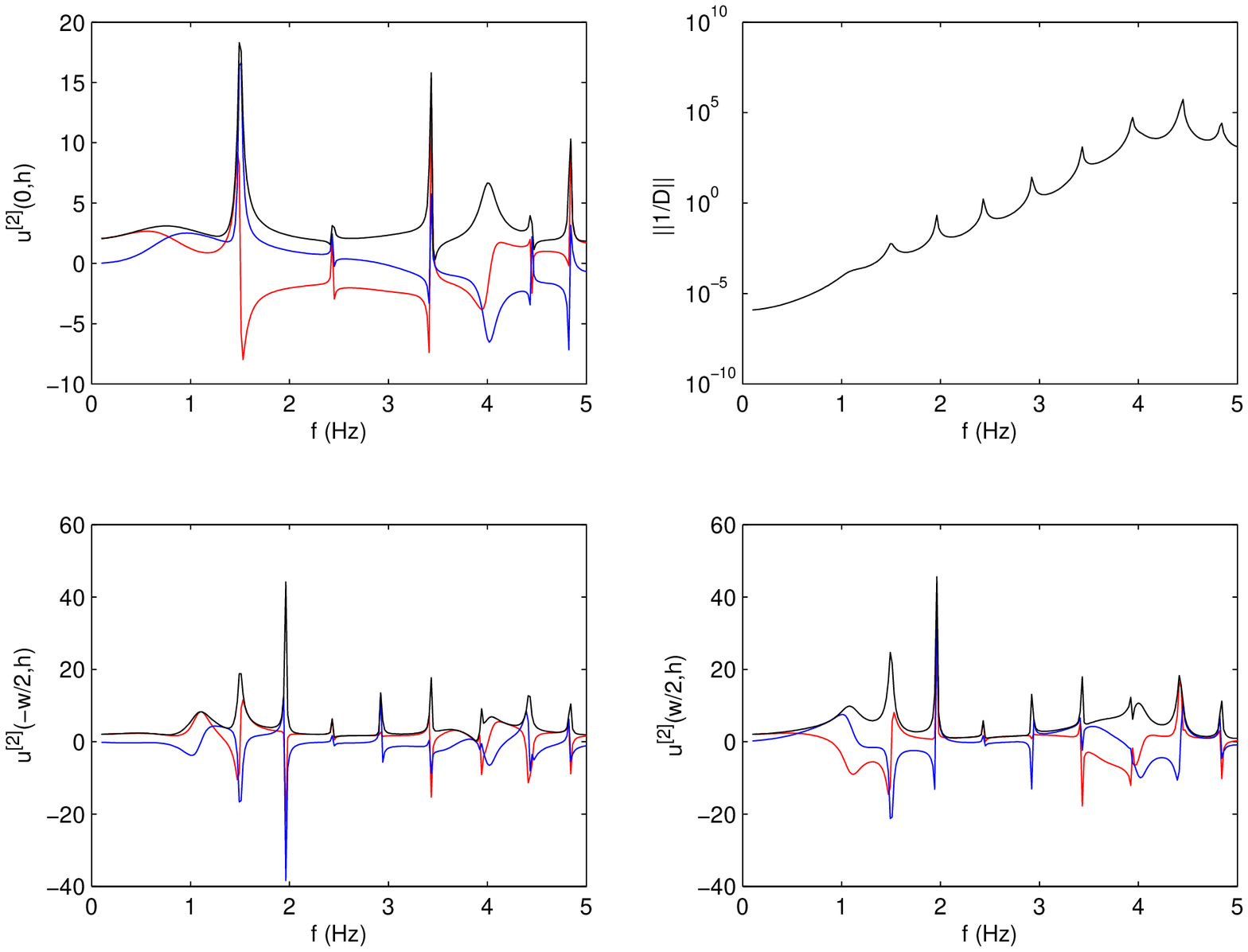}
      \label{a}
                         }
    \subfloat[]{
      \includegraphics[width=0.5\textwidth]{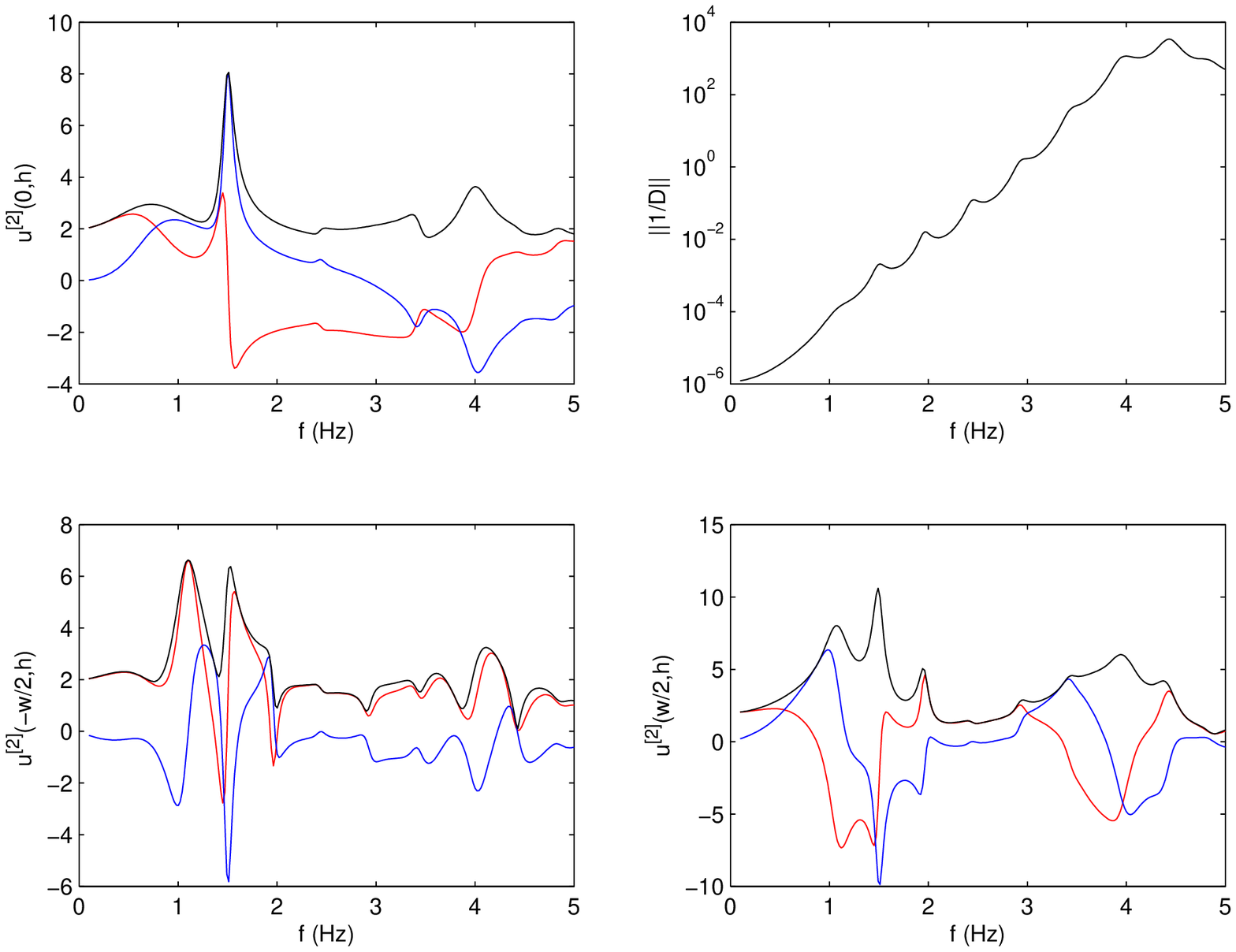}
      \label{b}
                         }

    \caption{
(a): $\beta^{[1]''}=\beta^{[2]''}=0~ms^{-1}$.\\
(b): $\beta^{[1]''}=\beta^{[2]''}=-20~ms^{-1}$.
}
    \label{be-020}
  \end{center}
\end{figure}
\clearpage
\newpage
\subsection{Case $\beta^{[1]'}=\beta^{[2]'}=1000~ms^{-1}$}
\begin{figure}[h]
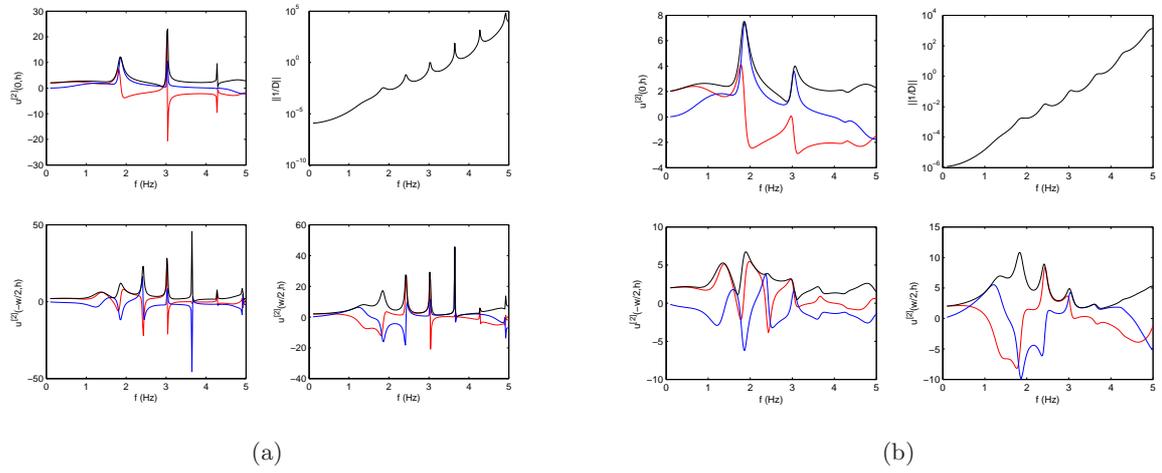

  \begin{center}
    \subfloat[]{
      \includegraphics[width=0.475\textwidth]{recthill2layer_13-110919-1854.eps}
      \label{a}
                         }
    \subfloat[]{
      \includegraphics[width=0.5\textwidth]{recthill2layer_13-110919-1556.eps}
      \label{b}
                         }

    \caption{
(a): $\beta^{[1]''}=\beta^{[2]''}=0~ms^{-1}$.\\
(b): $\beta^{[1]''}=\beta^{[2]''}=-25~ms^{-1}$.
}
    \label{be-030}
  \end{center}
\end{figure}
\clearpage
\newpage
\subsection{Discussion}\label{discbeta}
Figs. \ref{be-010}-\ref{be-030} show that:\\
(i) the transfer functions  are very different at the frequency locations of their maxima  and also quite different at the three points of the free-surface boundary of the protuberance;
(ii) put in another way,   a resonance peak (i.e., a response peak at the location of a resonance frequency $f^{R}$, the latter being determined from the position of a local maximum of $1/|D(f)\|$) at a given $f^{R}$ is not always present at one or the other of the spatial locations of the 3-point transfer functions, but, as explained in \cite{wi20b}, this is simply the result of: a) how many, and the degree of coupling to, excited resonant modal coefficients occur at this frequency, as well as b) the geometrical factors that modulate the sum of these coefficients at that frequency;\\
(iii) if we concentrate our attention on the different response maxima for the lossy configurations (of greatest interest in practical applications) then  their strengths, for the wavespeeds $\beta^{[1]'}=\beta^{[2]'}=600,~800,~1000~ms^{-1}$, are  8, 6.5 and 6.5 at the left-hand edge, 6.5, 8, and 7.5 at the center, and 9, 10.2 and 11 at the right-hand edge of the top segment of the protuberance, which seems (verb that is employed because the following conclusion might change with different choices of shear modulus and aspect ratio of the protuberance) to imply that the maximal response (regardless of its frequency of occurrence) increases as the wavespeed $\beta^{[1]'}=\beta^{[2]'}$ increases;\\
(iv) the maximal response, for all the  choices of the wavespeed, is greater at the right-hand edge of the top segment of the protuberance than at the center and left-hand edge of this segment, which is related to the fact that the angle of incidence is $60^{\circ}$. Consequently, the center of the top segment is not the best location for predicting the maximal response  to obliquely-incident  seismic waves.
\newpage
\section{Variation of the shear modulus within the protuberance}\label{discmu}
In this set of figures we assume:
 $h_{1}=h_{2}=75~m$, $w=750~m$, $\mu^{[0]}=6.85~GPa$, $\beta^{[0]}=1629.4~ms^{-1}$,$\beta^{[1]}=\beta^{[1]}=1000~ms^{-1}$, $\theta^{i}=80^{\circ}$. Furthermore,  $\mu^{[1]}=\mu^{[2]}$ takes the values $1,2,3~GPa$.

 In each of these figures:  the upper left, lower left-hand and lower right-hand panels depict $T(0,h;f)$, $T(-w/2,h;f)$ and $T(w/2,h;f)$ as a function of $f$, whereas the upper right-hand panel depicts $\|1/D(f_{R})\|$ as a function of $f$.  Red, blue and black curves relate to the real part, the imaginary part and the absolute value of a complex function.
\subsection{Case $\mu^{[1]}=\mu^{[2]}=1~GPa$}
\begin{figure}[h]
  \begin{center}
    \subfloat[]{
      \includegraphics[width=0.45\textwidth]{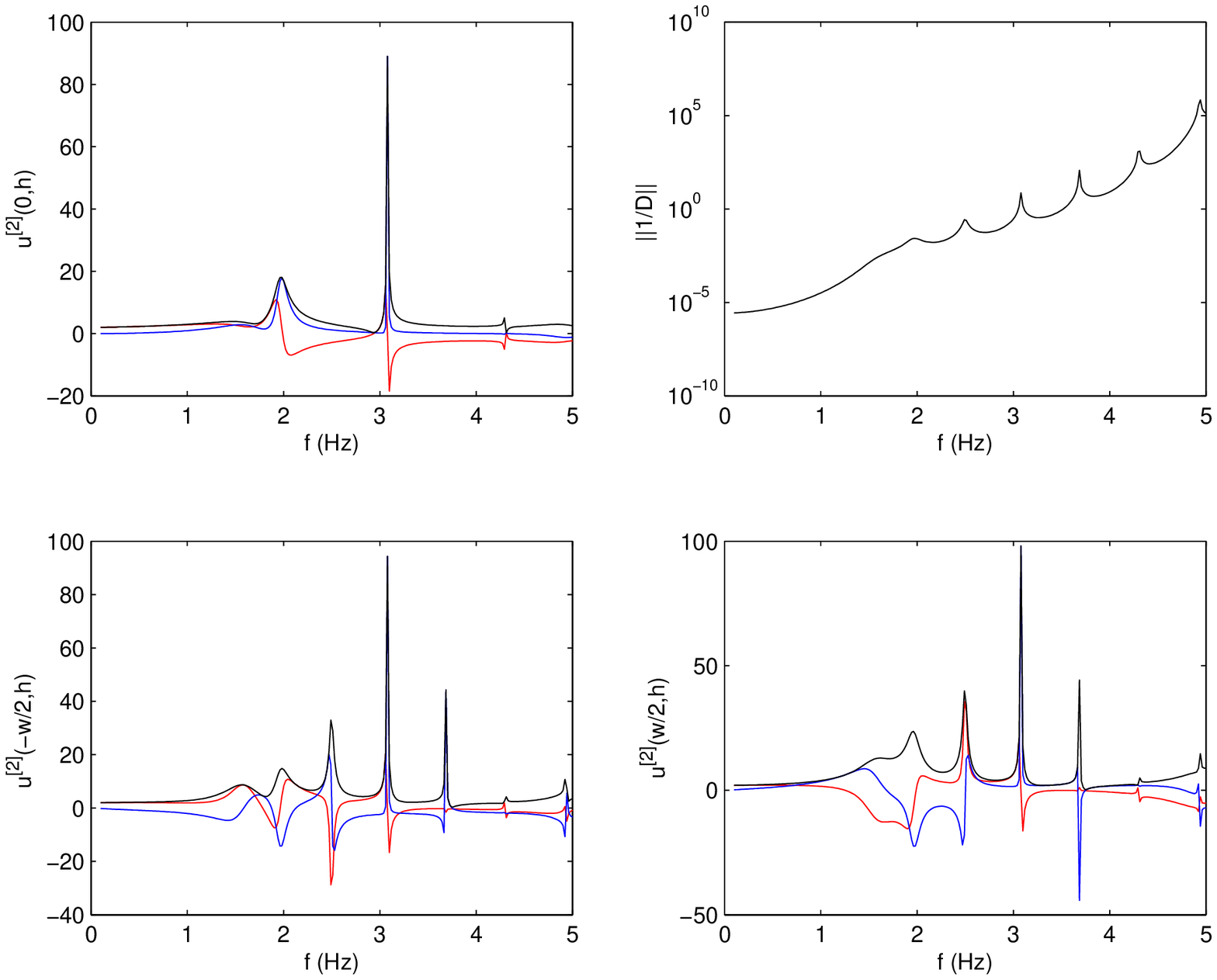}
      \label{a}
                         }
    \subfloat[]{
      \includegraphics[width=0.5\textwidth]{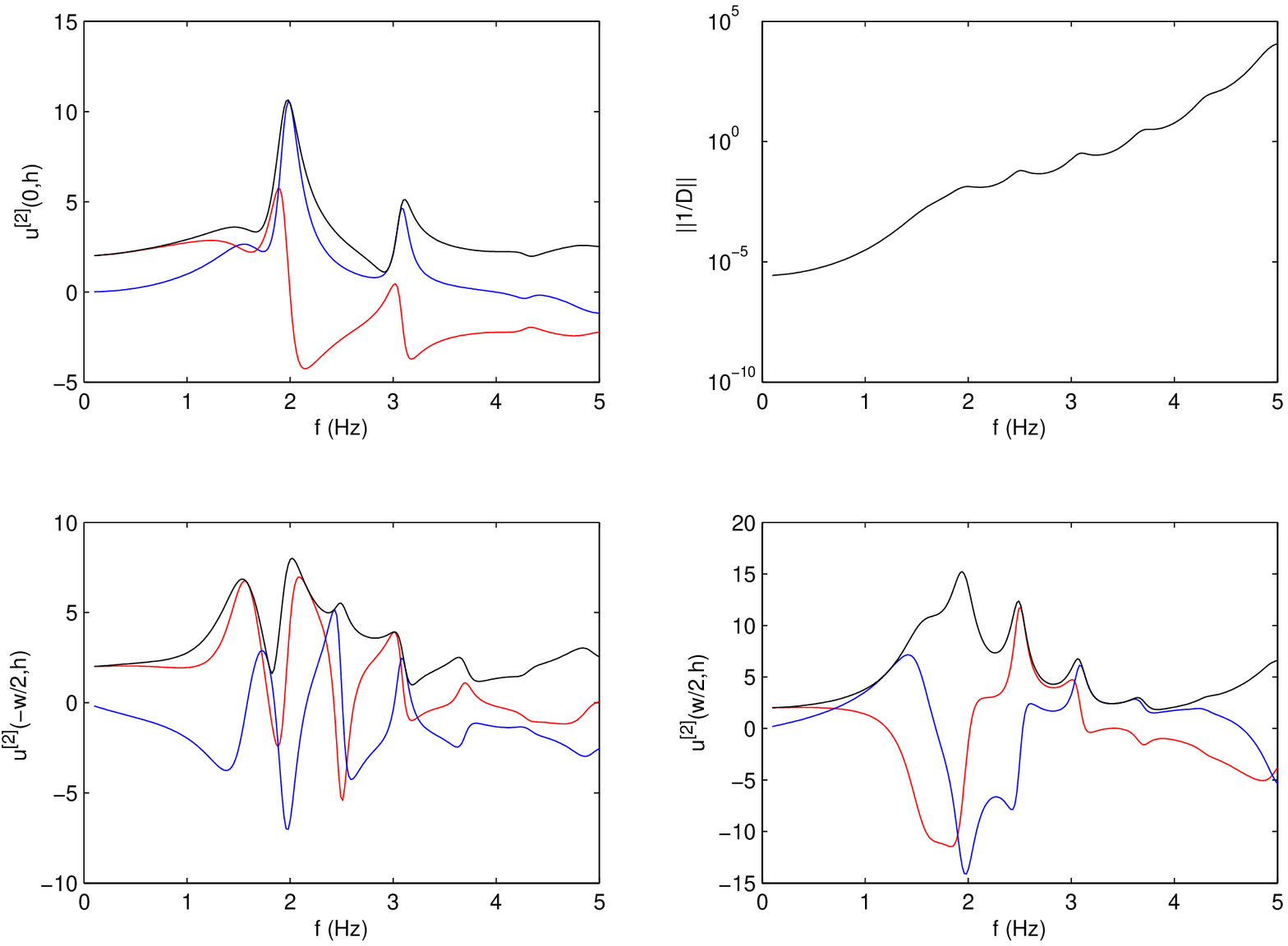}
      \label{b}
                         }

    \caption{
(a): $\beta^{[1]''}=\beta^{[2]''}=0~ms^{-1}$.\\
(b): $\beta^{[1]''}=\beta^{[2]''}=-25~ms^{-1}$.
}
    \label{mu10}
  \end{center}
\end{figure}
\clearpage
\subsection{Case $\mu^{[1]}=\mu^{[2]}=2~GPa$}
\begin{figure}[ht]
  \begin{center}
    \subfloat[]{
      \includegraphics[width=0.472\textwidth]{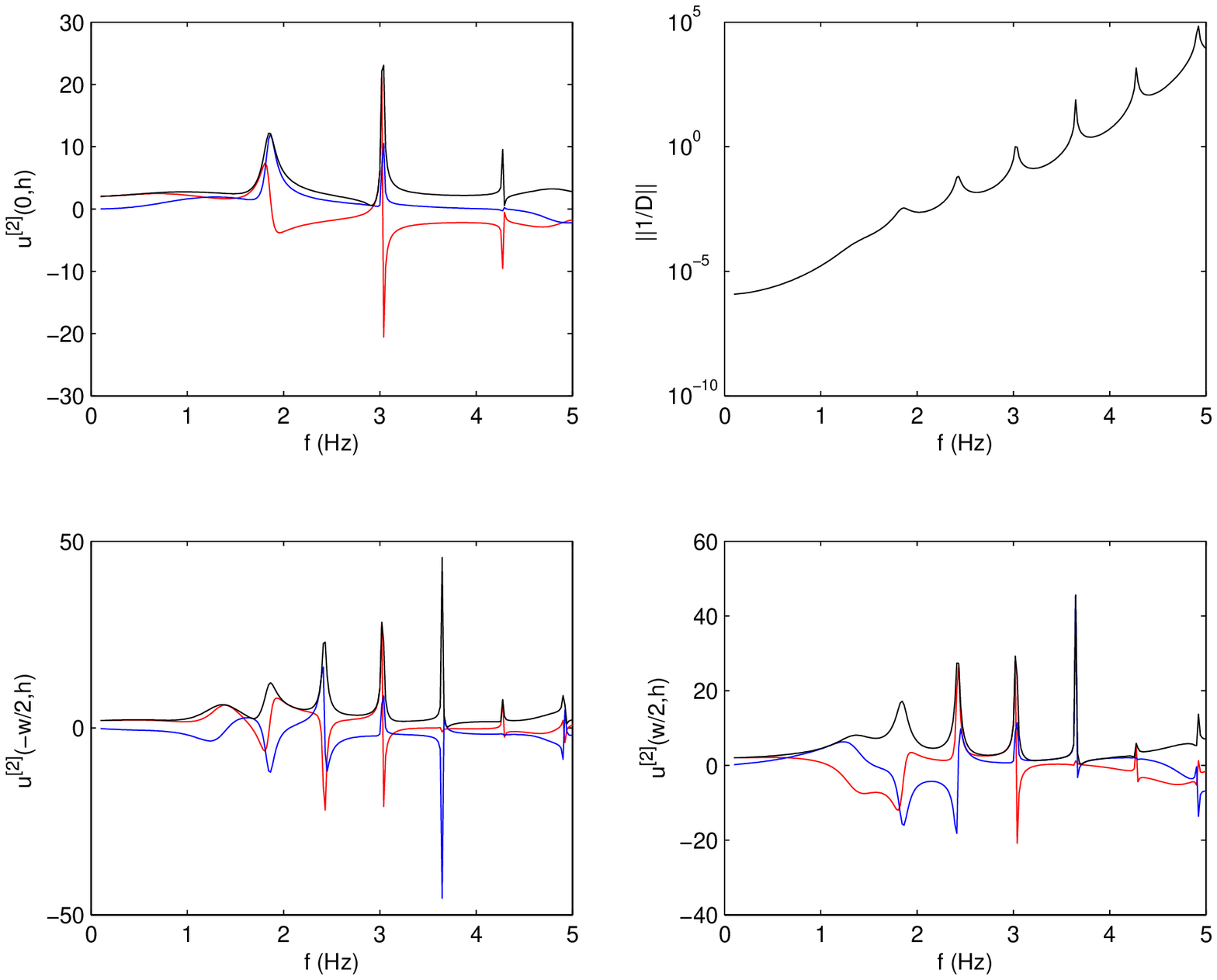}
      \label{a}
                         }
    \subfloat[]{
      \includegraphics[width=0.5\textwidth]{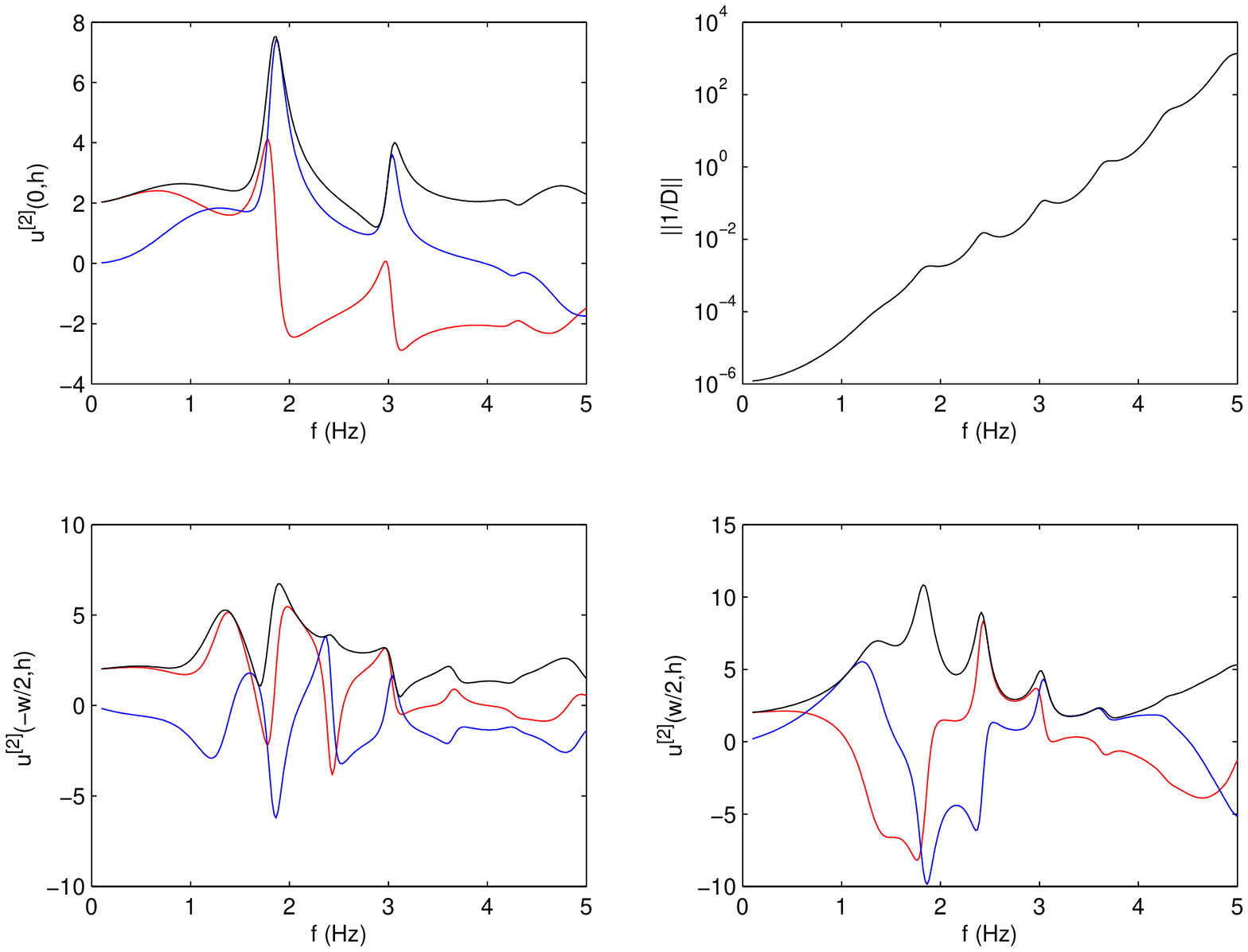}
      \label{b}
                         }

    \caption{
(a): $\beta^{[1]''}=\beta^{[2]''}=0~ms^{-1}$.\\
(b): $\beta^{[1]''}=\beta^{[2]''}=-25~ms^{-1}$.
}
    \label{mu20}
  \end{center}
\end{figure}
\clearpage
\subsection{Case $\mu^{[1]}=\mu^{[2]}=3~GPa$}
\begin{figure}[h]
  \begin{center}
    \subfloat[]{
      \includegraphics[width=0.472\textwidth]{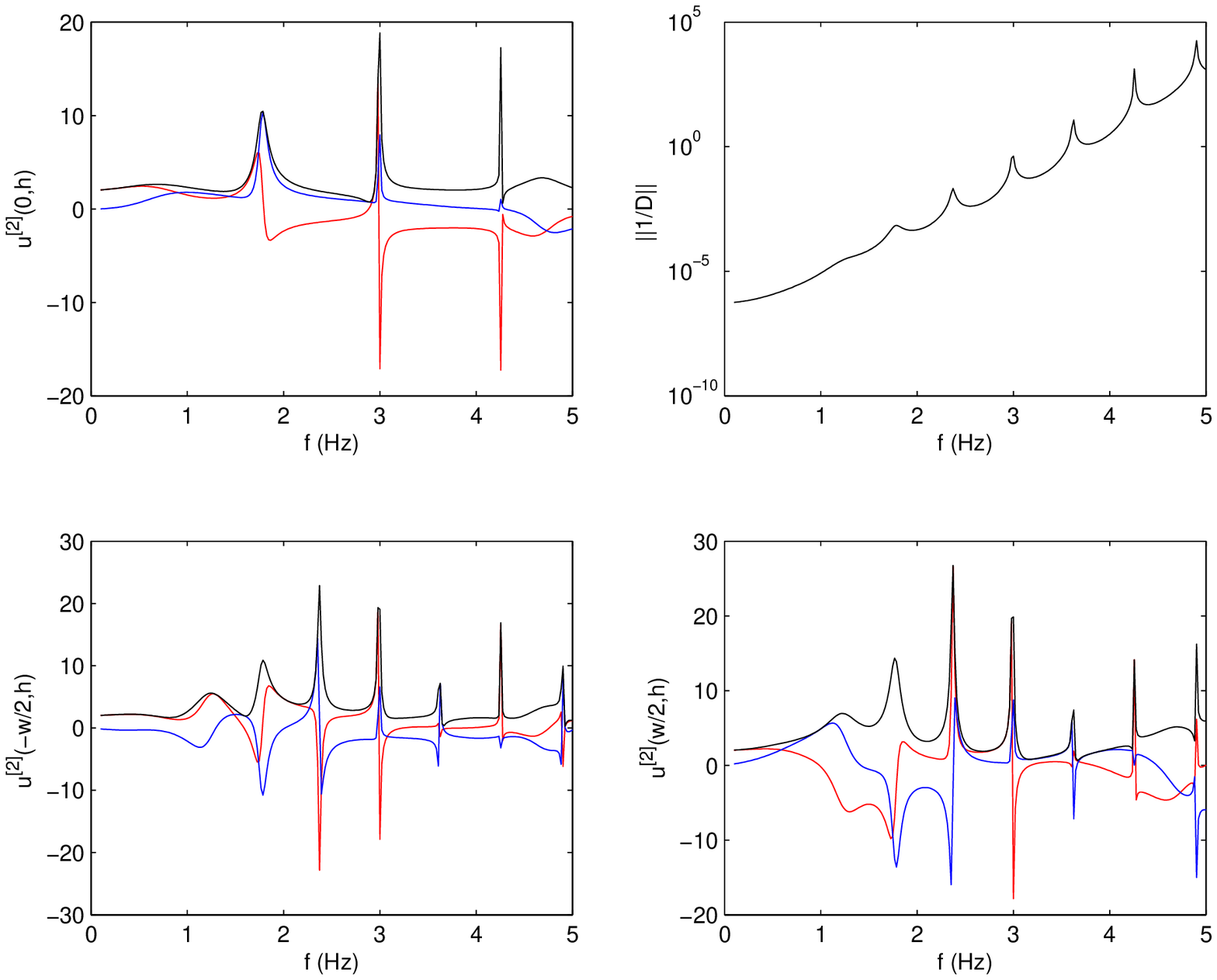}
      \label{a}
                         }
    \subfloat[]{
      \includegraphics[width=0.5\textwidth]{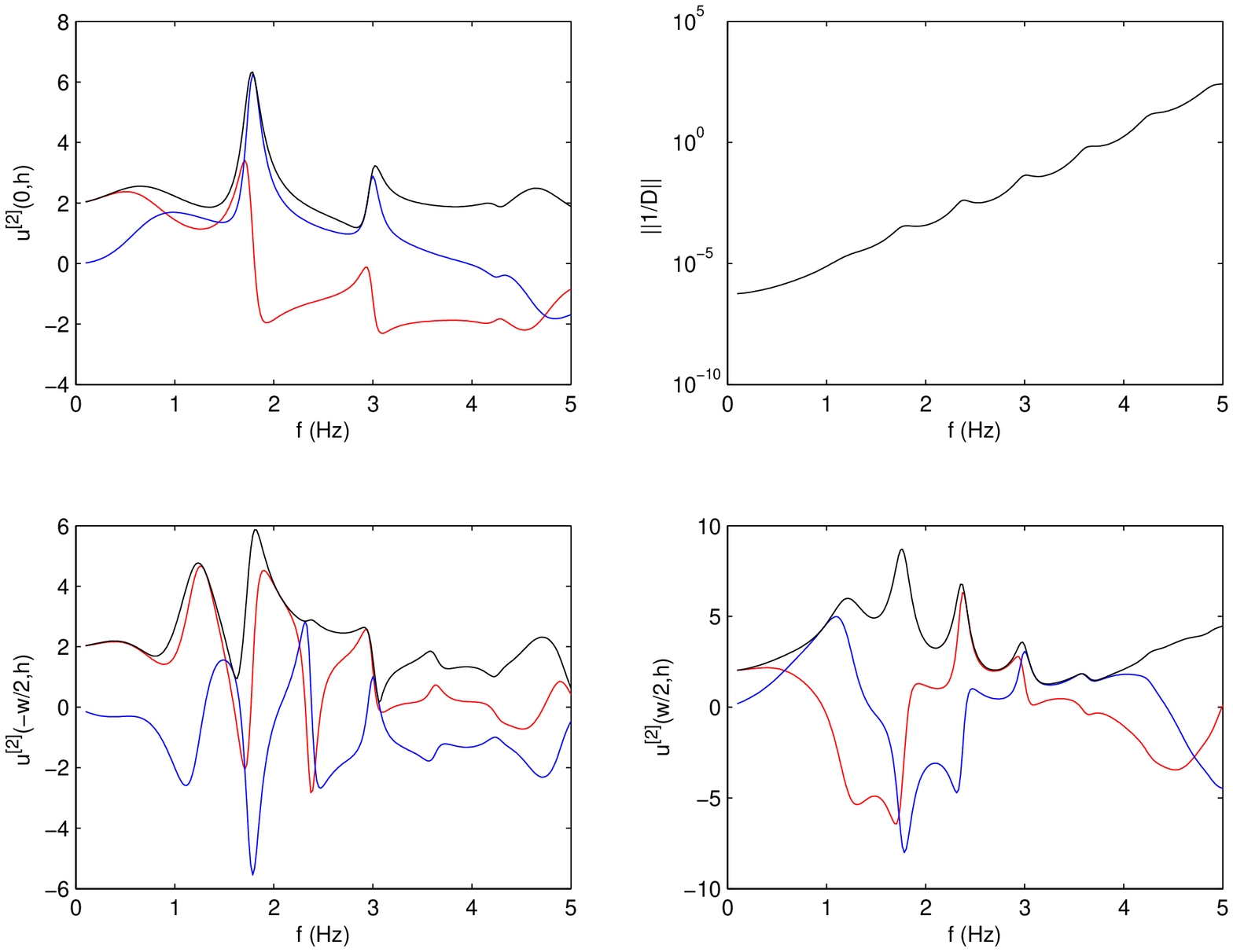}
      \label{b}
                         }

    \caption{
(a): $\beta^{[1]''}=\beta^{[2]''}=0~ms^{-1}$.\\
(b): $\beta^{[1]''}=\beta^{[2]''}=-25~ms^{-1}$.
}
    \label{mu30}
  \end{center}
\end{figure}
\clearpage
\newpage
\subsection{Discussion}
Figs. \ref{mu10}-\ref{mu30} show:\\
(i) same comment as (i) in sect. \ref{discbeta};\\
(ii) same comment as (ii) in sect \ref{discbeta};\\
(iii) that if we concentrate our attention on the different response maxima for the lossy configurations (of greatest interest in practical applications) then  their strengths, for the shear moduli $\mu^{[1]}=\mu^{[2]}=1,~2,~3~GPa$, are  8, 7 and 6 at the left-hand edge, 11.1, 7.5, and 6.7 at the center, and 16, 12 and 8 at the right-hand edge of the top segment of the protuberance, which seems (verb that is employed because the following conclusion might change with different choices of wavespeed and aspect ratio of the protuberance) to imply that the maximal response (regardless of its frequency of occurrence) decreases as the shear modulus $\mu^{[1]}=\mu^{[2]}$ increases;\\
(iv) that the maximal responses, for all the  choices of the shear modulus, is greater at the right-hand edge of the top segment of the protuberance than at the center and left-hand edge of this segment, which is related to the fact that the angle of incidence is $60^{\circ}$. Consequently, the center of the top segment is not the best location for predicting the maximal response due to obliquely-incident  seismic waves.
\section{Variation of the aspect ratio for constant $w$}\label{car}
 Some discussion has appeared in the scientific literature \cite{gb13,be14,bf14} as to the relative importance of the geometry of the stress-free surface on the one hand, and the structure of the subsurface on the other hand, to the  seismic response at a site (with one or several protuberances emerging from flat ground). Translated into the present context, this distinction is between the size and shape (essentially the aspect ratio since the protuberance is rectangular in its cross-section) on the one hand, and the composition of the protuberance (its structure is   monolayer or bilayer-like) and that (homogeneous) of the below-ground medium. The dominant opinion \cite{bf14} is that the subsurface composition is the main contributing factor.

 As most studies on this issue (especially those based on empirical evidence) do not (or cannot) distinguish between these relative contributions, we decided, in the present section, to change the geometry of the protuberance (more precisely, its aspect ratio $w/h$, with $w$ fixed)  while maintaining the composition constant (and uniform) throughout the subsurface, the latter comprising both the protuberance and the basement. Thus, in all the following figures, we chose: $w=500~m$, $\mu^{[0]}=\mu^{[1]}=\mu^{[2]}=6.85~GPa$ and  $\beta^{[0]}=\beta^{'[1]}=\beta^{'[2]}=1629.4~ms^{-1}$ with variable $h$ with  $\beta^{''[1]}=\beta^{''[2]}=0~ms^{-1}$ in the lossless medium case and $\beta^{''[1]}=\beta^{''[2]}=-20~ms^{-1}$  in the lossy medium case.

Since, what appeared to us to be of greatest interest is how the variation of shape affects the {\it resonant} seismic response, we decided to carry out the comparisons for the first four ($f^{R}_{1},~f^{R}_{2},~f^{R}_{3},~f^{R}_{4}$) resonant frequencies, while keeping in mind that these resonant frequencies change with changing aspect ratio. Note, that if we are to share the dominant opinion, the  strong (i.e., resonant) response in the hill or mountain should change weakly or not at all as its aspect ratio changes for constant composition. The purpose of the following figures is to find out if this opinion is justified.
\subsection{Examples of displacement transfer function maps at resonant frequencies}
\begin{figure}[ht]
  \begin{center}
    \subfloat[]{
      \includegraphics[width=0.33\textwidth]{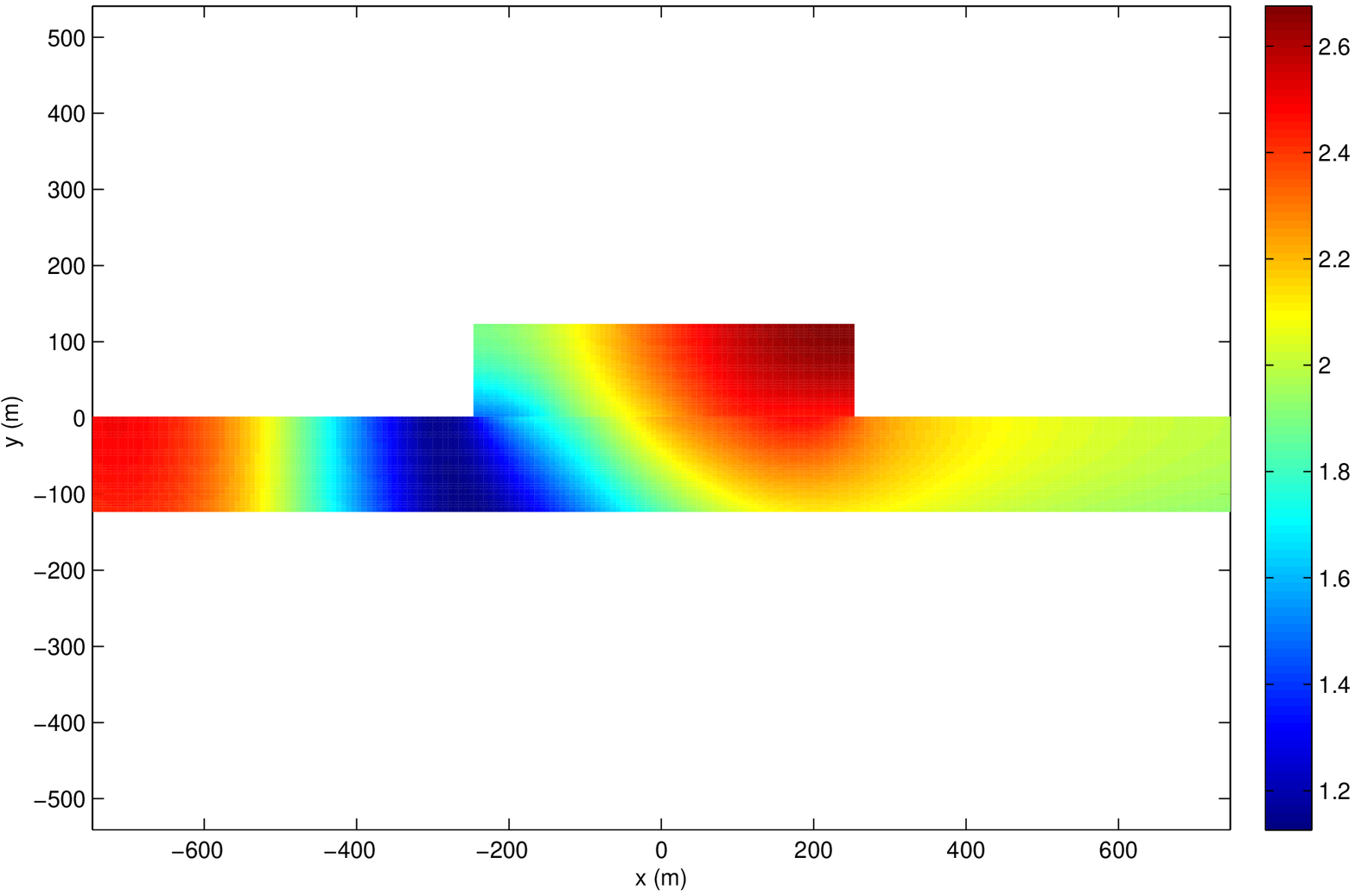}
      \label{a}
                         }
    \subfloat[]{
      \includegraphics[width=0.31\textwidth]{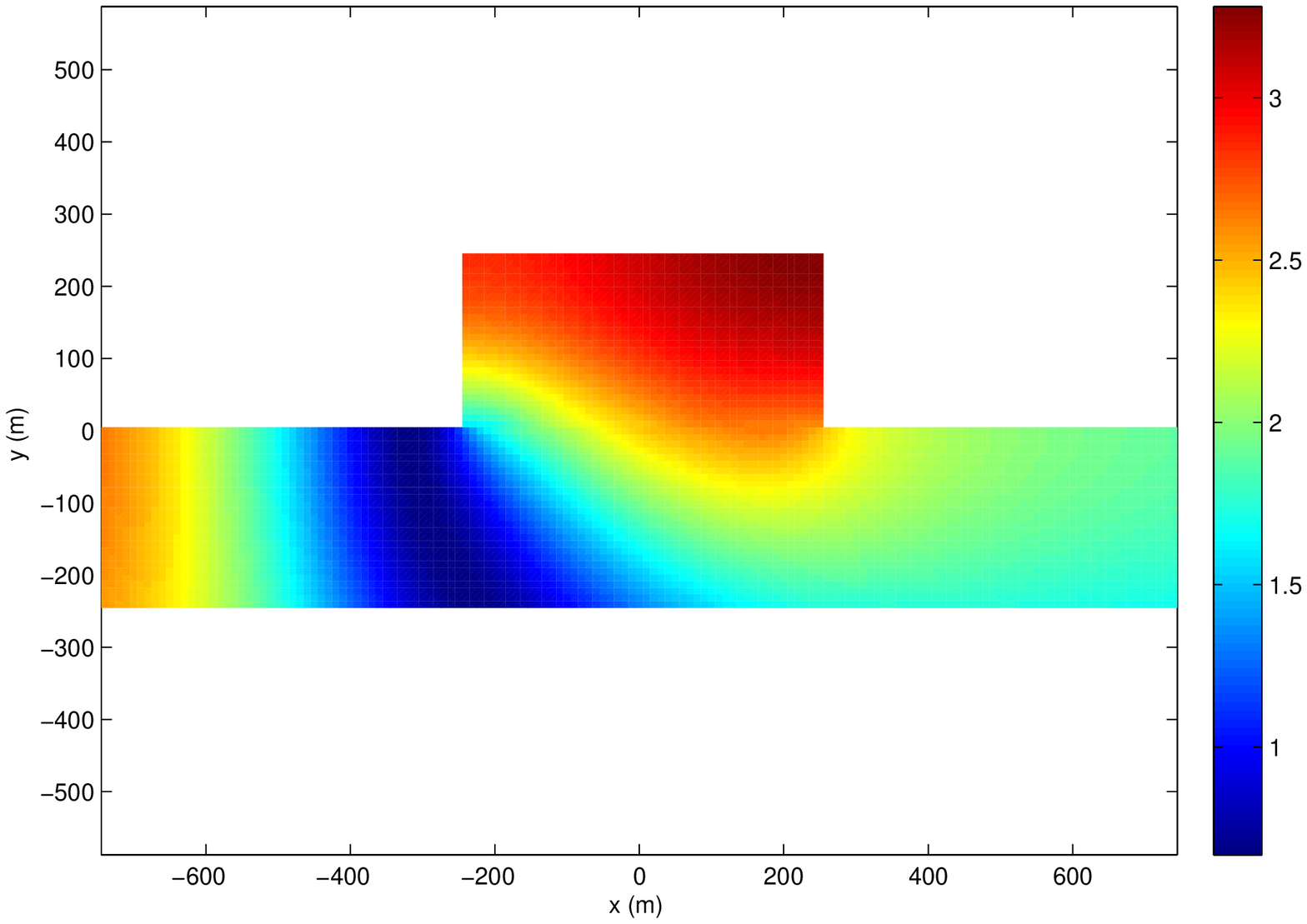}
      \label{b}
                         }
 \subfloat[]{
      \includegraphics[width=0.32\textwidth]{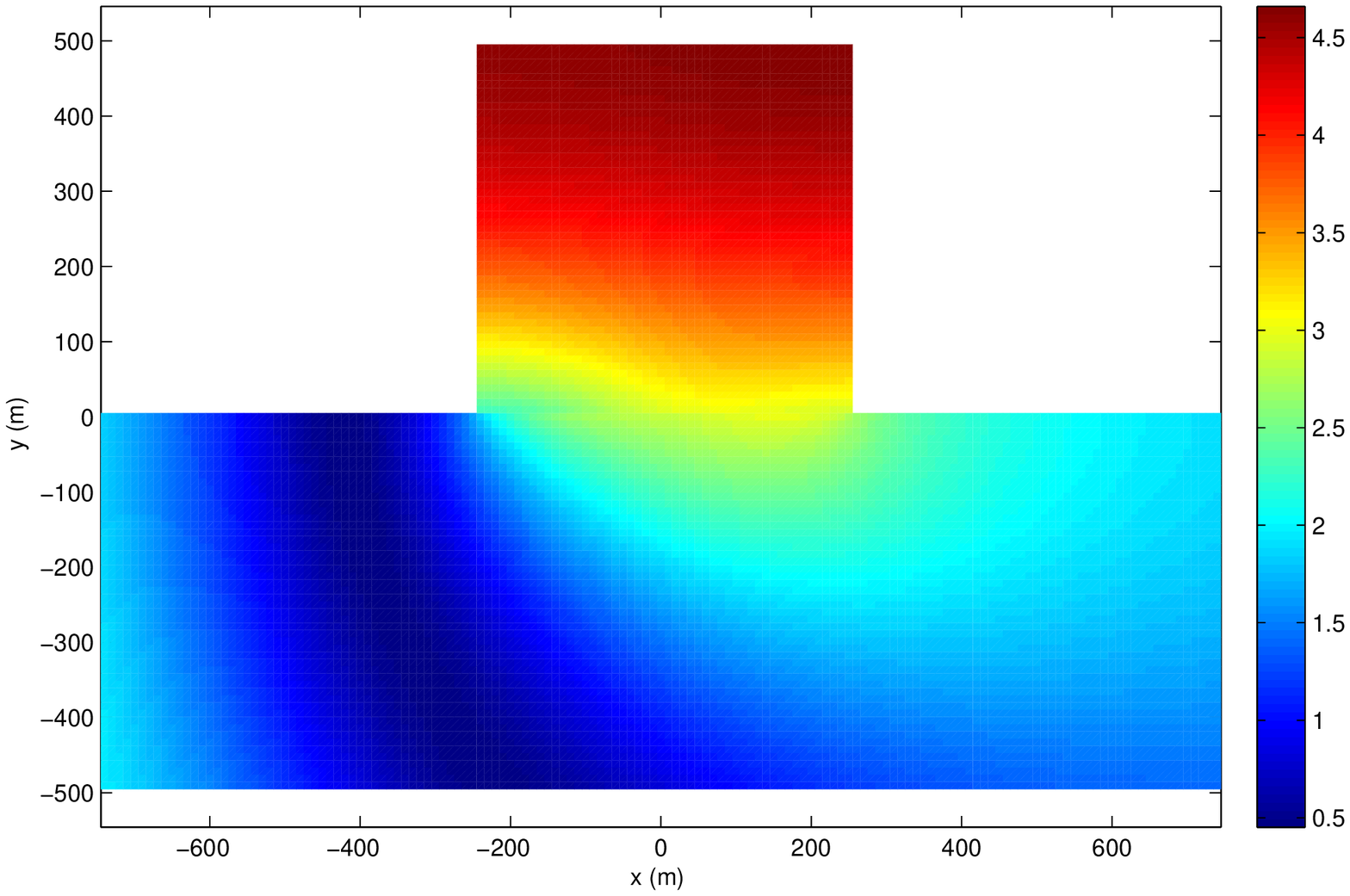}
      \label{c}
                        }
    \caption{
(a): $\theta^{i}=60^{\circ}$, $h=125~m$, $f^{R}_{1}=0.9840~Hz$, Max=2.9(2.62).\\
(b): $\theta^{i}=60^{\circ}$, $h=250~m$, $f^{R}_{1}=0.7592~Hz$, Max=3.2(3.2).\\
(c): $\theta^{i}=60^{\circ}$, $h=500~m$, $f^{R}_{1}=0.4845~Hz$, Max=4.6(4.3).
}
    \label{car03}
  \end{center}
\end{figure}

\begin{figure}[ht]
  \begin{center}
    \subfloat[]{
      \includegraphics[width=0.31\textwidth]{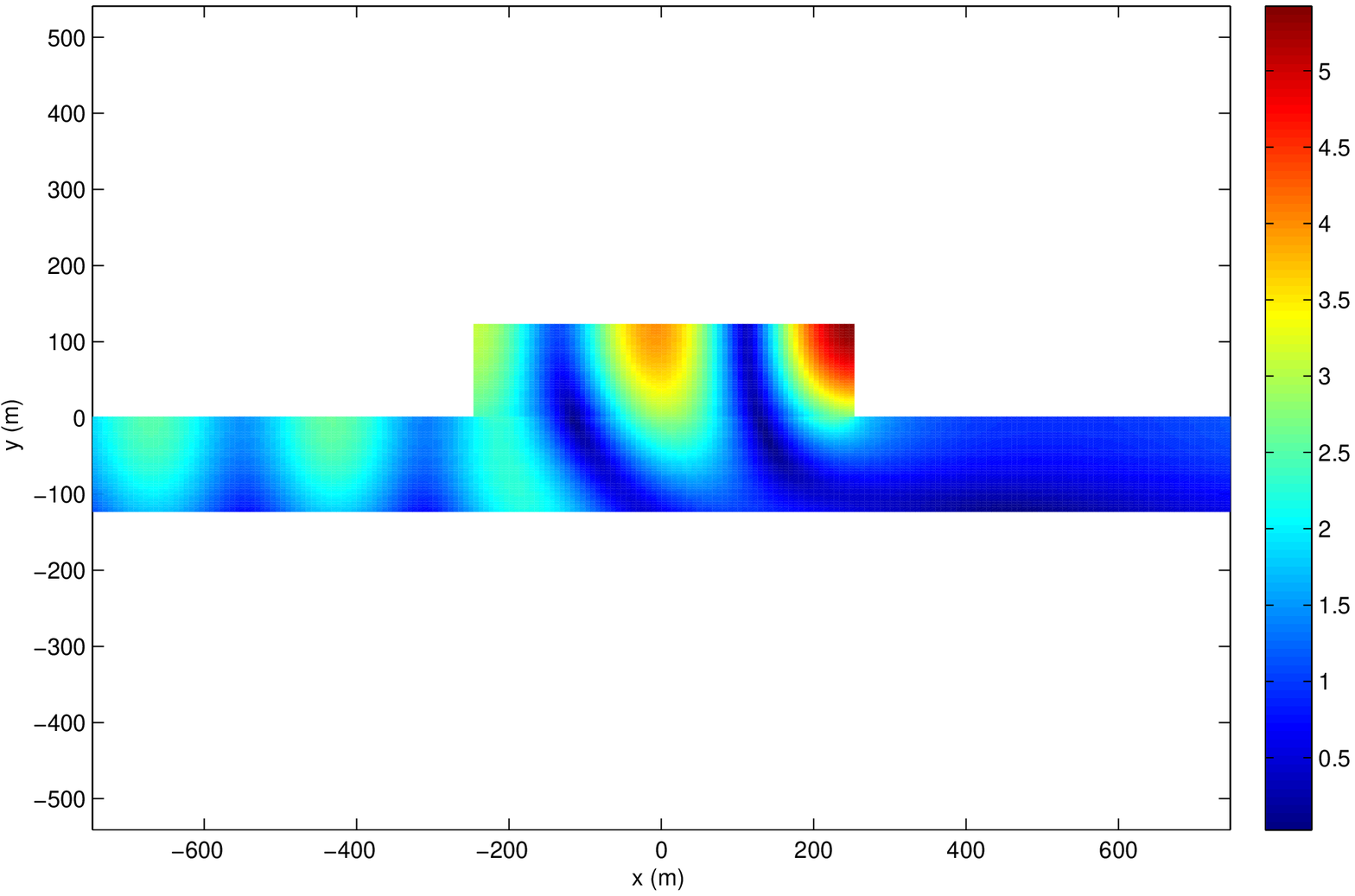}
      \label{a}
                         }
    \subfloat[]{
      \includegraphics[width=0.33\textwidth]{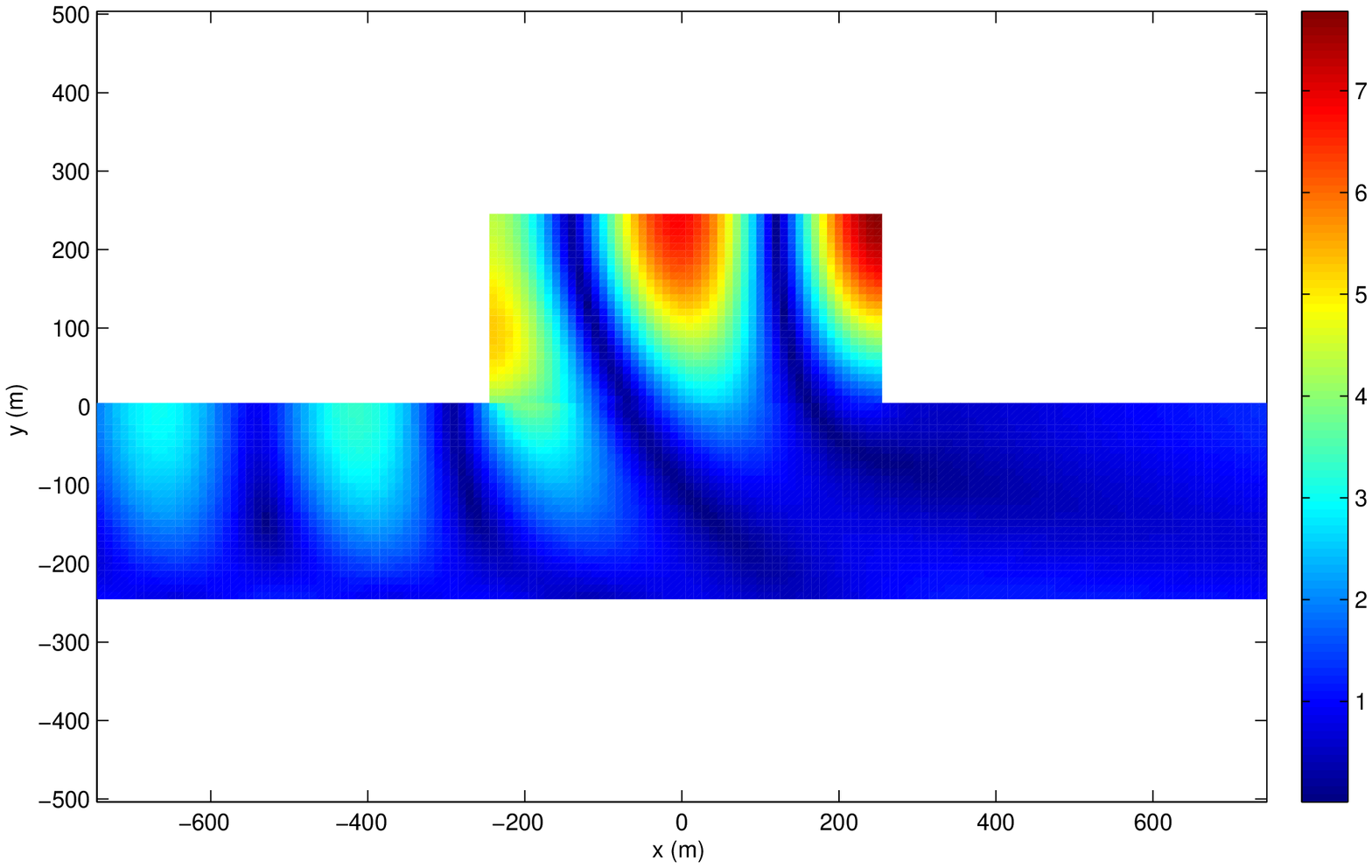}
      \label{b}
                         }
 \subfloat[]{
      \includegraphics[width=0.29\textwidth]{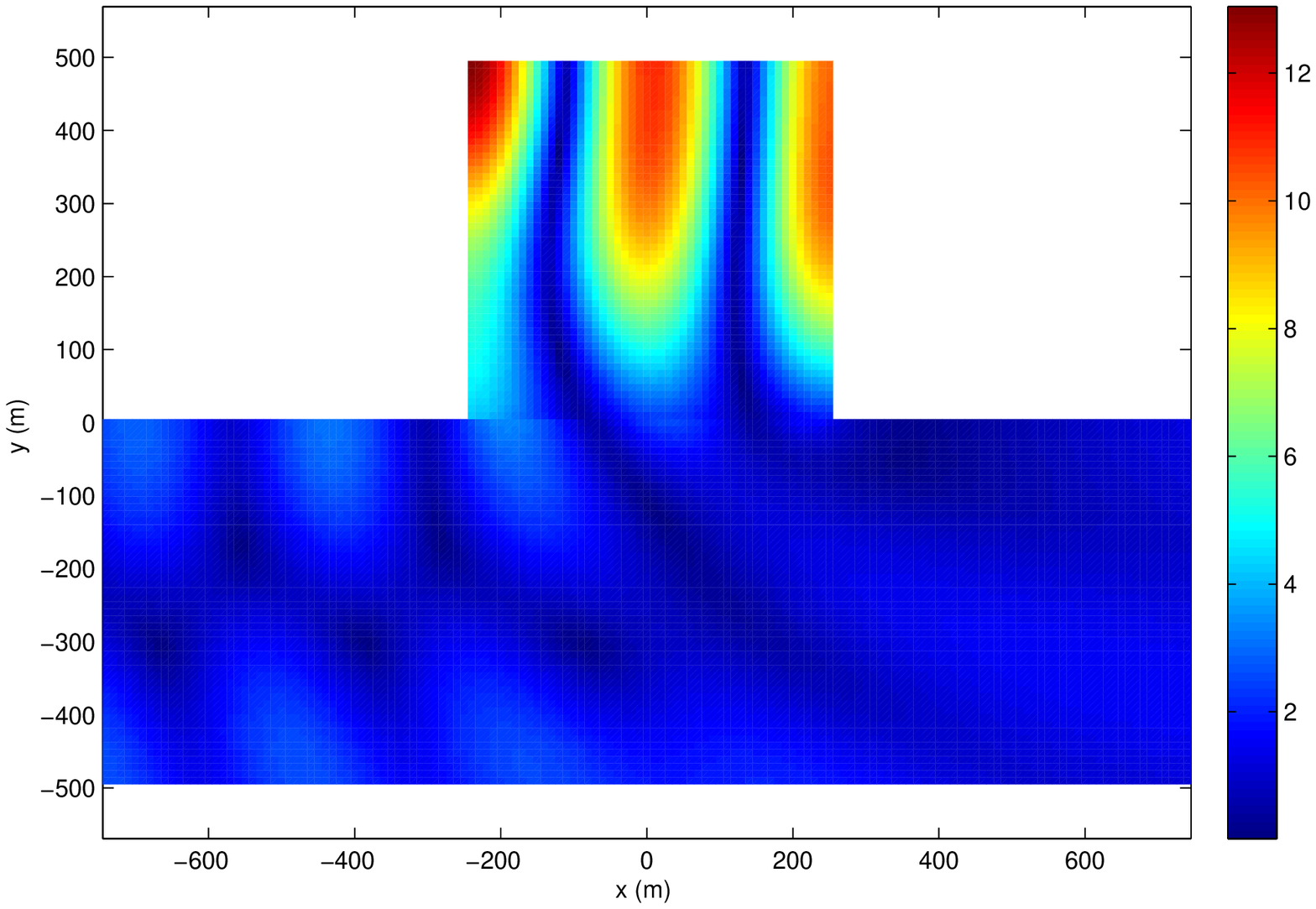}
      \label{c}
                        }
    \caption{
(a): $\theta^{i}=60^{\circ}$, $h=125~m$, $f^{R}_{3}=3.681~Hz$, Max=5.4(4.7).\\
(b): $\theta^{i}=60^{\circ}$, $h=250~m$, $f^{R}_{3}=3.482~Hz$, Max=7.5(5.5).\\
(c): $\theta^{i}=60^{\circ}$, $h=500~m$, $f^{R}_{3}=3.332~Hz$, Max=13(4.2).
}
    \label{car09}
  \end{center}
\end{figure}

Figs. \ref{car03} and \ref{car09} depict the displacement field maps at the first and third resonant frequencies for $\theta^{i}=60^{\circ}$ incidence. The left-hand, middle and right-hand panels pertain to $h=125,~250,~500~m$ respectively. Since the figures are reduced considerably making it difficult to read the numbers, we have indicated, in their captions, the maximum value of $\|u\|$ (within the protuberance, and always for $a^{i}=1$, so that $\|u\|$  is identical (not unit-wise) to the modulus of the transfer function) by the symbol 'Max'. We have also indicated the maximum value of the displacement field within the protuberance (the number within parenthesis after the first number which applies to the lossless case) when there exists a constant (in all the treated cases) material loss which is accounted for by $\beta^{[1]}=\beta^{[2]}=-20~ms^{-1}$.
\clearpage
\newpage
\subsection{Examples of the displacement transfer function maps at off-resonance frequencies}
\begin{figure}[ht]
  \begin{center}
    \subfloat[]{
      \includegraphics[width=0.22\textwidth]{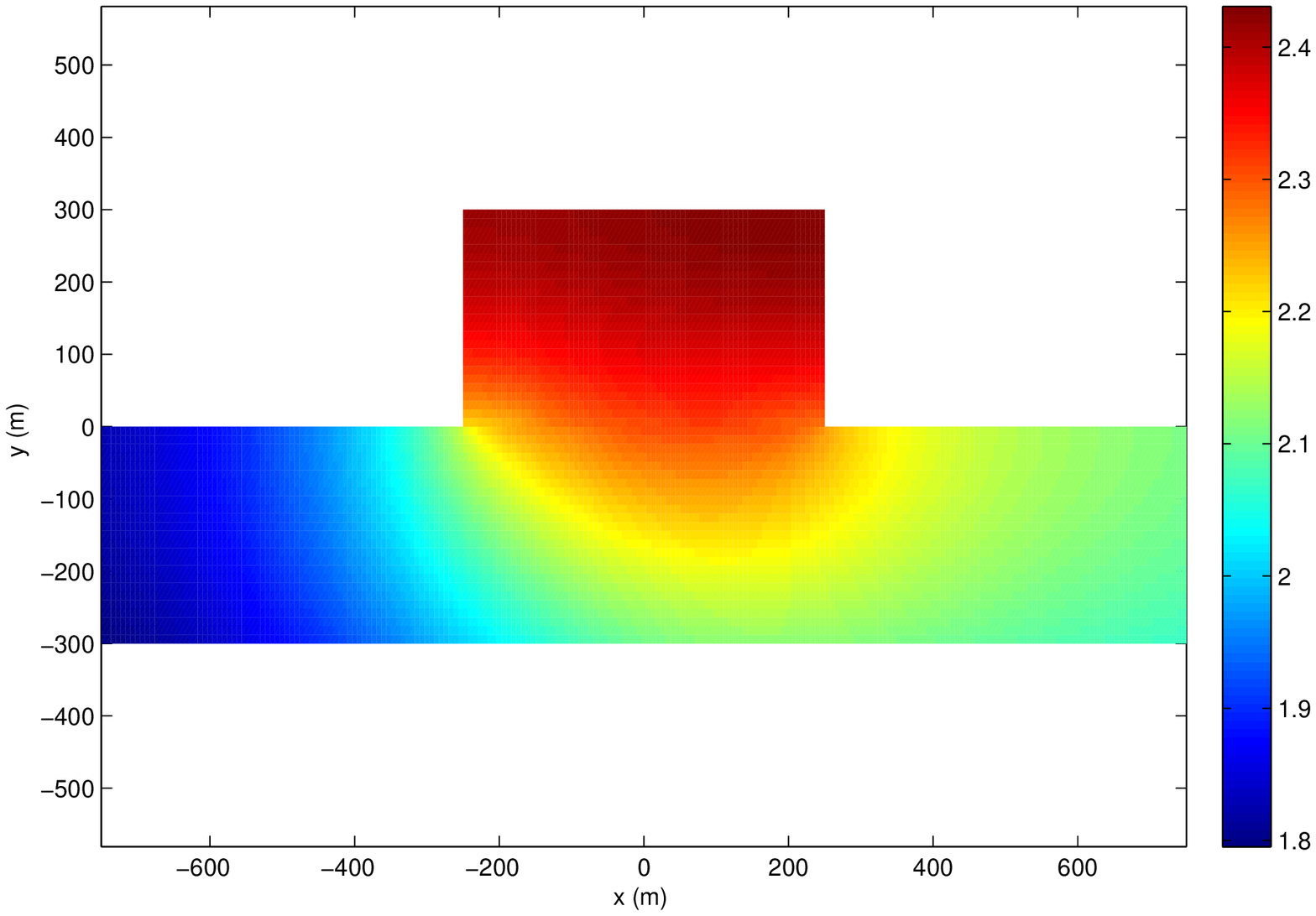}
      \label{a}
                         }
    \subfloat[]{
      \includegraphics[width=0.22\textwidth]{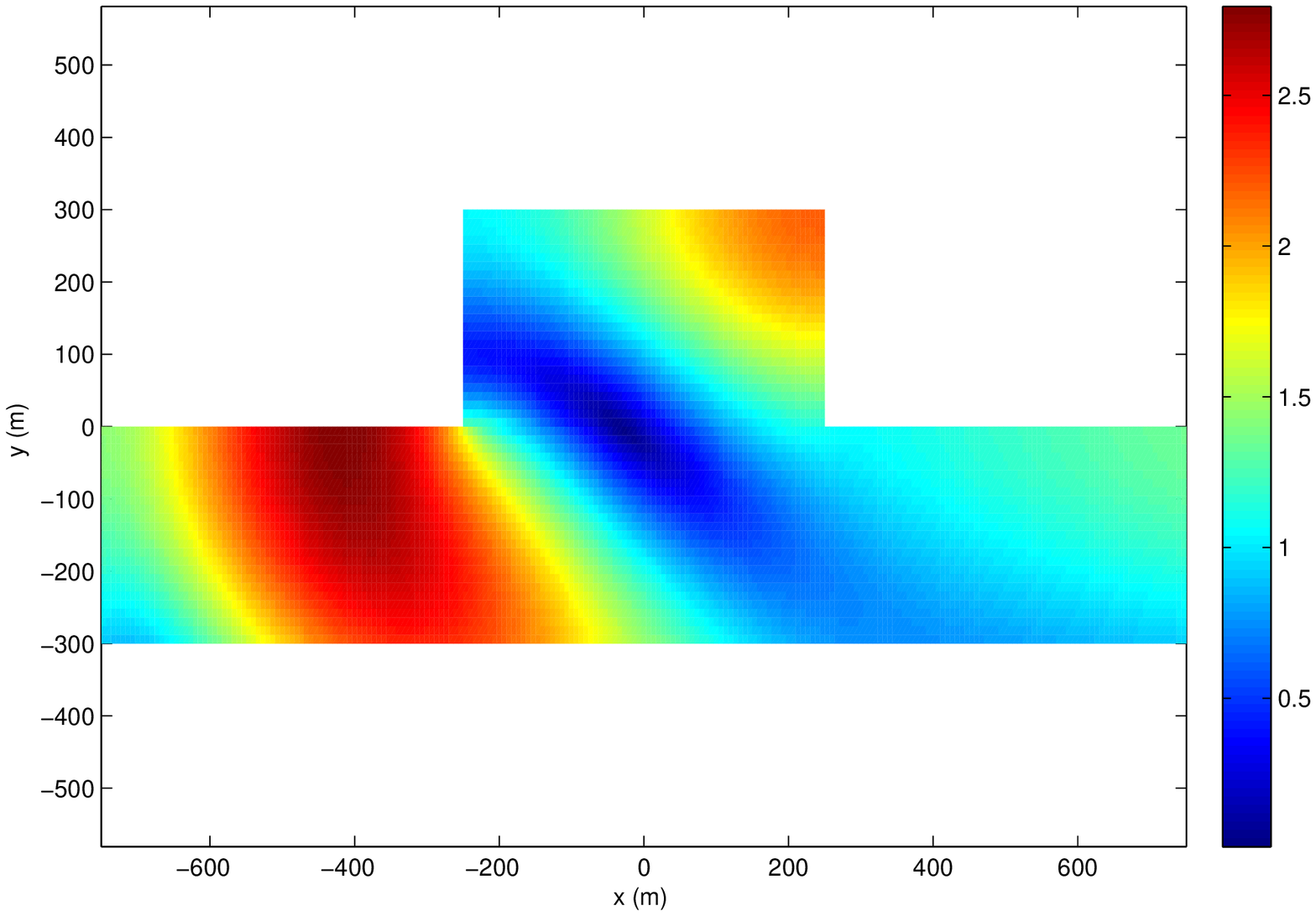}
      \label{b}
                         }
 \subfloat[]{
      \includegraphics[width=0.22\textwidth]{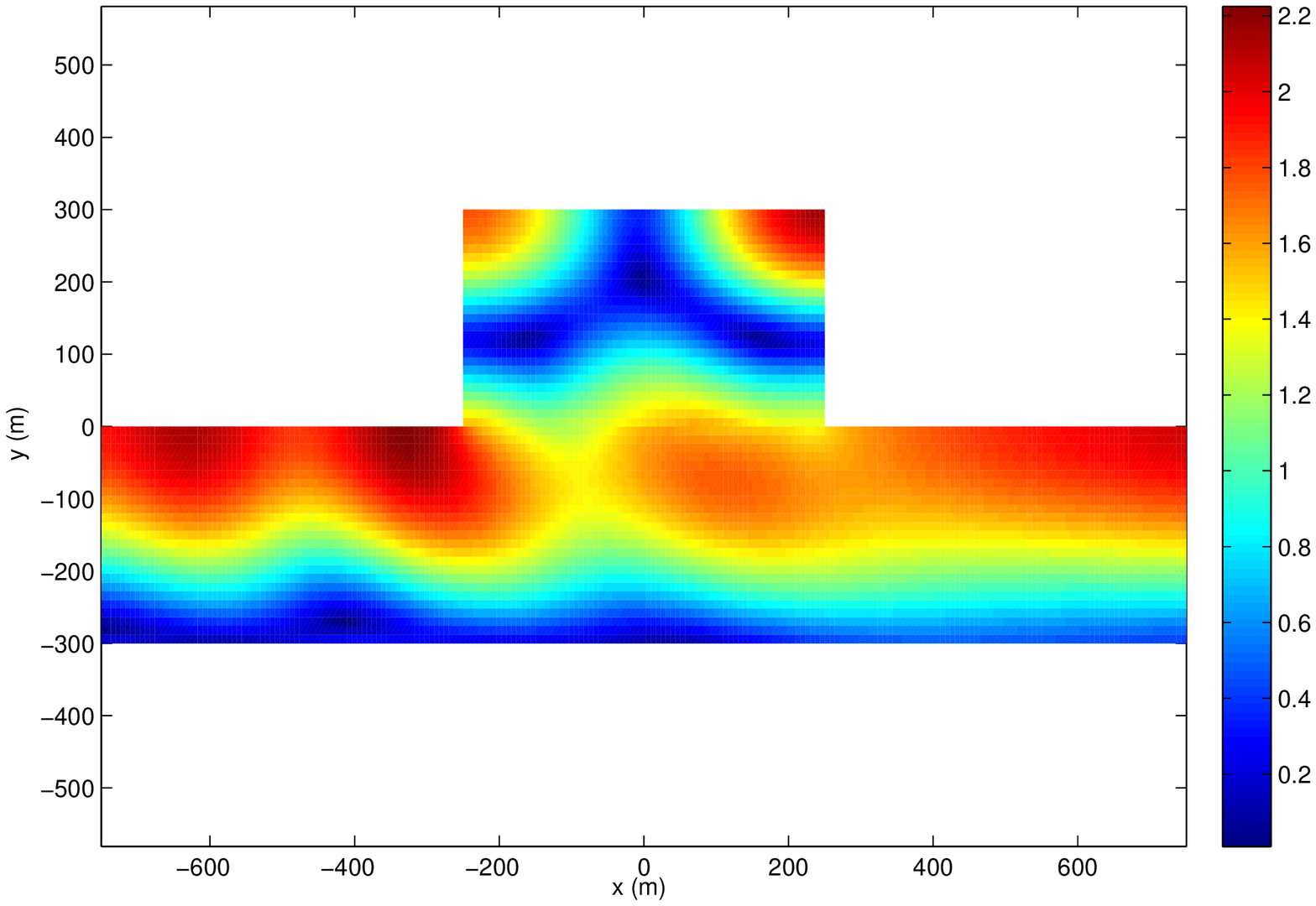}
      \label{c}
                        }
 \subfloat[]{
      \includegraphics[width=0.22\textwidth]{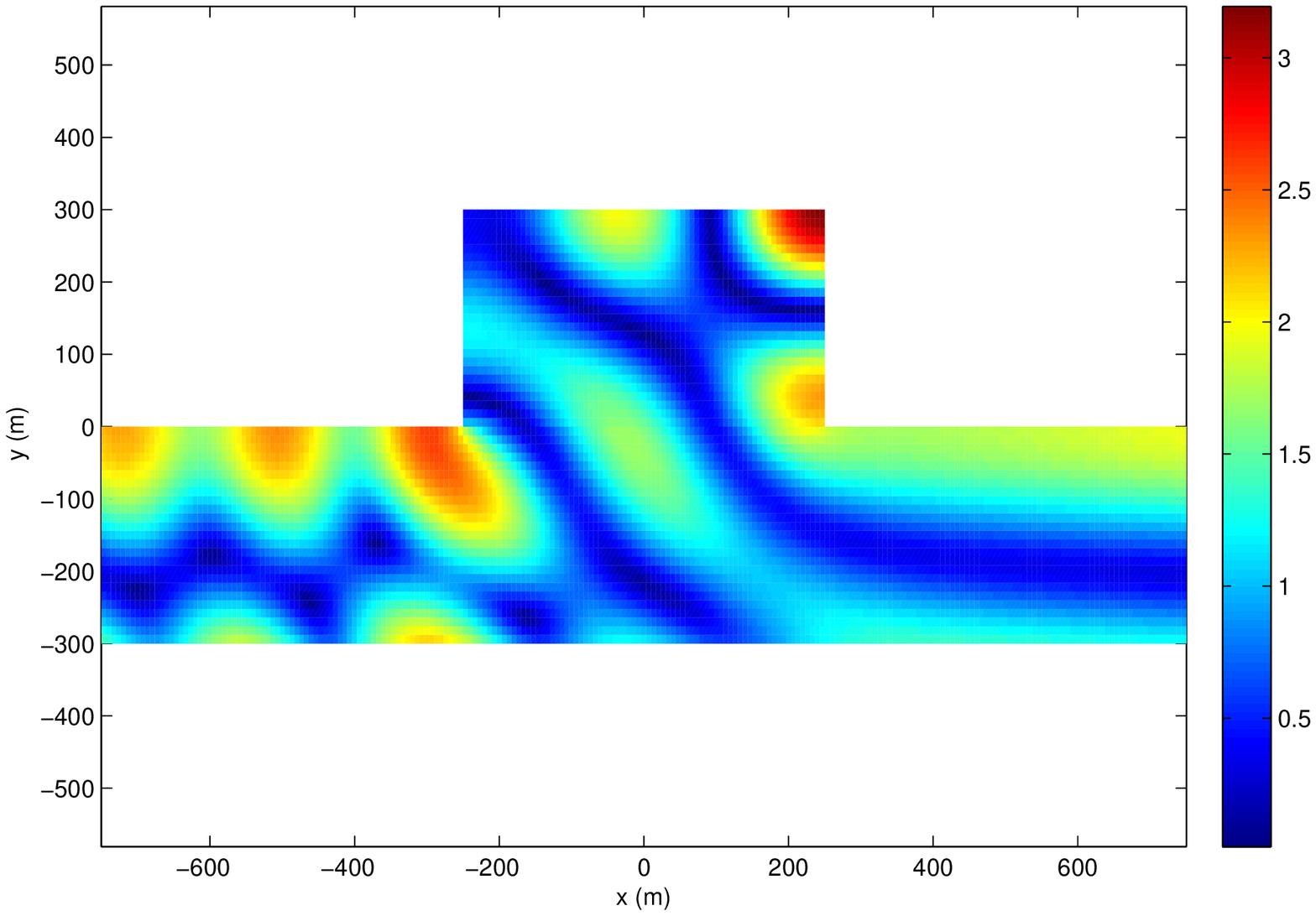}
      \label{d}
                        }
    \caption{
(a): $\theta^{i}=60^{\circ}$, $h=300~m$,  $f=0.3~Hz$, Max=2.45.\\
(b): $\theta^{i}=60^{\circ}$, $h=300~m$,  $f=1.3~Hz$, Max=2.7.\\
(c): $\theta^{i}=60^{\circ}$, $h=300~m$,  $f=2.8~Hz$, Max=2.2.\\
(d): $\theta^{i}=60^{\circ}$, $h=300~m$,  $f=4.0~Hz$, Max=3.1.
}
    \label{or3}
  \end{center}
\end{figure}
\clearpage
\newpage
Fig. \ref{or3} shows, by comparison with figs. \ref{car03} and \ref{car09}, that:\\
 (a)  large-scale seismic response in a mountain (like that in a hill) is possible only at, or in the neighborhood of, (shape-) resonant frequencies,\\
 (b) at resonance frequencies, the field is concentrated within the protuberance, so as to be maximal  either at hot spots (HS) (each of the latter defined as  a bull's eye target-like feature with a core that is dark orange-to-red on the scale of the color map of e.g.  fig. \ref{car09}) or (e.g.  as in fig. \ref{car03}) the field is maximal and spread out horizontally and in uniform manner at the top of the protuberance (UTHS) standing for 'uniform top hot spot', and\\
 (c) at off-resonance frequencies, the field in the underground, and/on the free-surface flanks of the mountain (i.e., $u^{[0]}(|x|>w/2,y\le 0)$, is often larger than the field at a point (usually the topmost or midpoint) of the summit  of the protuberance (e.g., $u^{[2]}(0,h)$ so that if, as is often the case in the seismic engineering community \cite{fa91,az93,bm14}, the ratio $\mathfrak{R}=\frac{\|u^{[2]}(0,h)\|}{\|u^{[0]}(|x|>w/2,0)\|}$ is adopted as the indication of amplified (for $\mathfrak{R}>1$ or de-amplified (for $\mathfrak{R}<1$) response, then one sees, at off-resonance frequencies,  that this ratio indicates almost systematically-large amplification although the field is almost systematically un-amplified within and on top of our rectangular protuberance.
\subsection{Analysis of the transfer function maps}
Let us denote the transfer function {\it within the protuberance} of width $w$ and height $h$, for a given order $j$ (=1,2,3,4) of resonance, angle of incidence $\theta^{i}$, by $\mathcal{T}(f^{R}_{j},\theta^{i},w,h)$ and by 'Max' the maximum value of $\mathcal{T}(f^{R}_{j},\theta^{i},w,h)$.

The object is here to compare the three functions $\mathcal{T}(f^{R}_{j},\theta^{i},w,h^{k})~;~k=1,2,3$. Since this  would require the comparison of a large number of graphs each of which occupies a great deal of storage space, we decided rather to carry out both a   coarse qualitative and a succinct quantitative comparison of  $\mathcal{T}$ for each $h$ triplet. The method of qualitative comparison (the results of which occupy the column 'comments' in the table) is explained in a few subsequent lines, whereas the method of quantitative comparison is simply that of the comparison of three numbers (the results of which occupy the 'Max' columns), also explained in the next lines.

In the table below, we depict the  maxima (designated by the symbol 'Max', and taken from the color maps of $T$) of the transfer function $T(x,y;f)~;(x,y)\in\Omega_{1}+\Omega_{2}$, at both resonant (R) and non-resonant (NR) frequencies, for the three incident angles:  $\theta^{i}=~0^{\circ},~40^{\circ},~80^{\circ}$. This is done for both the lossless and lossy hills (only for the largest incident angle).

The comments (amounting to the above-mentioned qualitative comparison) in this table refer mostly to the number ($\ge 1$) of  'hot spots' (HS), or the (necessarily-single) 'uniform top hot spot' (UTHS), within the hill at the resonance frequencies. Since, as in fig. \ref{or3}, relative to  the response map at off-resonant frequencies, the transfer function is close to its value (=2) in the absence of the hill, we do not qualify it  (i.e., either HS or UTHS) at all. Note that the qualifications HS or UTHS are relative to the non-lossy hills, and are essentially applicable as well to the corresponding lossy hills.
%
\begin{center}
\begin{tabular}{|c|c|c|c|c|c|c|}
  \hline
  $f_{j}^{R}$ &  $\theta^{i}(^{\circ})$ & $h(m)$ & $f(Hz)$ & lossless  & lossy   & comments\\
  or $f^{NR}$ &                         &        &         & Max       & Max     & \\
  \hline
  $f_{1}^{R}$ & 0  & 125 & 0.9840 & 2.45  &  & UTHS\\
  "           & "  & 250 & 0.7592 & 3.30  &  & UTHS\\
  "           & "  & 500 & 0.4845 & 4.60  &  & UTHS\\
  \hline
  "           & 30 & 125 & 0.9841 & 2.90  &  & 1 HS\\
  "           & "  & 250 & 0.7592 & 3.30  &  & 1HS\\
  "           & "  & 500 & 0.4845 & 4.60  &  & UTHS\\
  \hline
  "           & 60 & 125 & 0.9841 & 2.90  & 2.62 & 1 HS\\
  "           & "  & 250 & 0.7592 & 3.20  & 3.20 & 1 HS\\
  "           & "  & 500 & 0.4845 & 4.60  & 4.30 & UTHS\\
  \hline
  \hline
  $f_{2}^{R}$ & 0  & 125 & 2.233  & 2.60  &  & 0 HS \\
  "           & "  & 250 & 1.958  & 2.60  &  & 0 HS \\
  "           & "  & 500 & 1.758  & 2.20  &  & UTHS\\
  \hline
  "           & 30 & 125 & 2.233  & 3.70  &  & 1 HS\\
  "           & "  & 250 & 1.958  & 5.70  &  & 1 HS\\
  "           & "  & 500 & 1.758  & 8.50  &  & 1 HS\\
  \hline
  "           & 60 & 125 & 2.233  & 4.40  & 4.2 & 2 HS\\
  "           & "  & 250 & 1.958  & 6.70  & 5.7 & 2 HS\\
  "           & "  & 500 & 1.758  & 11.5  & 7.0 & 2 HS\\
  \hline
  \hline
  $f_{3}^{R}$ & 0  & 125 & 3.681  & 4.10  &  & 2 HS\\
  "           & "  & 250 & 3.482  & 2.70  &  & 3 HS\\
  "           & "  & 500 & 3.332  & 2.80  &  & 5 HS\\
  \hline
  "           & 30 & 125 & 3.681  & 4.10  &  & 1 HS\\
  "           & "  & 250 & 3.482  & 5.20  &  & 1 HS\\
  "           & "  & 500 & 3.332  & 6.40  &  & 3 HS\\
  \hline
  "           & 60 & 125 & 3.681  & 5.40  & 4.7 & 2 HS\\
  "           & "  & 250 & 3.482  & 7.50  & 5.5 & 2 HS\\
  "           & "  & 500 & 3.332  & 13.0  & 4.2 & 3 HS\\
  \hline
  \hline
  $f_{4}^{R}$ & 0  & 125 & 5.255  & 2.30  &  & 0 HS \\
  "           & "  & 250 & 5.030  & 2.60  &  & 0 HS \\
  "           & "  & 500 & 4.980  & 3.10  &  & 0 HS \\
  \hline
  "           & 30 & 125 & 5.255  & 4.10  &  & 1 HS\\
  "           & "  & 250 & 5.030  & 5.60  &  & 2 HS\\
  "           & "  & 500 & 4.980  & 6.40  &  & 1 HS\\
  \hline
  "           & 60 & 125 & 5.255  & 5.60  & 3.2 & 2 HS\\
  "           & "  & 250 & 5.030  & 7.10  & 3.6 & 4 HS\\
  "           & "  & 500 & 4.980  & 9.30  & 3.7 & 4 HS\\
  \hline
  \hline
  $f^{NR}$    & 0  & 300 & 0.3    & 2.45  &  & \\
  "           & "  & "   & 1.3    & 2.70  &  & \\
  "           & "  & "   & 2.8    & 2.10  &  & \\
  "           & "  & "   & 4.0    & 2.70  &  & \\
  \hline
  "           & 30 & "   & 0.3    & 2.43  &  & \\
  "           & "  & "   & 1.3    & 2.70  &  & \\
  "           & "  & "   & 2.8    & 2.60  &  & \\
  "           & "  & "   & 4.0    & 3.60  &  & \\
  \hline
  "           & 60 & "   & 0.3    & 2.45  &  & \\
  "           & "  & "   & 1.3    & 2.70  &  & \\
  "           & "  & "   & 2.8    & 2.20  &  & \\
  "           & "  & "   & 4.0    & 3.10  &  & \\
  \hline

\end{tabular}
\newline
\end{center}
%
\subsection{Discussion of the results in  the table}
The results in the above table call for the following comments:\\\\
(1) as one would expect, $f_{j}^{R}$ decreases as $h$ increases, more so for low $j$ than for high $j$;\\
(2) the resonant response is generally greatest at the highest incident angle;\\
(3) the responses are {\it qualitatively} the same (already perceptible in figs. \ref{car03} and \ref{car09}) for all three values of $h$ and  most incident angles and resonance frequencies;\\
(4) however, there are significative {\it quantitative} differences in response as a function of $h$, with, in most cases,  an increase of 'Max' with $h$, principally at the larger incident angles;\\
(5) the introduction of material losses in the protuberance can result in important reductions of resonant response, particularly for the largest $h$;\\
(5) no significant amplification of response is obtained at off-resonance frequencies.\\

Hence we conclude, on the basis of these results, and recalling that they were obtained for fixed composition of the configuration, that variations of the aspect ratio (for constant $w$) produce little qualitative, but large quantitative, changes of resonant response within the protuberance. The question of whether these changes are larger or smaller than those due to compositional variations cannot be answered on the sole basis of these results.
\section{Variation of the aspect ratio for constant $h$}
In this set of figures and table we assume:
 $h_{1}=h_{2}=75~m$, $\mu^{[0]}=6.85~GPa$, $\beta^{[0]}=1629.4~ms^{-1}$, $\mu^{[1]}=\mu^{[1]}=2~GPa$, $\beta^{[1]'}=\beta^{[2]'}=1000~ms^{-1}$. Furthermore $w$ takes on the values $500~m$, 750~m$, 1000~m$ and, for each $w$, we analyze the field maps for the first four resonant frequencies and three incident angles $\theta^{i}=0,~40,~80^{\circ}$.

 The so-called lossless cases corresponds to  $\beta^{[1]''}=\beta^{[2]''}=0~ms^{-1}$ and the lossy cases to $\beta^{[1]''}=\beta^{[2]''}=-15~ms^{-1}$.
 \clearpage
 \newpage
\subsection{Examples of  field maps at the first resonance frequency of each protuberance}
%
\subsubsection{Lossless case, first resonance frequency, $\theta^{i}=0^{\circ}$}
\begin{figure}[ht]
  \begin{center}
    \subfloat[]{
      \includegraphics[width=0.33\textwidth]{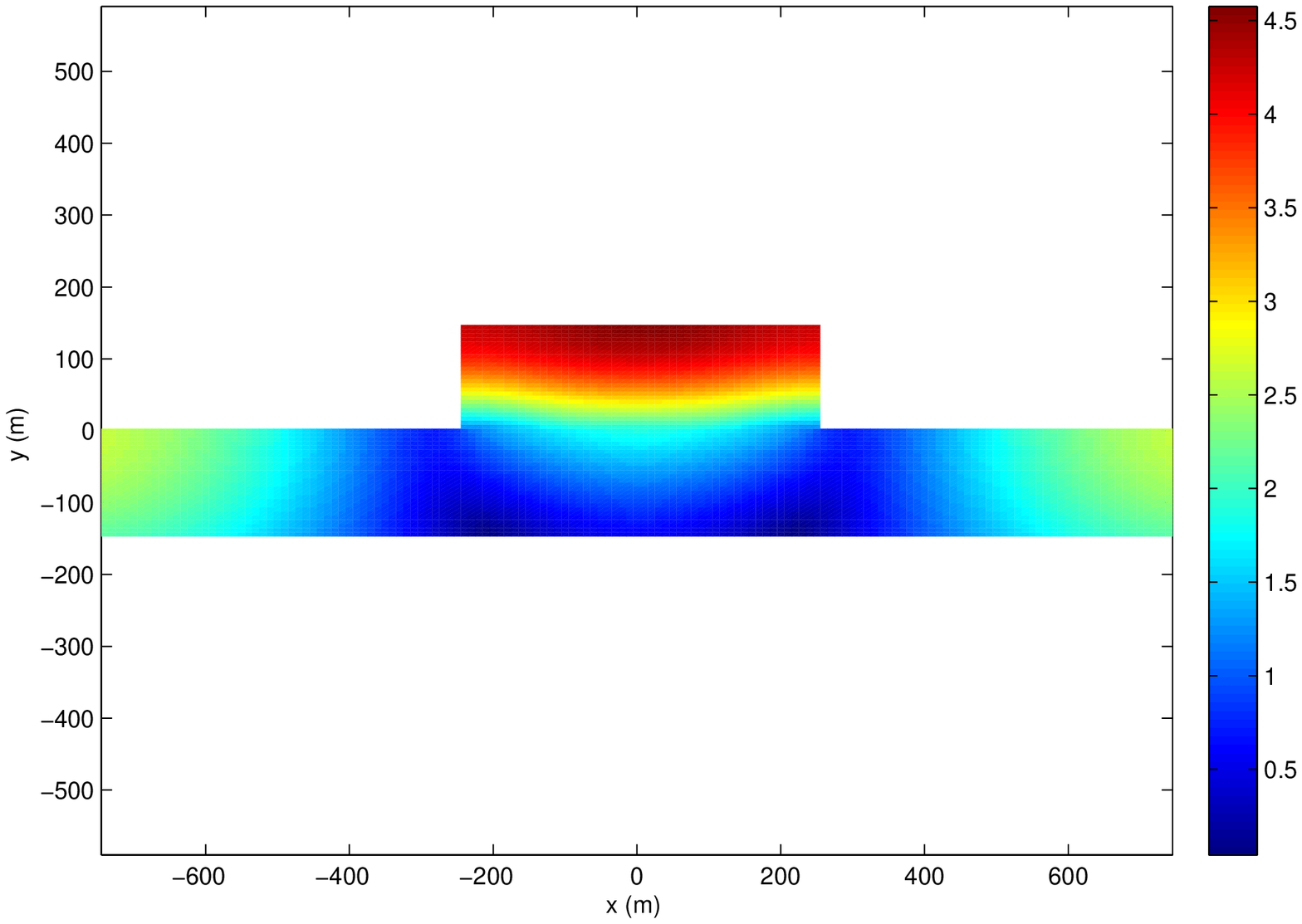}
      \label{a}
                         }
    \subfloat[]{
      \includegraphics[width=0.32\textwidth]{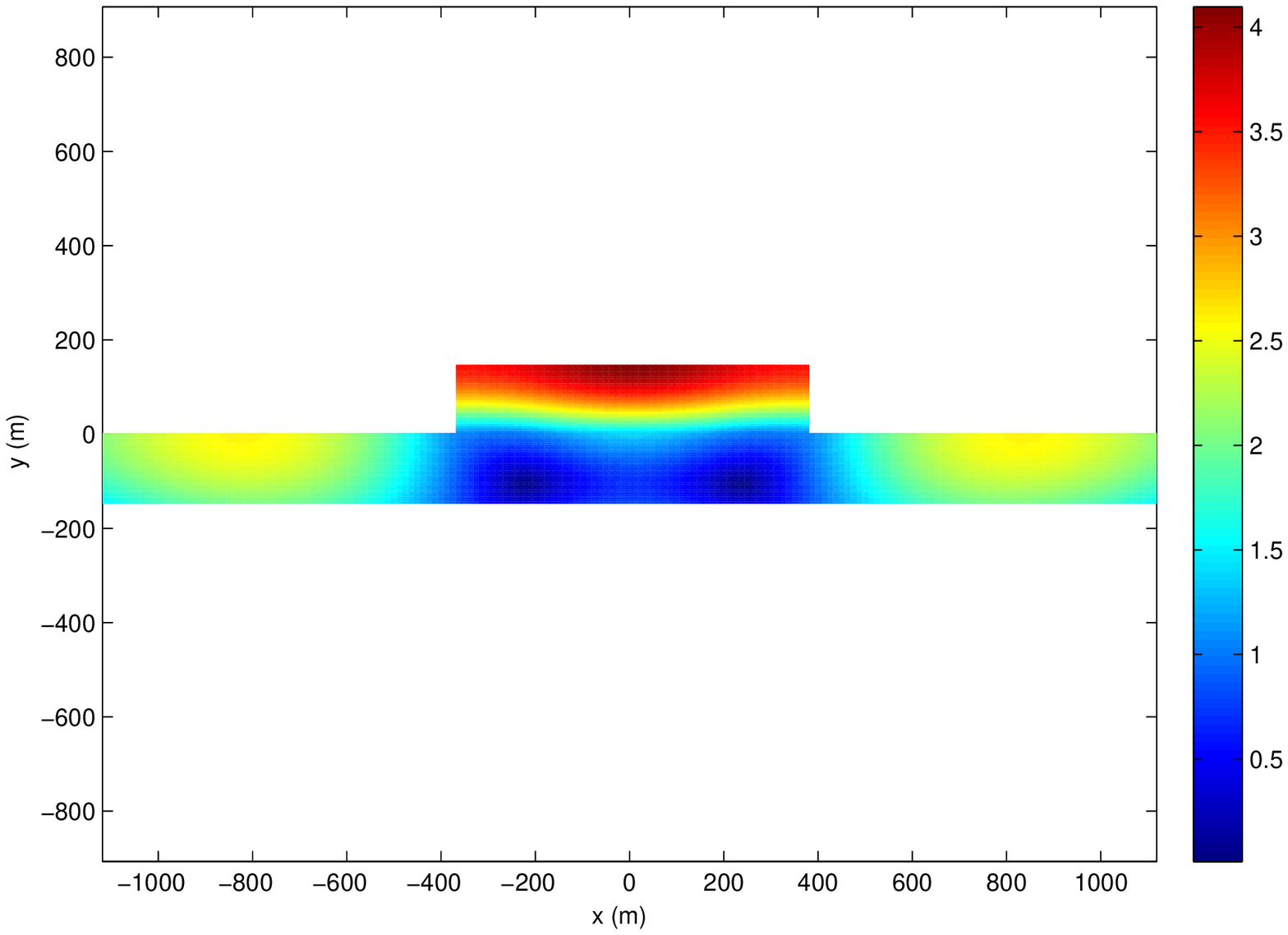}
      \label{b}
                         }
 \subfloat[]{
      \includegraphics[width=0.33\textwidth]{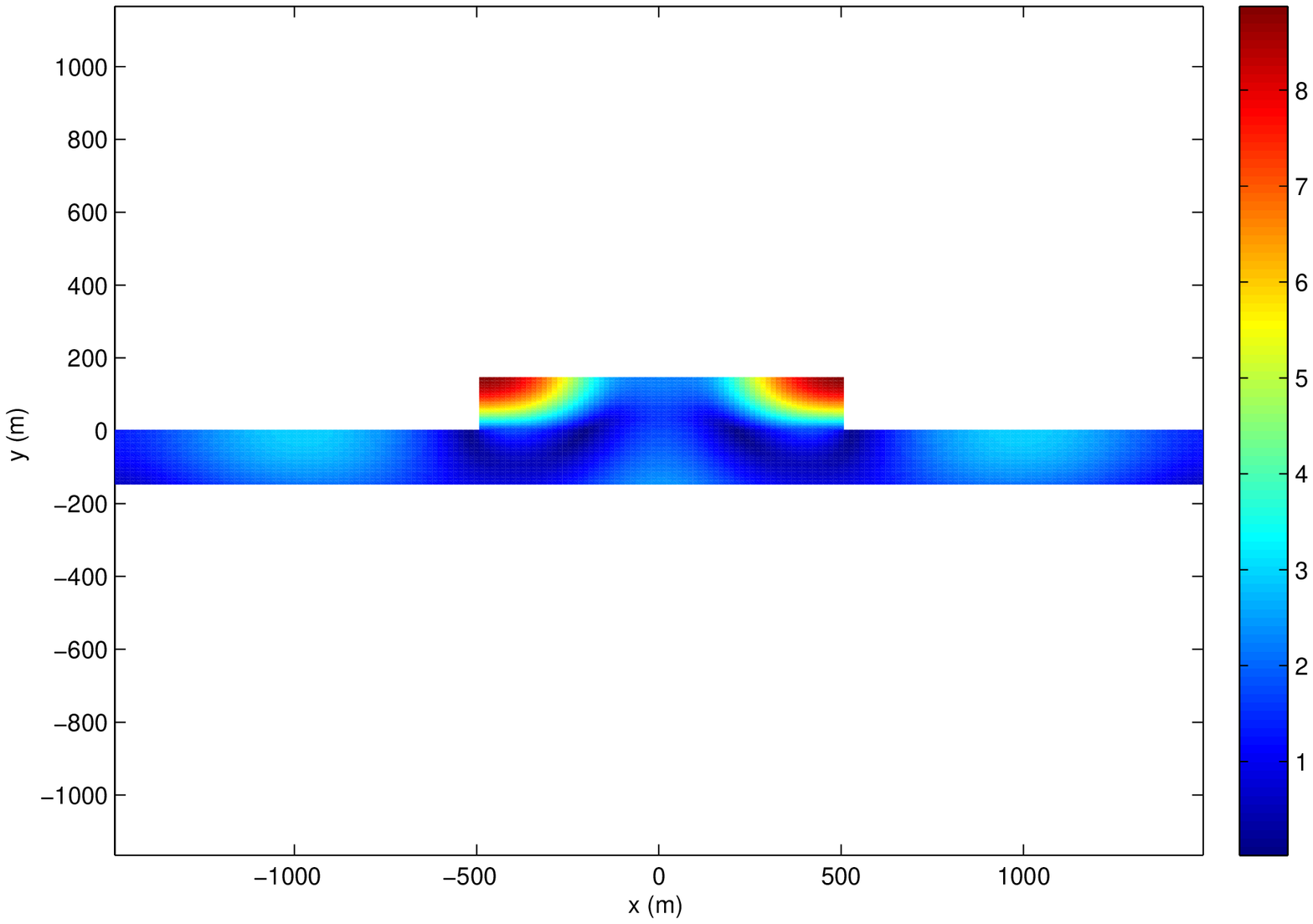}
      \label{c}
                        }
    \caption{
(a): $\theta^{i}=0^{\circ}$, $w=500~m$, $f_{1}^{R}=1.280~Hz$, Max=4.5.\\
(b): $\theta^{i}=0^{\circ}$, $w=750~m$, $f_{1}^{R}=1.380~Hz$, Max=4.1.\\
(c): $\theta^{i}=0^{\circ}$, $w=1000~m$, $f_{1}^{R}=1.608~Hz$, Max=8.7.
}
    \label{carh01}
  \end{center}
\end{figure}
%
\subsubsection{Lossless and lossy cases, first resonance frequency, $\theta^{i}=80^{\circ}$}
\begin{figure}[ptb]
  \begin{center}
    \subfloat[]{
      \includegraphics[width=0.33\textwidth]{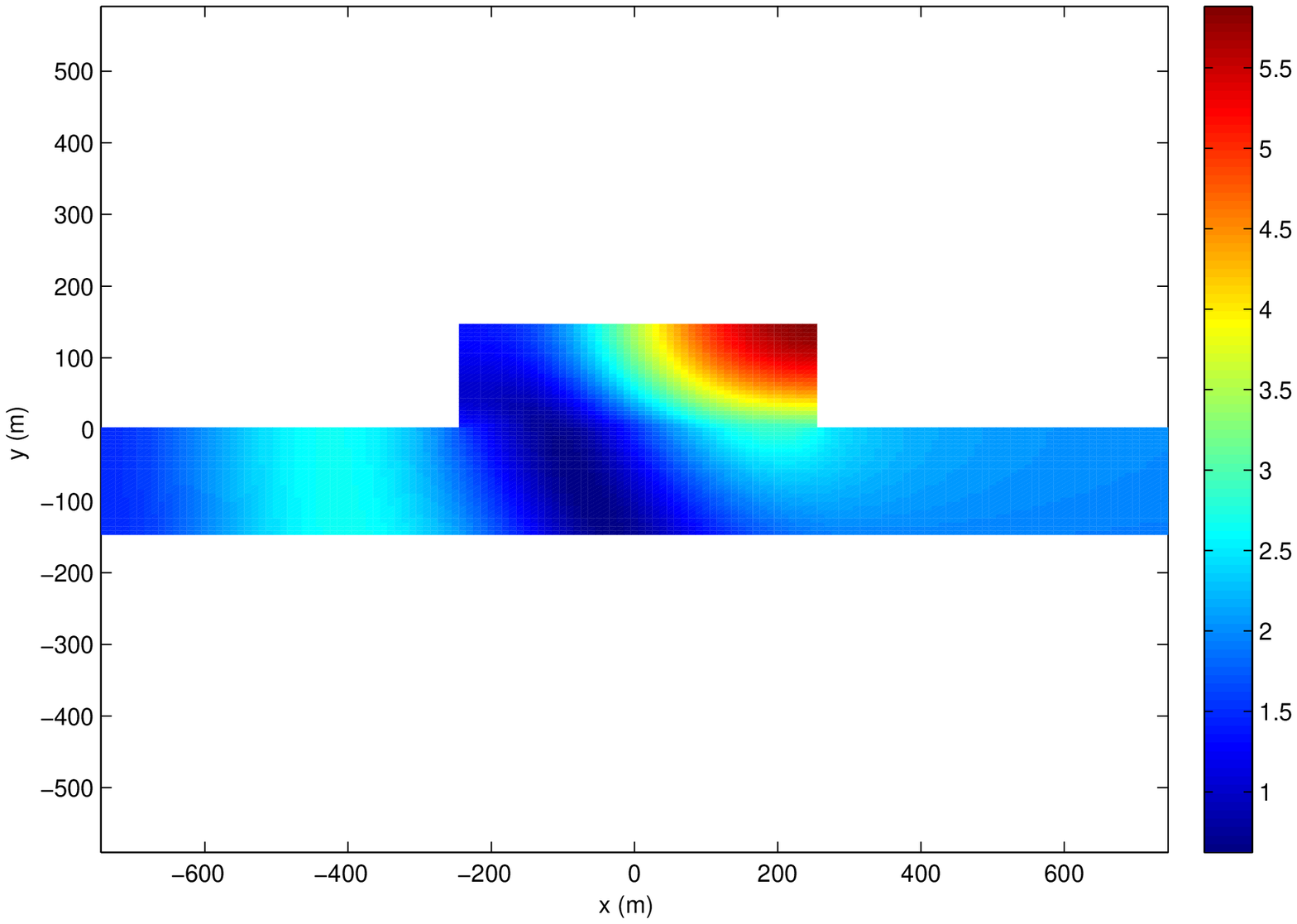}
      \label{a}
                         }
    \subfloat[]{
      \includegraphics[width=0.325\textwidth]{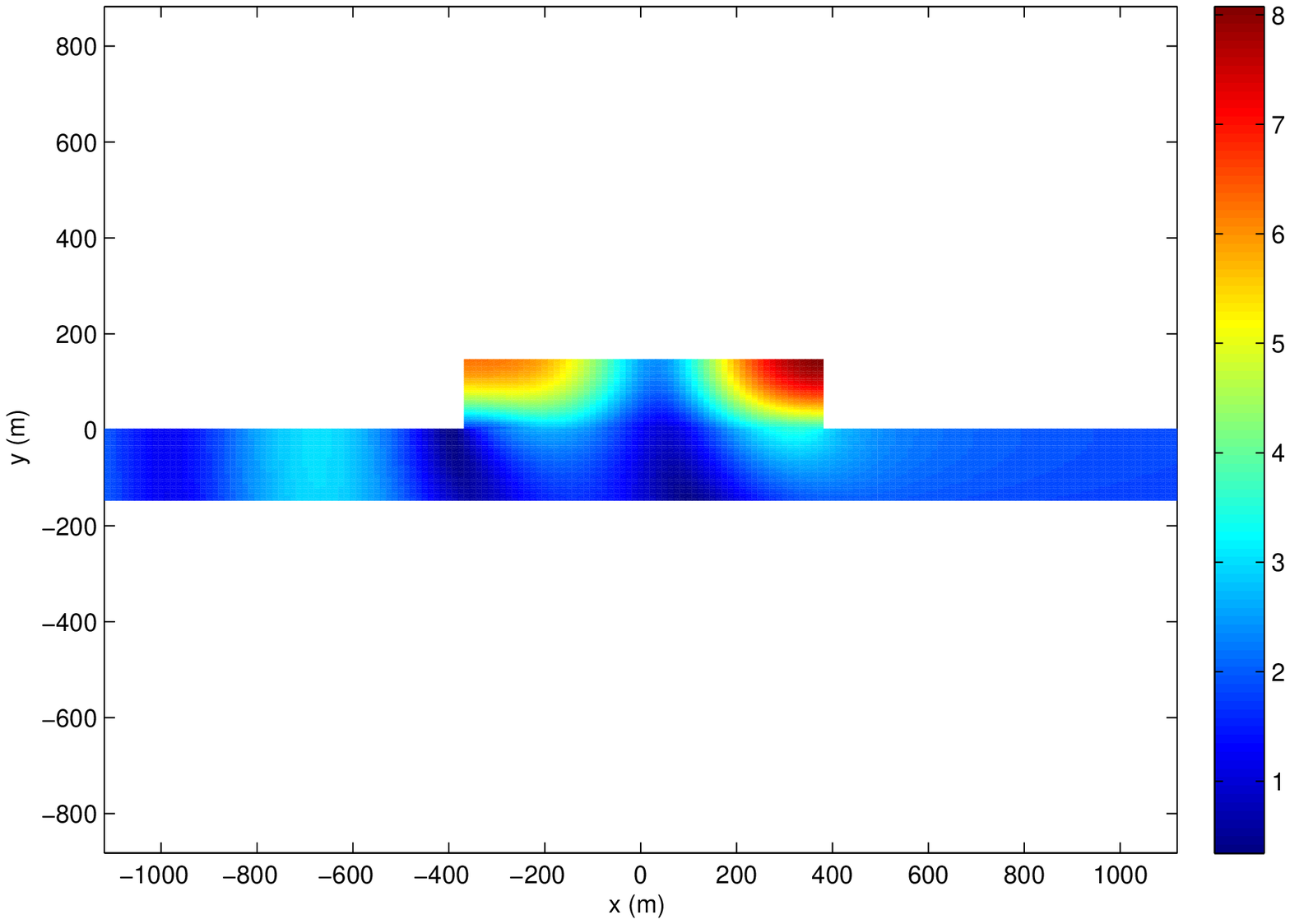}
      \label{b}
                         }
 \subfloat[]{
      \includegraphics[width=0.33\textwidth]{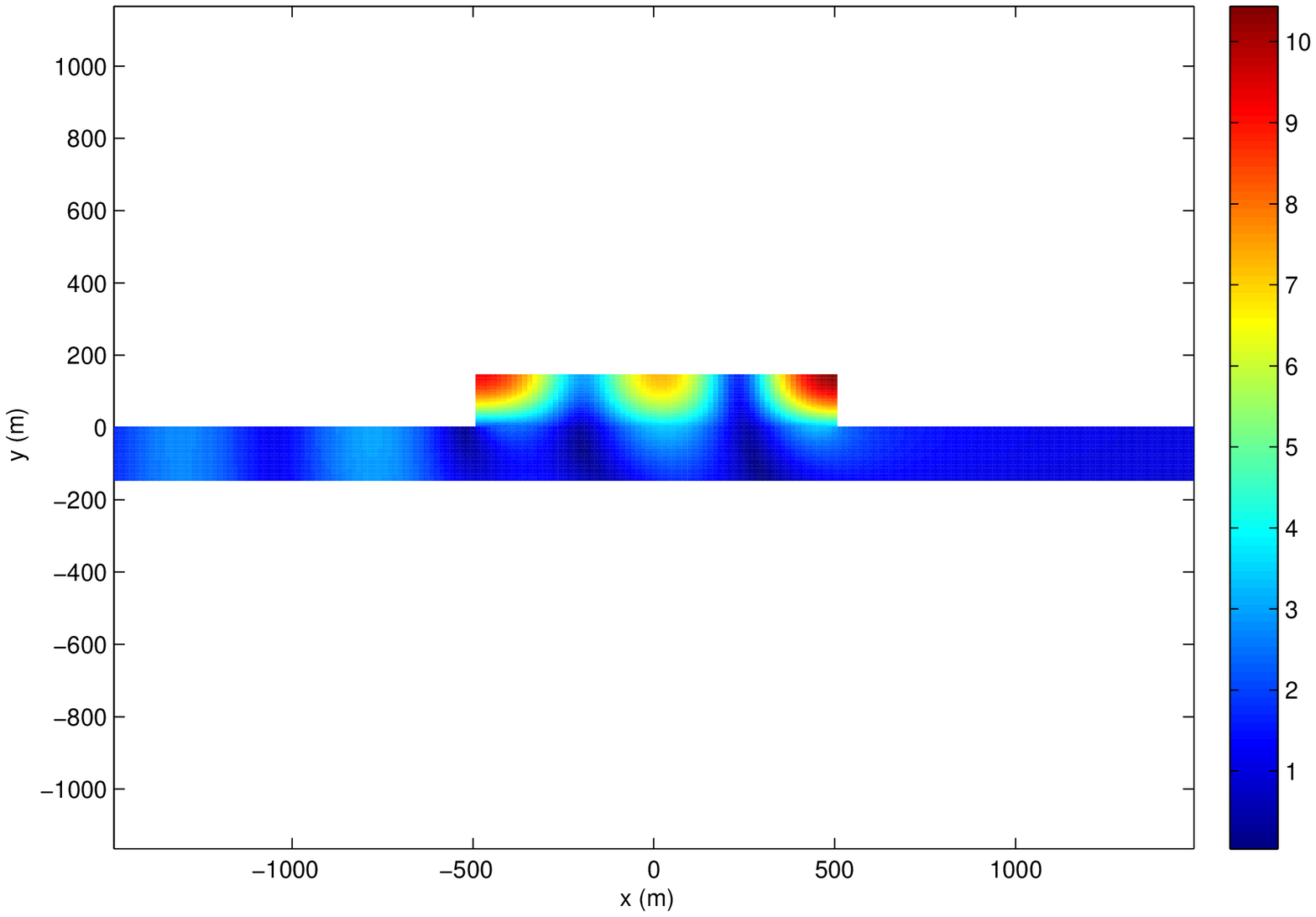}
      \label{c}
                        }
    \caption{
(a): Lossless case. $\theta^{i}=80^{\circ}$, $w=500~m$,  $f_{1}^{R}=1.280~Hz$, Max=5.7.\\
(b): Lossless case. $\theta^{i}=80^{\circ}$, $w=750~m$,  $f_{1}^{R}=1.380~Hz$, Max=8.\\
(c): Lossless case. $\theta^{i}=80^{\circ}$, $w=1000~m$, $f_{1}^{R}=1.608~Hz$, Max=10.3.
}
    \label{carh02}
  \end{center}
\end{figure}
\begin{figure}[ptb]
  \begin{center}
    \subfloat[]{
      \includegraphics[width=0.324\textwidth]{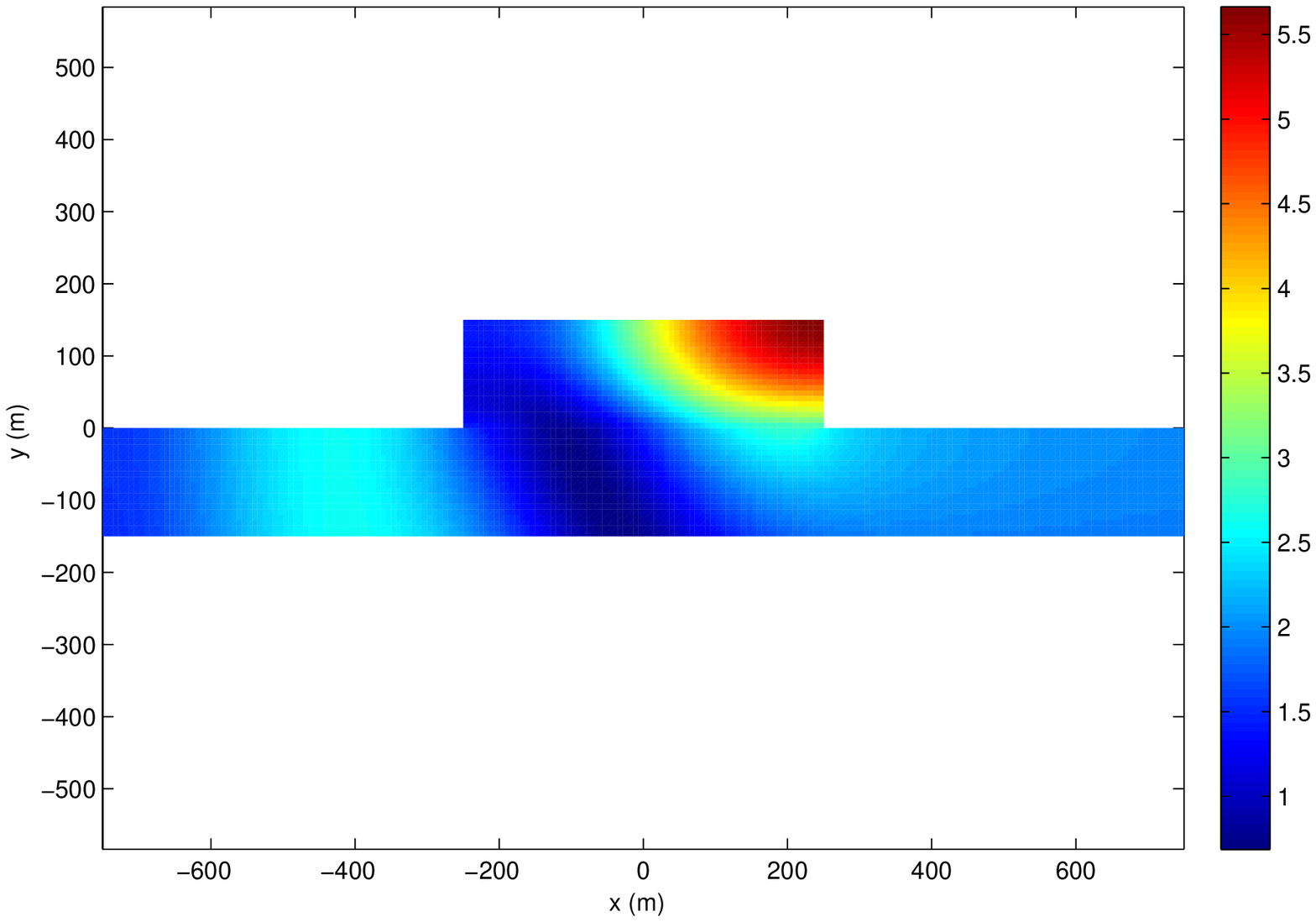}
      \label{a}
                         }
    \subfloat[]{
      \includegraphics[width=0.328\textwidth]{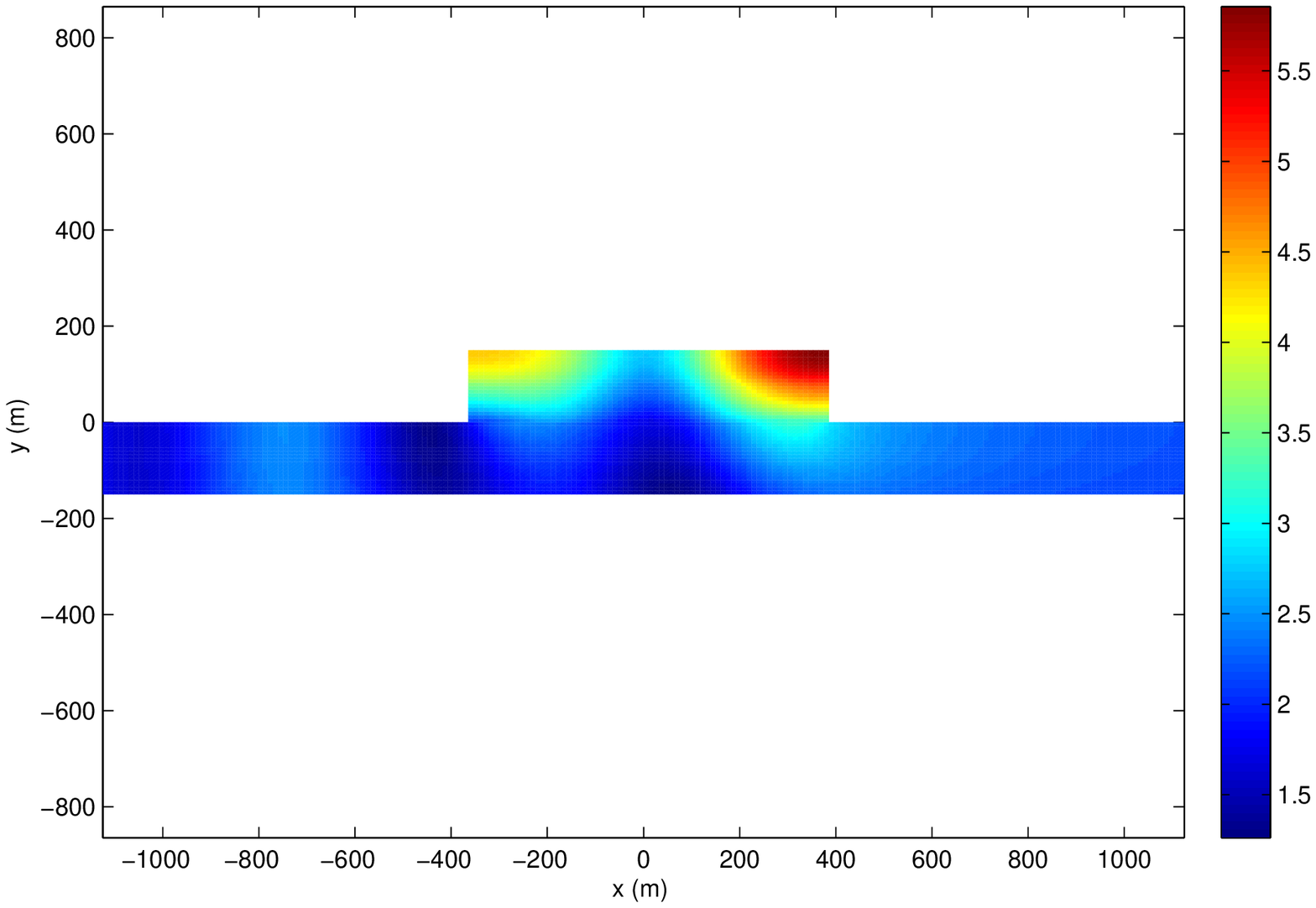}
      \label{b}
                         }
 \subfloat[]{
      \includegraphics[width=0.33\textwidth]{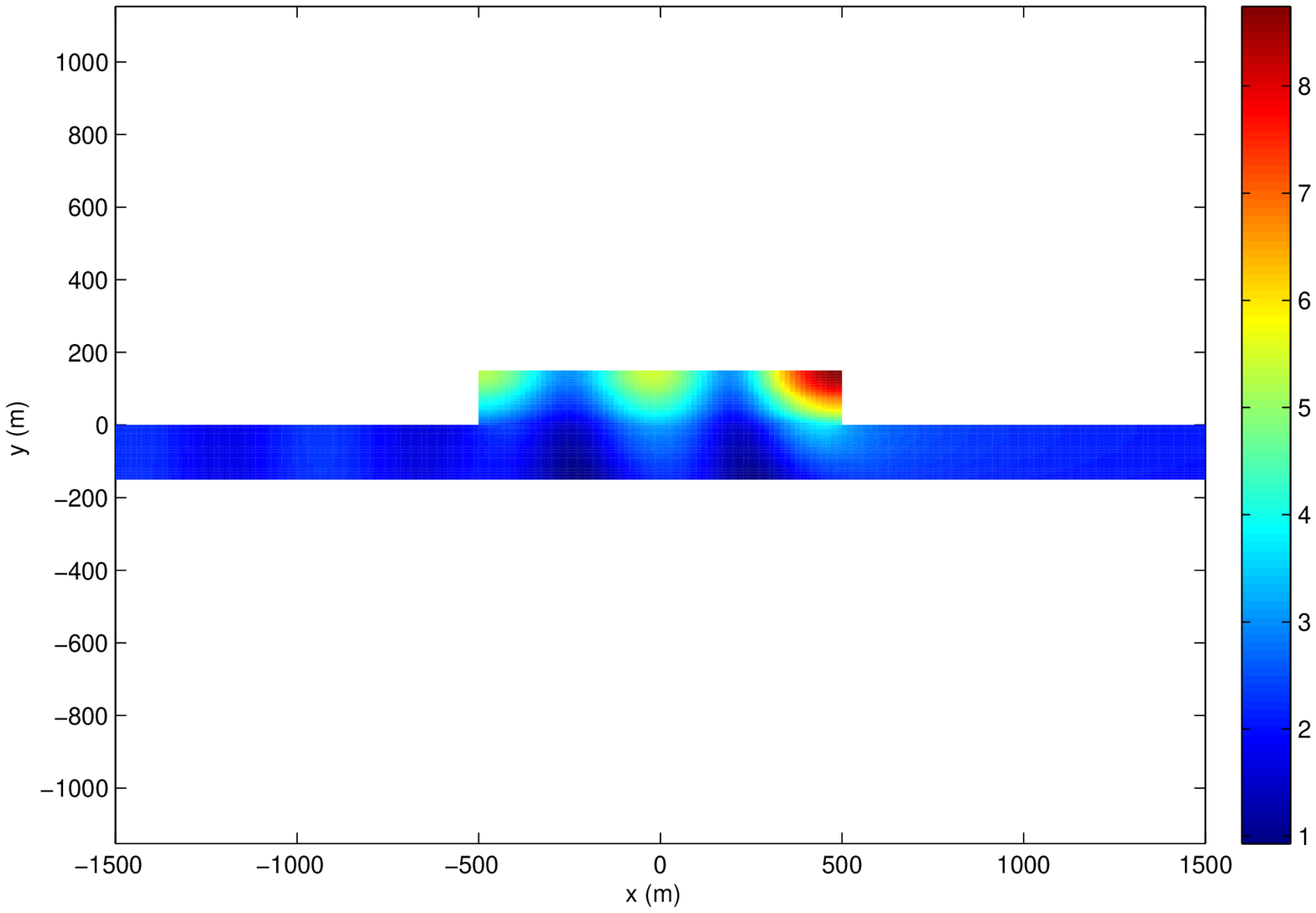}
      \label{c}
                        }
    \caption{
(a): Lossy case. $\theta^{i}=80^{\circ}$, $w=500~m$,  $f_{1}^{R}=1.280~Hz$, Max=5.6 .\\
(b): Lossy case. $\theta^{i}=80^{\circ}$, $w=750~m$,  $f_{1}^{R}=1.380~Hz$, Max=5.7\\
(c): Lossy case. $\theta^{i}=80^{\circ}$, $w=1000~m$,  $f_{1}^{R}=1.608~Hz$, Max=8.5.
}
    \label{carh03}
  \end{center}
\end{figure}
\clearpage
\newpage
In figs. \ref{carh01}- \ref{carh03} we observe that:\\
(a) at resonance frequencies, the field is concentrated within the protuberance (near its summit), so as to be maximal  at hot spots (HS) or or along a horizontal strip (UTHS);\\
(b) the number and intensity of the HS (the latter appearing to be arranged in near-periodic, spatially-horizontal manner) seem to increase with $w$   and incident angle.

\subsection{Analysis of the transfer function maps}
In the table below, we depict the  maxima (designated by the symbol 'Max', and taken from the color maps of $T$) of the transfer function $T(x,y;f)~;(x,y)\in\Omega_{1}+\Omega_{2}$, at the resonant (R)  frequencies, for both the lossless and lossy hills. This is done for the three incident angles:  $\theta^{i}=~0^{\circ},~40^{\circ},~80^{\circ}$.

The comments in this table refer to the HS and UTHS  within the lossless or lossy protuberance at the resonance frequencies.\\
\begin{center}
\begin{tabular}{|c|c|c|c|c|c|c|}
  \hline
  $f_{j}^{R}$ &  $\theta^{i}(^{\circ})$ & $w(m)$ & $f(Hz)$ & lossless  & lossy   & comments\\
  or $f^{NR}$ &                         &        &         & Max       & Max  & \\
  \hline
  $f_{1}^{R}$ & 0  & 500 & 1.280 & 4.6  &  & UTHS\\
  "           & "  & 750 & 1.380 & 4.1  &  & UTHS\\
  "           & "  & 1000 & 1.608 & 8.7  &  & 2 HS\\
  \hline
  "           & 40 & 500 & 1.280 & 5.6  &  & 1 HS\\
  "           & "  & 750 & 1.380 & 6.3  &  & 2 HS\\
  "           & "  & 1000 & 1.608 & 5.2  &  & 1 HS\\
  \hline
  "           & 80 & 500 & 1.280 & 5.7 & 5.6 & 1 HS(1 HS)\\
  "           & "  & 750 & 1.380 & 8.0 & 5.7 & 1 H(1 HS)\\
  "           & "  & 1000 & 1.608 & 10.3 & 8.5 & 2 HS(1 HS)\\
  \hline
  \hline
  $f_{2}^{R}$ & 0  & 500 & 1.640  & 4.1  &  & UTHS \\
  "           & "  & 750 & 1.850  & 12.3  &  & 2 HS \\
  "           & "  & 1000 & 1.975  & 3.7  &  & UTHS\\
  \hline
  "           & 40 & 125 & 1.640  & 10.5  &  & 1 HS\\
  "           & "  & 250 & 1.850  & 8.2  &  & 1 HS\\
  "           & "  & 500 & 1.975  & 9.2  &  & 2 HS\\
  \hline
  "           & 80 & 500 & 1.640  & 13.0 & 8.0 & 1 HS(1 HS)\\
  "           & "  & 750 & 1.850  & 17.0 & 10.0 & 2 HS(1 HS)\\
  "           & "  & 1000 & 1.975  & 22.0 & 10.0 & 2 HS(1 HS)\\
  \hline
  \hline
  $f_{3}^{R}$ & 0  & 500 & 2.440  & 8.70  &  & 2 HS\\
  "           & "  & 750 & 2.420  & 2.70  &  & UTHS\\
  "           & "  & 1000 & 2.420   & 14.50  &  & 3 HS\\
  \hline
  "           & 40 & 500 & 2.440 & 12.4  &  & 2 HS\\
  "           & "  & 750 & 2.420  & 8.80  &  & 2 HS\\
  "           & "  & 1000 & 2.420  & 26.00  &  & 3 HS\\
  \hline
  "           & 80 & 500 & 2.440  & 23.30  & 9.9 & 3 HS(1 HS)\\
  "           & "  & 750 & 2.420  & 30.00  & 8.7 & 4 HS(1 HS)\\
  "           & "  & 1000 & 2.420  & 42.00  & 7.2 & 5 HS(3 HS)\\
  \hline
  \hline
  $f_{4}^{R}$ & 0  & 500 & 3.337  & 1.95  &  & 0 HS \\
  "           & "  & 750 & 3.030  & 8.00  &  & 3 HS \\
  "           & "  & 1000 & 2.874  & 2.24  &  & 0 HS \\
  \hline
  "           & 40 & 500 & 3.337  & 8.70  &  & 3 HS\\
  "           & "  & 750 & 3.030  & 32.00  &  & 5 HS\\
  "           & "  & 1000 & 2.874  & 35.00  &  & 6 HS\\
  \hline
  "           & 80 & 500 & 3.337  & 43.00  & 8.4 & 4 HS(1 HS)\\
  "           & "  & 750 & 3.030  & 72.00  & 6.5 & 5 HS(3 HS)\\
  "           & "  & 1000 & 2.874  & 25.00  & 5.2 & 6 HS (5 HS)\\
  \hline
\end{tabular}
\newline
\end{center}
%
\subsection{Discussion of the results in the table}
The results in the above table call for the following comments:\\\\
(1) $f_{j}^{R}$ increases as $w$ increases for low $j$, and then decreases with increasing $w$ for  for high $j$;\\
(2) the resonant response is generally greatest at the highest incident angle;\\
(3) the responses are {\it qualitatively} different  for all three values of $w$ and  most incident angles and resonance frequencies;\\
(4) moreover, there are significative {\it quantitative} differences in response as a function of $w$, with, in most cases an increase of 'Max'  with $w$, principally at the larger incident angles;\\
(5) the introduction of material losses in the protuberance results in important reductions of resonant response, particularly for the largest $w$ and/or resonance order $j$.\\

Hence we conclude, on the basis of these results, and recalling that they were obtained for fixed composition of the configuration, that variations of the aspect ratio (for constant $h$) produce substantial qualitative, as well as large quantitative, changes of resonant response within the protuberance. Once again, the question of whether these changes are larger or smaller than those due to compositional variations cannot be answered on the sole basis of these results.

\section{Effect of layering within a hill}\label{comp}
The possibility that  a 'weathered' region at the top, and more generally the subsurface velocity (and other mechanical attributes) structure, of a hill or mountain may significantly modify the seismic response of the convex (or concave) boundary feature has frequently been evoked in the scientific literature \cite{ba95,bc09,br00,br11,bu08,bf14,bt85,gb88,gr09,jo89,lc09,my87,mk06,ri03,wd18,ag05,bm14}. This is as expected because the subsurface geological structure of the medium below flat ground, as well as the geological structure of the sediment filler of basins and valleys are known to greatly modify the seismic response on the ground \cite{jo89,se11,sk05,sk08,vc99,wi95,ag05}.

Herein, we address this issue via a model of the geological structure of the hill or mountain consisting of two horizontal layers of height and width $h_{1},~w$ and $h_{2},~w$ that fill ($h=h_{1}+h_{2}$) the convex boundary feature and are in welded contact one with the other, the lower layer being in welded contact with the underground medium. If the lower layer has the same mechanical properties as those of the underground, the upper layer constitutes a (primitive) model of a 'weathered' region at the top.

Note that the solids filling these two layers are both assumed to be lossless in the following computations.
The common parameters are: $w=500~m$, $\mu^{[0]}=6.85~GPa$, $\beta^{[0]}=1629.4~ms^{-1}$, $\mu^{[1]}=4.85~GPa$, $\beta^{[1]}=1329.4-i0~ms^{-1}$, $\mu^{[2]}=2~GPa$, $\beta^{[2]}=1000-i0~ms^{-1}$.
\subsection{$h_{1}=0~m$, $h_{2}=150~m$ hill}
See the first graph in each of figs. \ref{carh01}-\ref{carh03}.
\subsection{$h_{1}=50~m$, $h_{2}=100~m$ hill}
By adopting~ the~~ method ~~of \cite{wi20b} we~~ find that the first~ three ~~  resonances ~occur ~at:
 $1.346,~1.788,$ $~2.641~Hz$. A figure not shown here gives the map of seismic response (more precisely, $\|T(x,y;f^{R})\|=
 \|u(x,y;f^{R})\|; (x,y)\in\Omega_{0}+\Omega_{1}+\Omega_{2}$  at $f_{3}^{R}=2.641~Hz$ for three incident angles. The salient features of these maps are resumed in the table in sect. \ref{tablelay}.
\clearpage
\newpage
\subsection{$h_{1}=75~m$, $h_{2}=75~m$ hill}
\begin{figure}[ht]
  \begin{center}
    \subfloat[]{
      \includegraphics[width=0.36\textwidth]{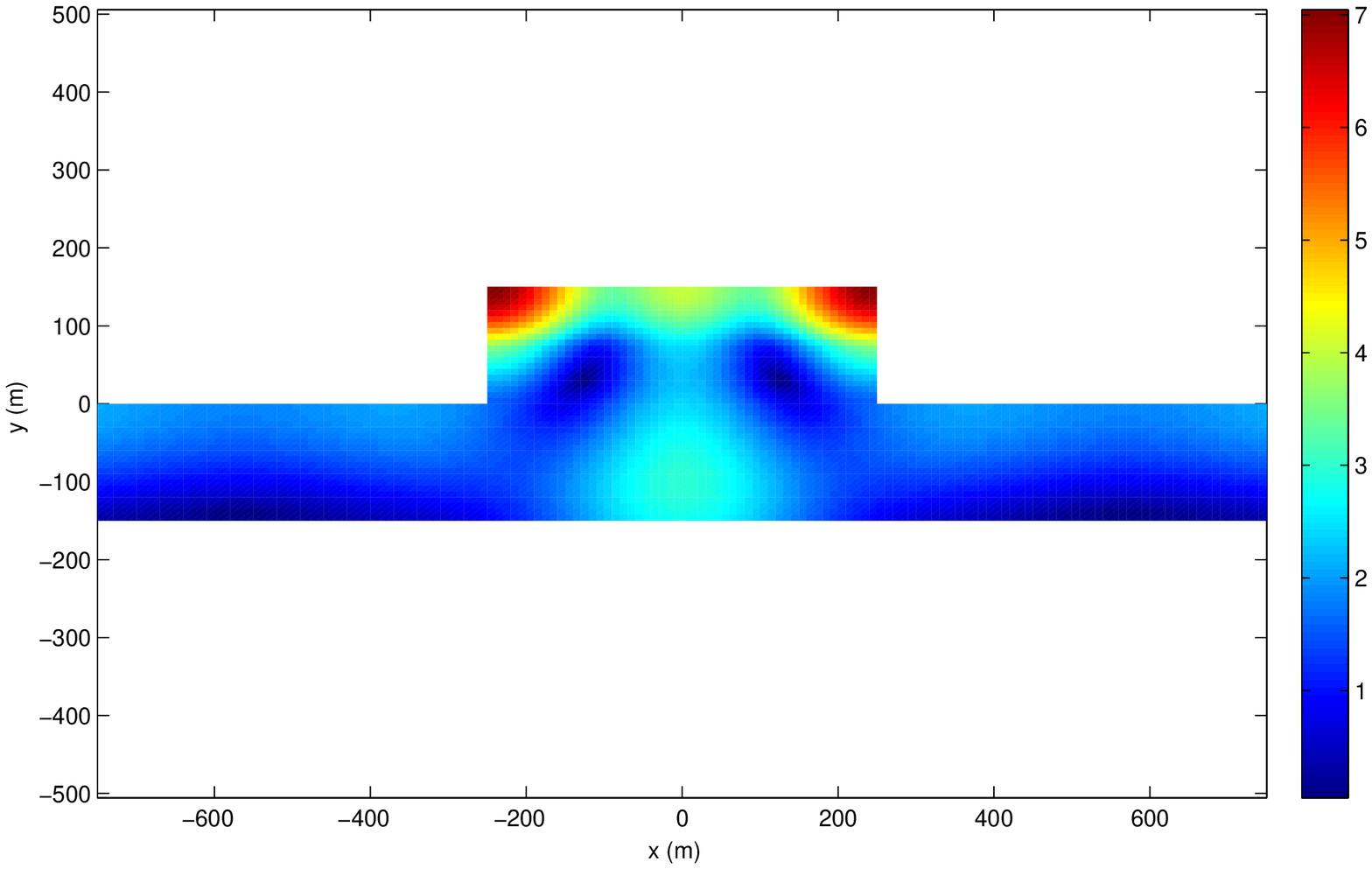}
      \label{a}
                         }
    \subfloat[]{
      \includegraphics[width=0.315\textwidth]{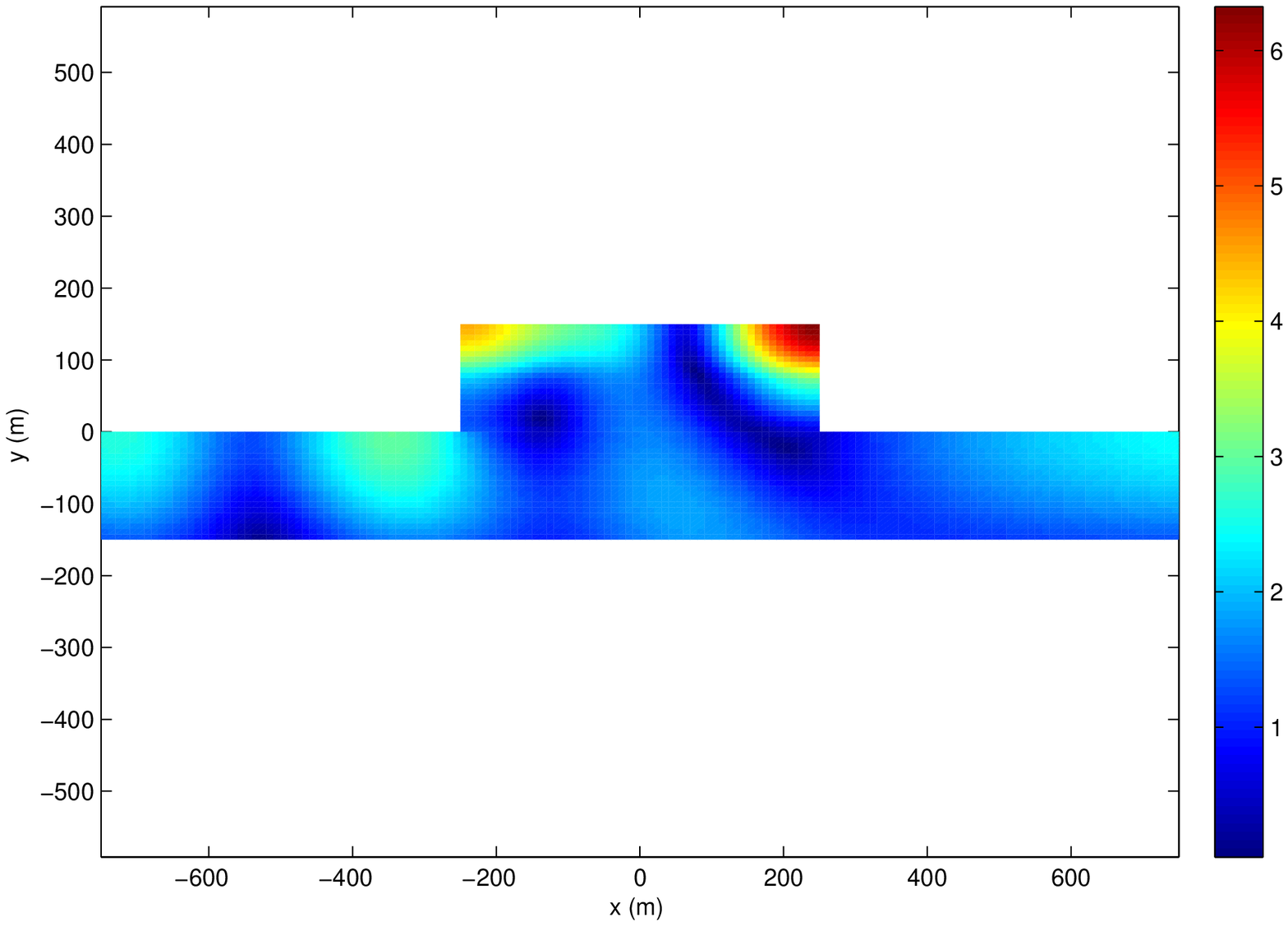}
      \label{b}
                         }
 \subfloat[]{
      \includegraphics[width=0.32\textwidth]{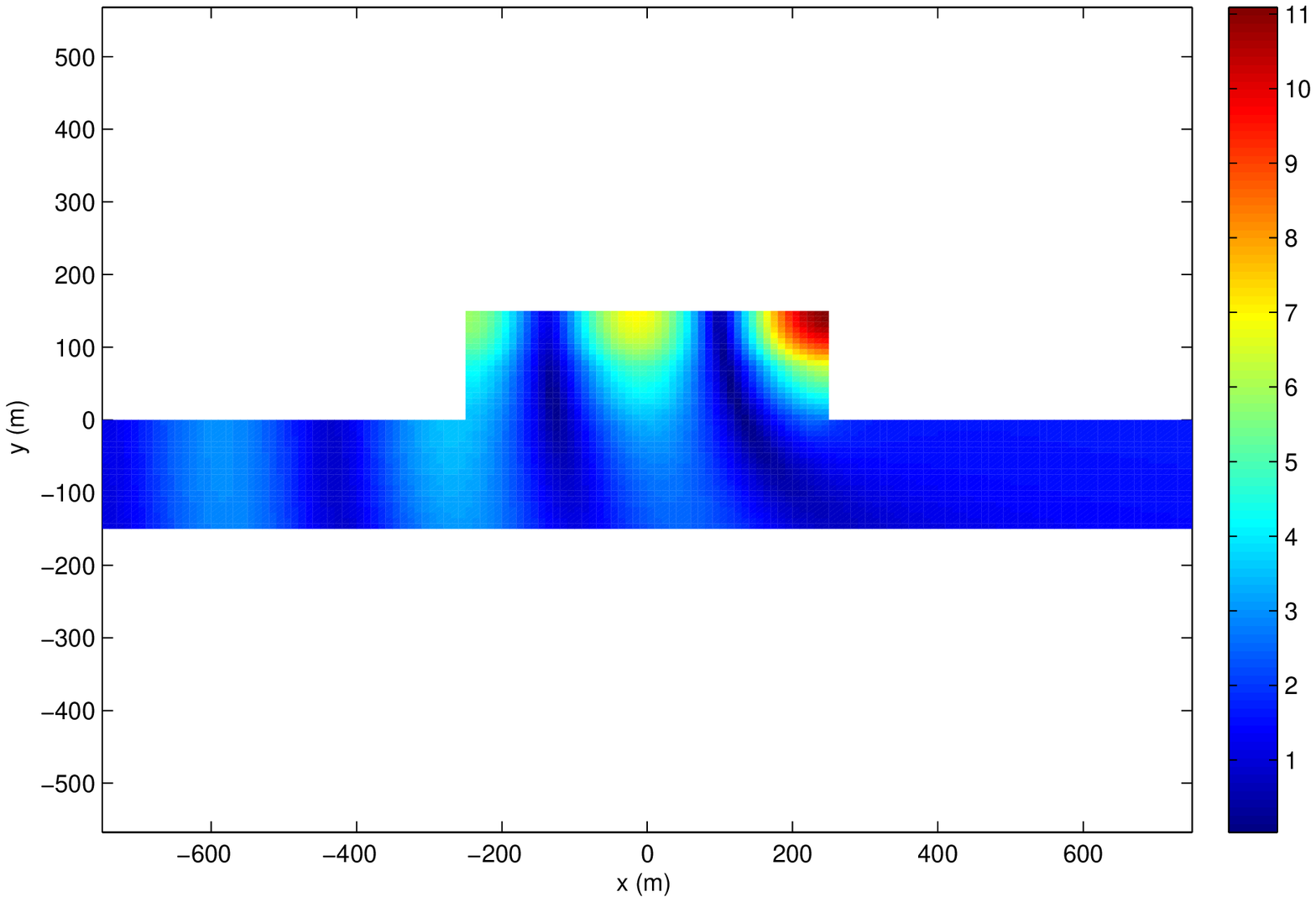}
      \label{c}
                        }
    \caption{
(a): $T_{3}^{R}(x,y,f=2.662~Hz)$ for $\theta^{i}=0^{\circ}$, $h_{1}=75~m$, $h_{2}=75~m$\\
(b): $T_{3}^{R}(x,y,f=2.662~Hz)$ for $\theta^{i}=40^{\circ}$,  $h_{1}=75~m$, $h_{2}=75~m$\\
(c): $T_{3}^{R}(x,y,f=2.662~Hz)$ for $\theta^{i}=80^{\circ}$,  $h_{1}=75~m$, $h_{2}=75~m$.
}
    \label{ch-60-80}
  \end{center}
\end{figure}

By adopting the method of \cite{wi20b} we find ~that the~ first~ three resonances occur at:
 $1.328,~1.834,$ $~2.662~Hz$. Fig. \ref{ch-60-80} gives the seismic responses at $f_{3}^{R}=2.662~Hz$ for three incident angles.
%
\subsection{$h_{1}=100~m$, $h_{2}=50~m$ hill}
By adopting the method of \cite{wi20b} we find that the first three resonances occur at:
$1.883,~2.673,$ $~2.862~Hz$.  In a figure not shown here we give the seismic responses at $f_{3}^{R}=2.862~Hz$ for three incident angles. The salient features of these maps are resumed in the table in sect. \ref{tablelay}.
\clearpage
\newpage
\subsection{$h_{1}=150~m$, $h_{2}=0~m$ hill}
\begin{figure}[ht]
  \begin{center}
    \subfloat[]{
      \includegraphics[width=0.33\textwidth]{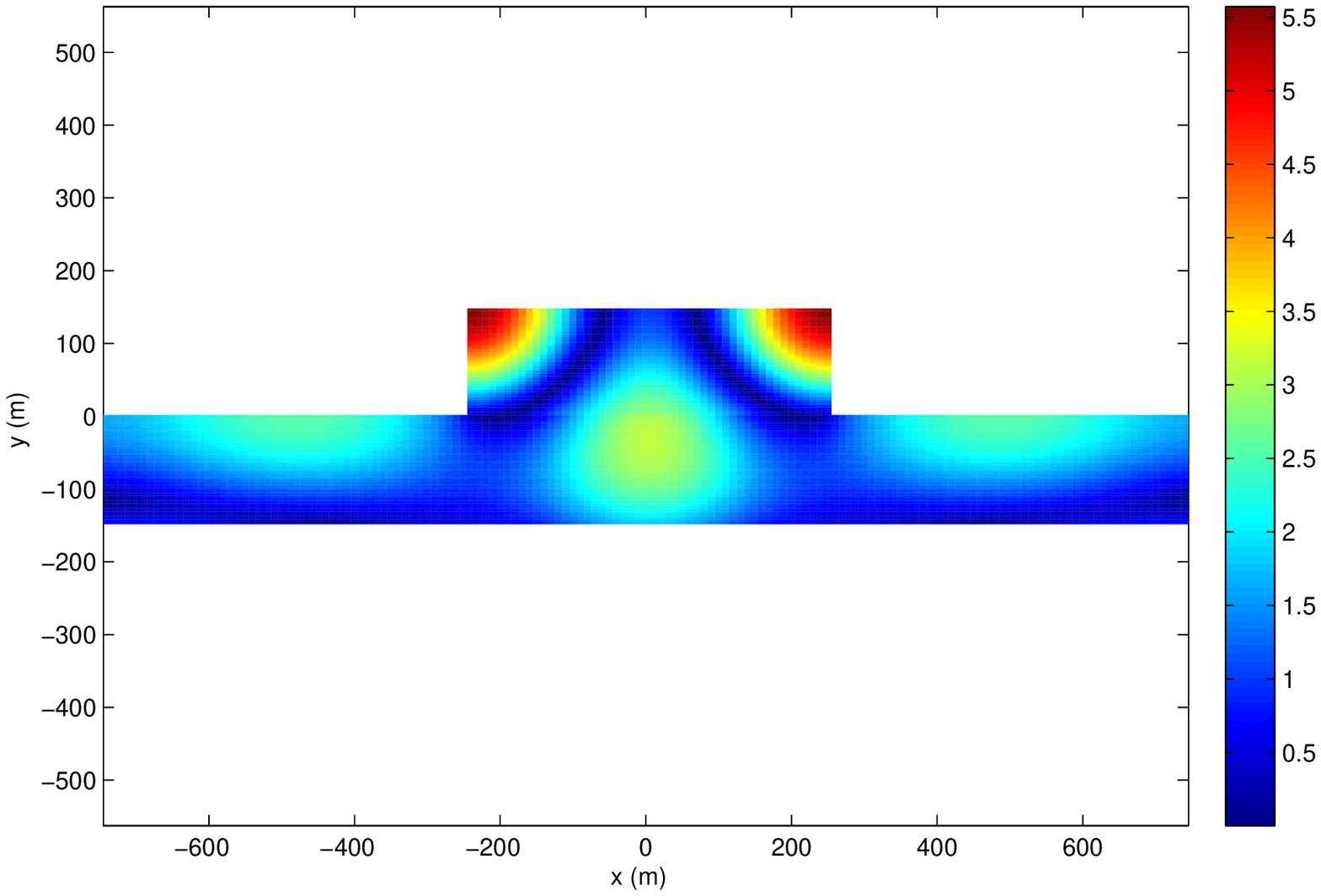}
      \label{a}
                         }
    \subfloat[]{
      \includegraphics[width=0.33\textwidth]{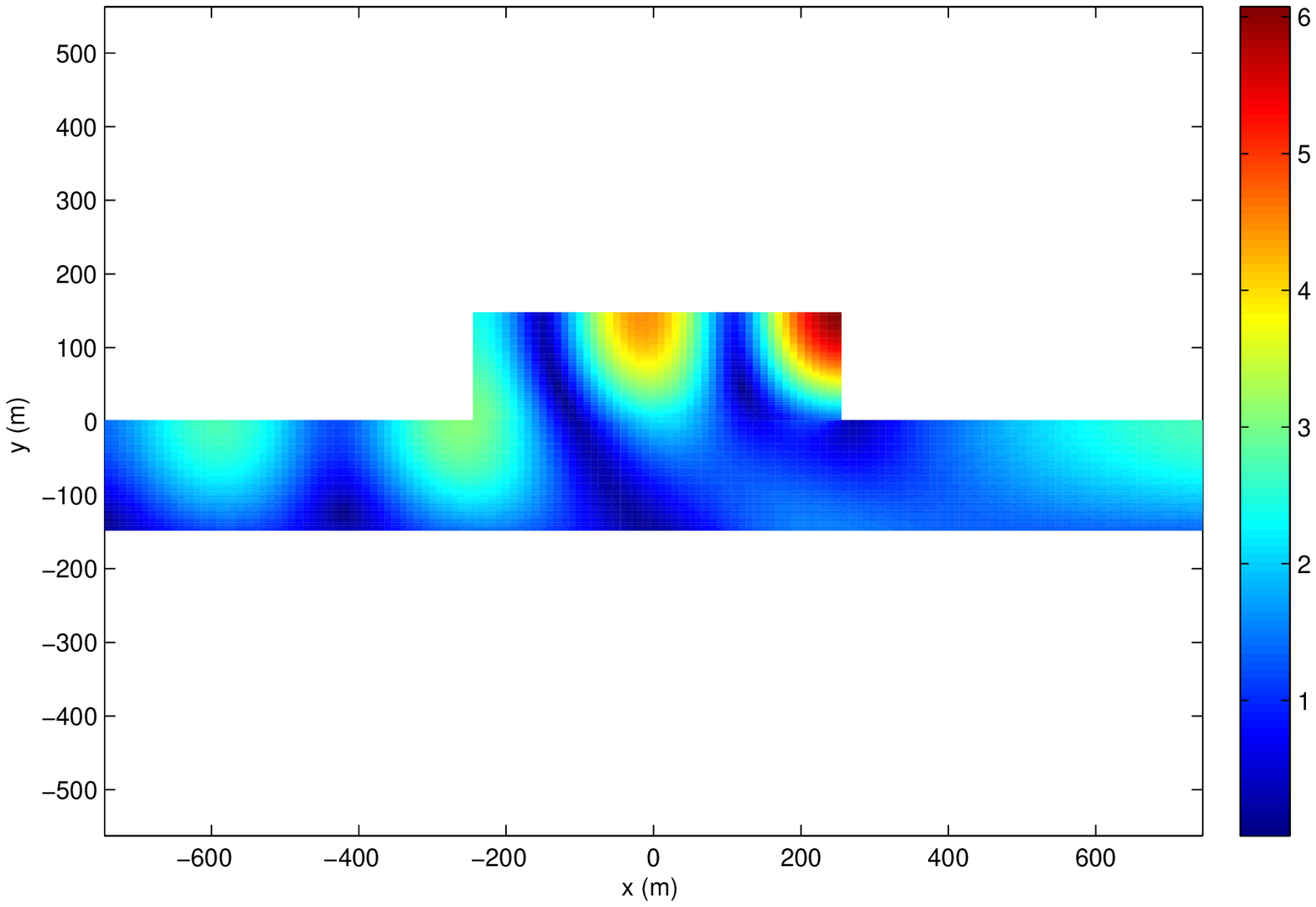}
      \label{b}
                         }
 \subfloat[]{
      \includegraphics[width=0.32\textwidth]{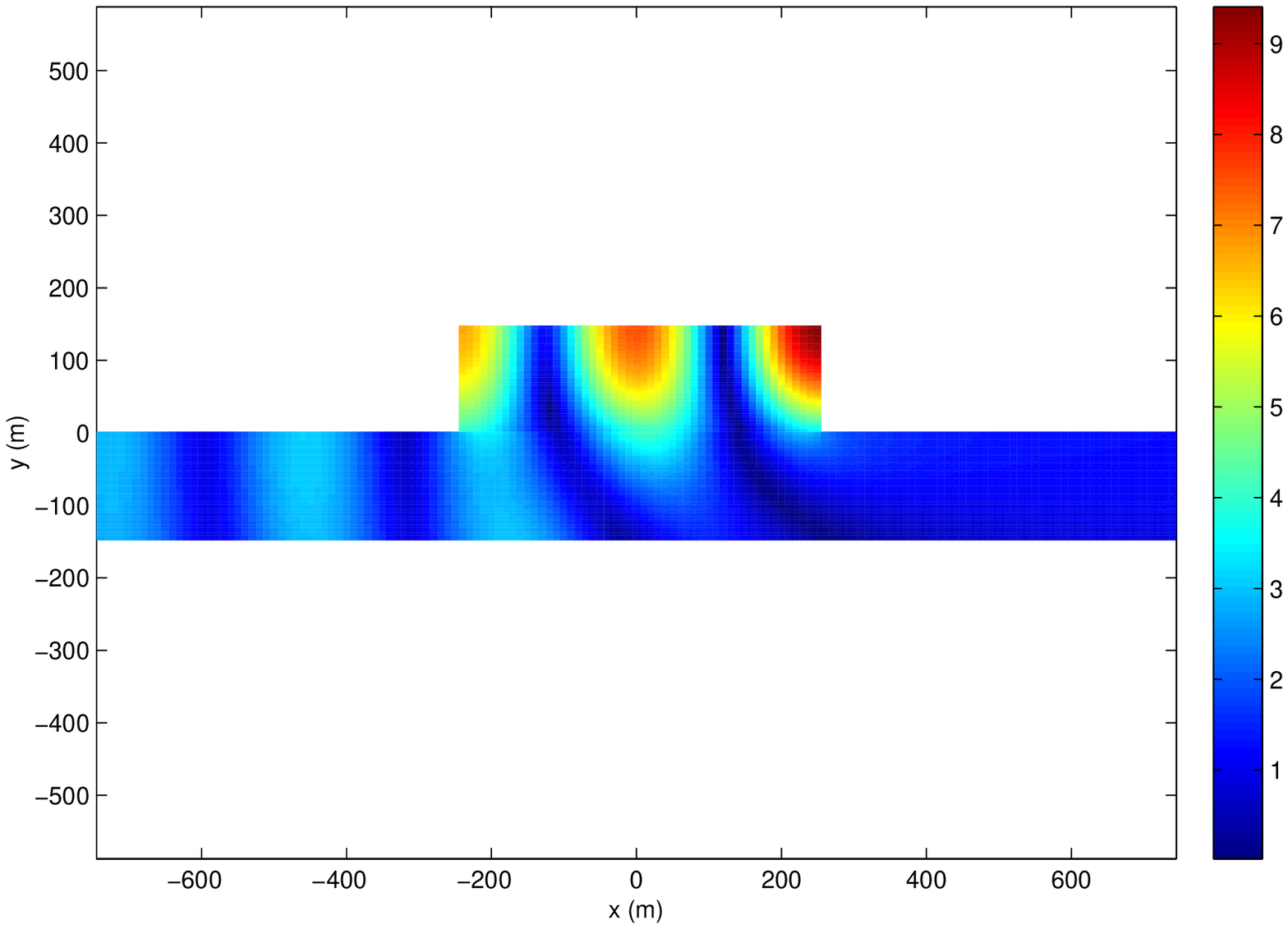}
      \label{c}
                        }
    \caption{
(a): $T_{3}^{R}(x,y,f=3.052~Hz)$ for $\theta^{i}=0^{\circ}$, $h_{1}=150~m$, $h_{2}=0~m$\\
(b): $T_{3}^{R}(x,y,f=3.052~Hz)$ for $\theta^{i}=40^{\circ}$,  $h_{1}=150~m$, $h_{2}=0~m$\\
(c): $T_{3}^{R}(x,y,f=3.052~Hz)$ for $\theta^{i}=80^{\circ}$,  $h_{1}=150~m$, $h_{2}=0~m$.
}
    \label{ch-140-160}
  \end{center}
\end{figure}

By adopting the methods of \cite{wi20b} we find that the first prominent resonances occur at:
$1.946,$ $~3.052~Hz$. Fig. \ref{ch-140-160} gives the seismic responses at $f_{3}^{R}=3.052~Hz$ for three incident angles.
%
\subsection{Comparison of the response maxima for the five composite, non-lossy hills}\label{tablelay}
It is not easy to synthesize the information conveyed in the preceding graphs, except perhaps by means of a single number for each graph. We chose this number to be 'Max' (the maximum of the transfer function maps read off the colorbars in the first graph in each of figs. \ref{carh01}-\ref{carh03}, the figures not shown here, as well as fig. \ref{ch-60-80} and fig. \ref{ch-140-160}).

In the table below, we depict 'Max' (columns 4-6) of displacement response within the composite, non-lossy, hill ($h_{1}$ the height of the lower layer and $h_{2}$ the height of the upper layer) for coupling to a plane wave of (third) resonant frequency $f_{R}$ at incident angles  $\theta^{i}=~0^{\circ},~40^{\circ},~80^{\circ}$.

Recall that the other parameters, common to all five hills, are: $w=500~m$, $\mu^{[0]}=6.85~GPa$, $\beta^{[0]}=1629.4~ms^{-1}$, $\mu^{[1]}=4.85~GPa$, $\beta^{[1]}=1329.4-i0~ms^{-1}$, $\mu^{[2]}=2~GPa$, $\beta^{[2]}=1000-i0~ms^{-1}$.
\\\\
\begin{center}
\begin{tabular}{|c|c|c|c|c|c|}
  \hline
  $h_{1}$(m) & $h_{2}(m)$ & $f_{3}^{R}(Hz)$ & Max for $\theta^{i}=~0^{\circ}$ & Max for $\theta^{i}=40^{\circ}$ & Max for $\theta^{i}=80^{\circ}$ \\
  \hline
  0 & 150 & 2.440 & 8.6 & 12.4 & 23.3 \\
  50 & 100 & 2.641 & 10.5 & 10.7 & 19 \\
  75 & 75 & 2.750 & 11.3 & 9.2 & 17 \\
  100 & 50 & 2.862 & 9.2 & 7.4 & 14 \\
  150 & 0 & 3.052 & 5.5 & 6 & 9.2 \\
  \hline
\end{tabular}
\end{center}
Several tendencies seem to emerge from this table:\\
(a) by comparing the last two lines in the table, it appears that a thin (i.e., $h_{2}=50~m$), soft, layer at the top of the  protuberance increases the maximal seismic response, at least  at the third resonance frequency,\\
(b) as the amount of lower (upper) layer material increases (decreases), the third resonance frequency increases,\\
(c) as the amount of lower upper) layer material increases (decreases), the resonant coupling for $0^{\circ}$ incidence first increases, to attain a maximum at $h_{1 }=h_{2}=75m$, and then decreases,\\
(d) as the amount of lower (upper) layer material increases (decreases), the resonant coupling for  $40^{\circ}$ incidence systematically decreases, which seems to contradict the hypothesis that increased weathering increases the level of resonant seismic response\\
(e) as the amount of lower (upper) layer material increases (decreases), the resonant coupling for $80^{\circ}$ incidence systematically decreases, with the exception of the case $h_{1}=h_{2}=75m$,\\
(f) the resonant coupling is usually (but not always) larger for $40^{\circ}$ incidence than for $0^{\circ}$ incidence,\\
(g) the resonant coupling is  always) larger for $80^{\circ}$ incidence than for $40^{\circ}$ incidence.
\subsection{Discussion}
Our results show, in agreement with \cite{gb88,br00,bc09,ag05,bm14}, that the (theoretical/numerical) prediction of the seismic response of a protuberance (e.g., hill or mountain) is meaningless unless a rather detailed knowledge of the subsurface (this meaning the protuberance and the underlying half space) composition (i.e., the geological features) of the protuberance as well as that of the medium underlying this feature (which we have not varied in this section) is available and incorporated in the computation. At the best, parametric studies, based on a systematic variation of the convex feature geology, should be carried out to evaluate the expected variability of response of a feature with given geometry.
\section{Variation of the homogeneous solid filling the bottom half space}\label{car}
The common parameters to the results in this section are: $\theta^{i}=60^{\circ}$, $w=500~m$, $h_{1}=h_{2}=250~m$, $\mu^{[1]}=\mu^{[2]}=\mu^{[2]}=6.85~GPa$ and  $\beta^{[1]}=\beta^{[2]}=1629.4~ms^{-1}$. We vary  $\mu^{[0]}$ and $\beta^{[0]}$ simultaneously corresponding to the transition from hard to very hard rock. In all the cases, the homogeneous media within both the protuberance and bottom half space are lossless.
\subsection{Three point transfer functions and determination of the resonant frequencies}\label{car}
\begin{figure}[ht]
  \begin{center}
    \subfloat[]{
      \includegraphics[width=0.33\textwidth]{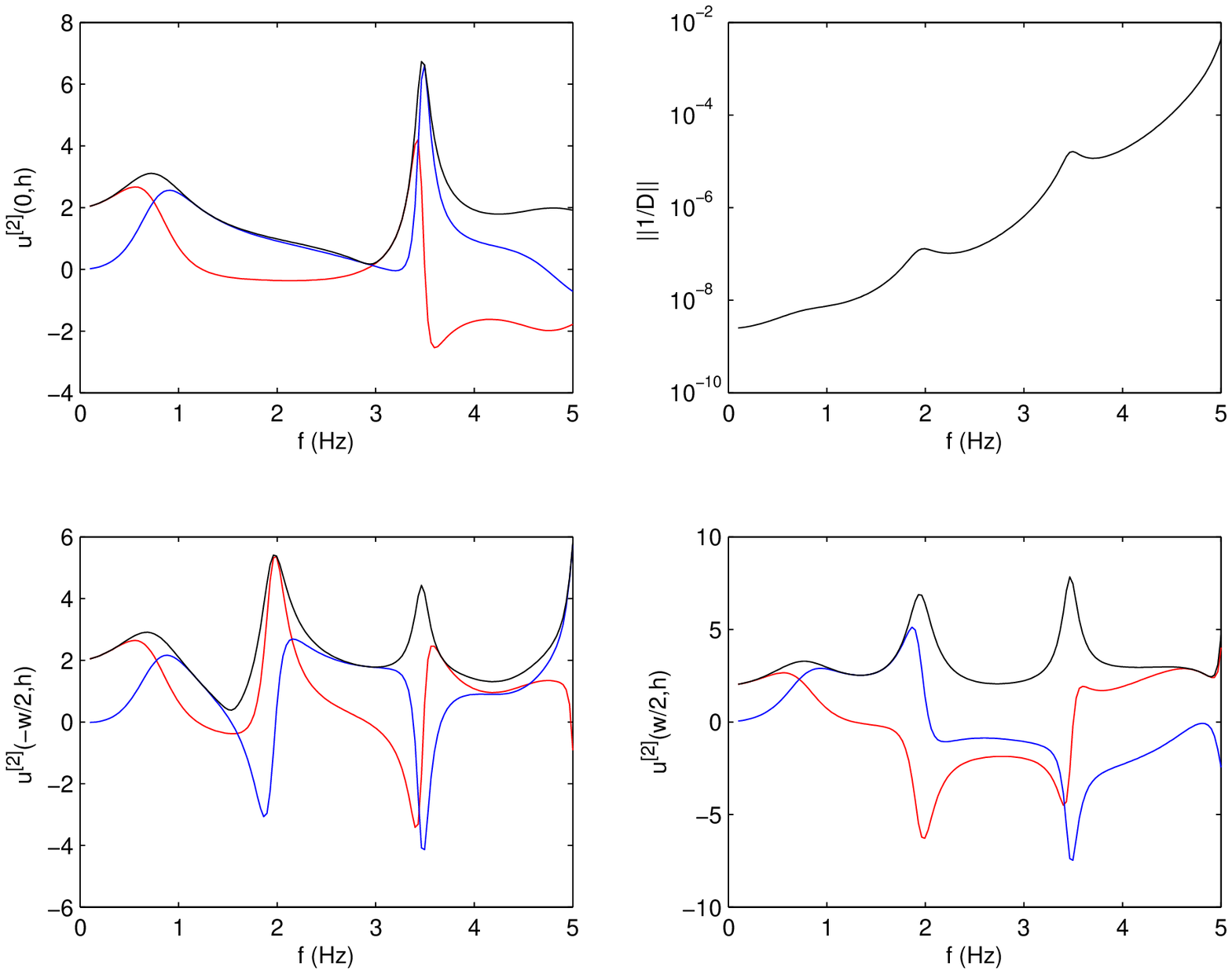}
      \label{a}
                         }
    \subfloat[]{
      \includegraphics[width=0.33\textwidth]{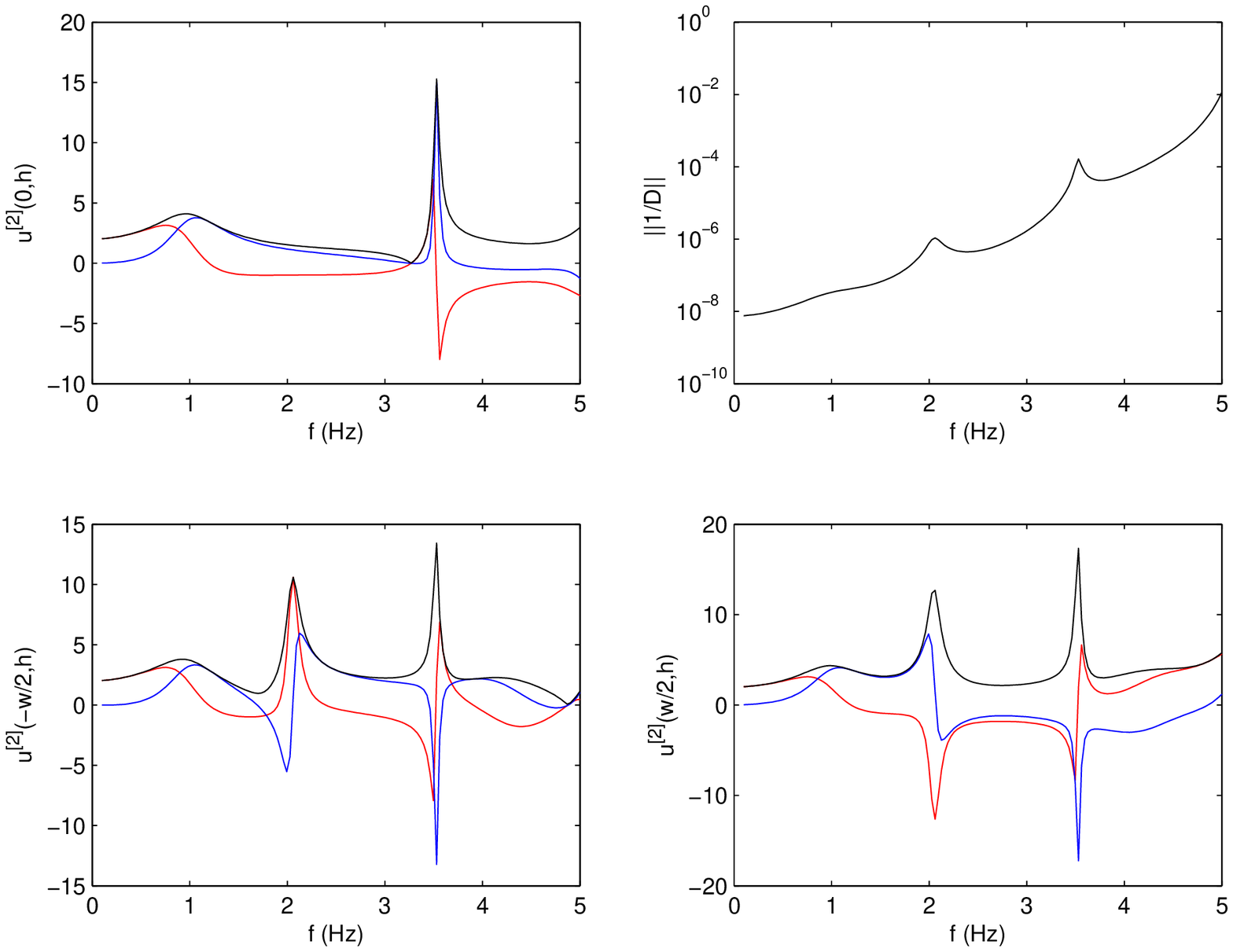}
      \label{b}
                         }
 \subfloat[]{
      \includegraphics[width=0.33\textwidth]{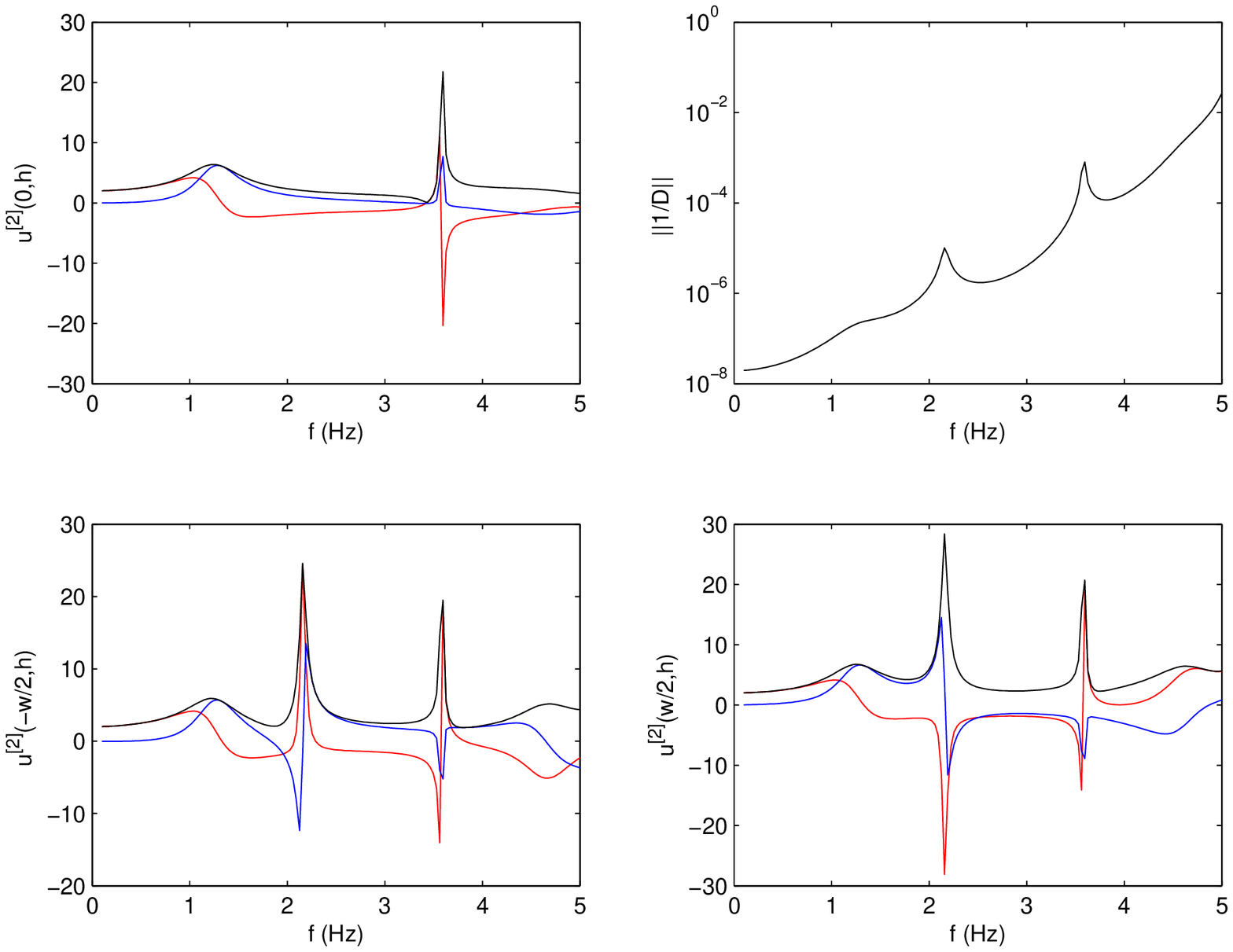}
      \label{c}
                        }
    \caption{
(a): $T(\mathbf{x},f)$ for $\mathbf{x}=(0,h)$, $\mathbf{x}=(-w/2,h)$, $\mathbf{x}=(w/2,h)$, in the (1,1), (2,1) and (2,2) panels respectively, and $1/\|D(f)\|$ in the (1,2) panel as a function of $f$. Case $\mu^{[0]}=6.85~GPa$ and $\beta^{[0]}=1629.4~ms^{-1}$.\\
(b): Same as (a) for case $\mu^{[0]}=12~GPa$ and $\beta^{[0]}=2500~ms^{-1}$.\\
(c): Same as (a) for case $\mu^{[0]}=25~GPa$ and $\beta^{[0]}=3500~ms^{-1}$.
}
    \label{ug-010}
  \end{center}
\end{figure}

In fig. \ref{ug-010} we depict the three-point transfer functions and $1/\|D(f)\|$  as a function of $f$ for the transition from hard to very hard rock undergrounds.
\clearpage
\newpage
The three-point transfer functions are observed to  be qualitatively similar from one type of underground rock to another , but the frequency positions and amplitudes of resonant response are different, with an increase of resonant frequencies and generally of the response at these frequencies for the transition from hard to very hard rock in the underground.
\subsection{The transfer function maps for the three undergrounds at their first resonance frequency}
\begin{figure}[ht]
  \begin{center}
    \subfloat[]{
      \includegraphics[width=0.32\textwidth]{recthill2layer_4-160719-1054.eps}
      \label{a}
                         }
    \subfloat[]{
      \includegraphics[width=0.345\textwidth]{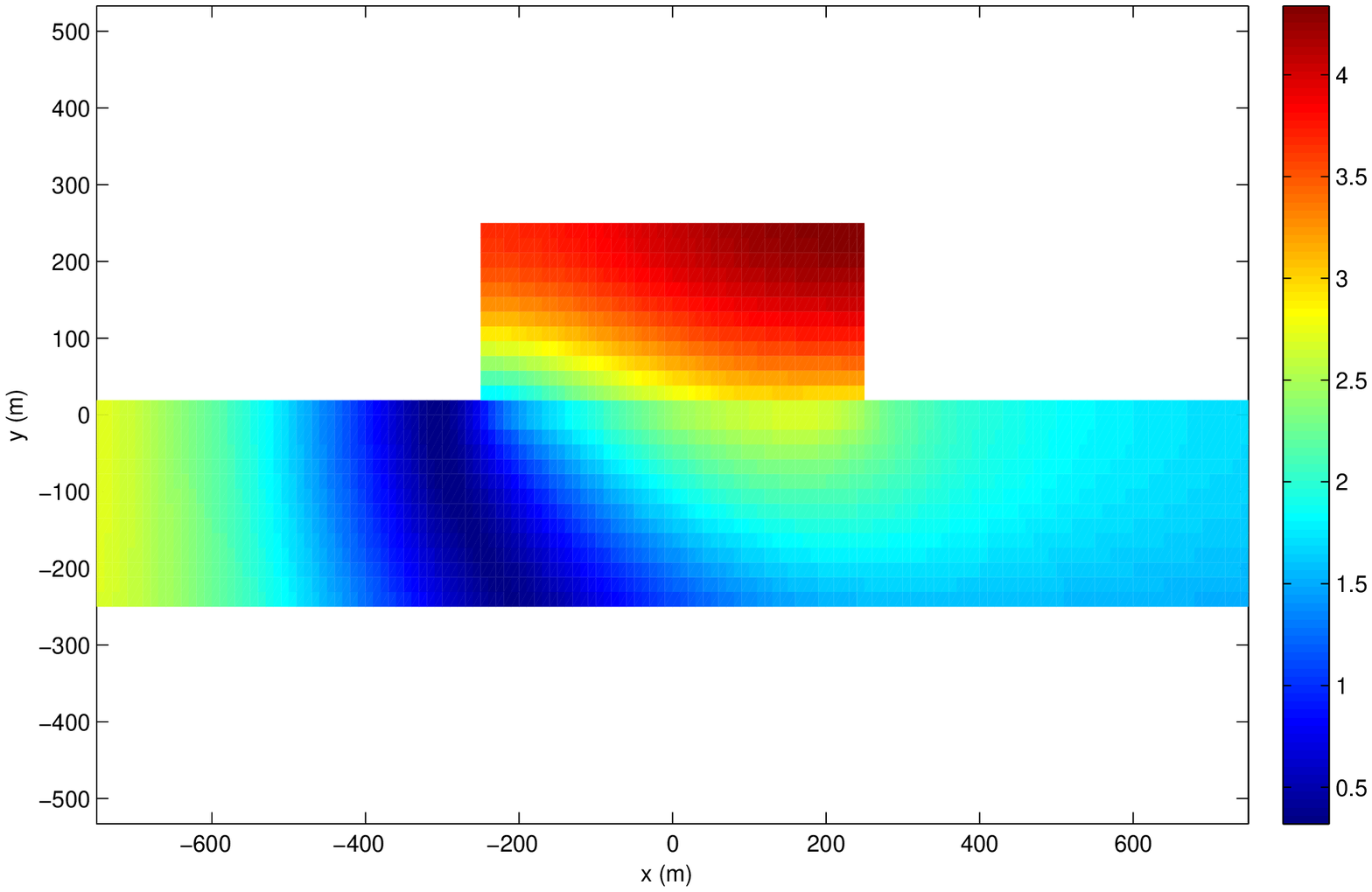}
      \label{b}
                         }
 \subfloat[]{
      \includegraphics[width=0.35\textwidth]{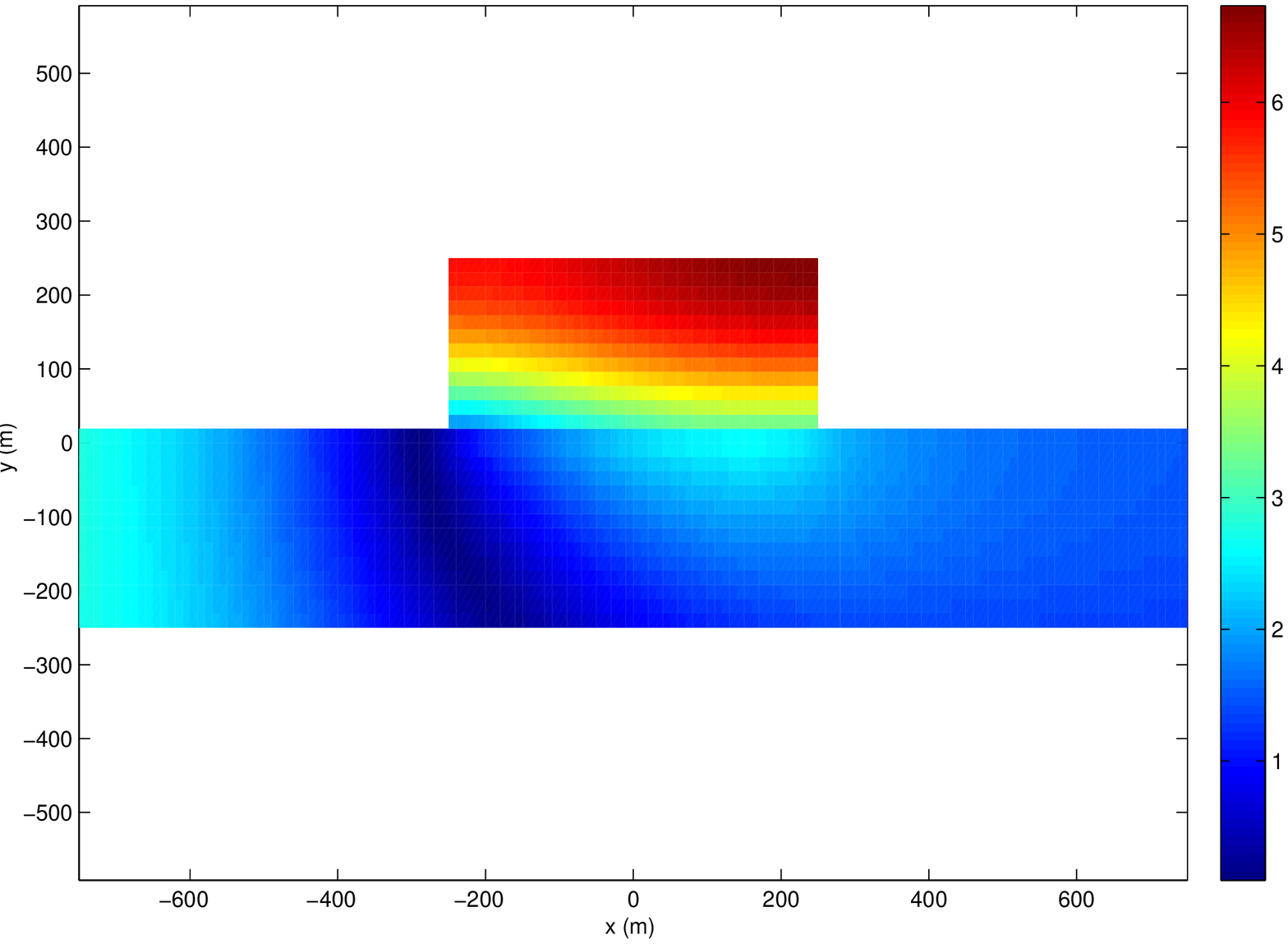}
      \label{c}
                        }
    \caption{
(a): $\|T(\mathbf{x},f_{1}^{R})\|$ for $\forall\mathbf{x}$ in the protuberance and in the upper portion of the underground. $\mathbf{x}=(0,h)$, $\mathbf{x}=(-w/2,h)$ and $\mathbf{x}=(w/2,h)$ in the (1,1), (2,1) and (2,2) panels respectively, and $1/\|D(f)\|$ in the (1,2) panel as a function of $f$. Case $\mu^{[0]}=6.85~GPa$ and $\beta^{[0]}=1629.4~ms^{-1}$ for which $f_{1}^{R}=0.7592~Hz$ and Max=3.2\\
(b): Same as (a) for case $\mu^{[0]}=12~GPa$ and $\beta^{[0]}=2500~ms^{-1}$  for which $f_{1}^{R}=1.015~Hz$ and Max=4.3\\
(c): Same as (a) for case $\mu^{[0]}=25~GPa$ and $\beta^{[0]}=3500~ms^{-1}$  for which $f_{1}^{R}=1.276~Hz$ and Max=6.7.\\
}
    \label{ug-020}
  \end{center}
\end{figure}
\clearpage
\newpage
\subsection{The transfer function maps for the three undergrounds at their third resonance frequency}
\begin{figure}[ht]
  \begin{center}
    \subfloat[]{
      \includegraphics[width=0.34\textwidth]{recthill2layer_4-150719-2024.eps}
      \label{a}
                         }
    \subfloat[]{
      \includegraphics[width=0.33\textwidth]{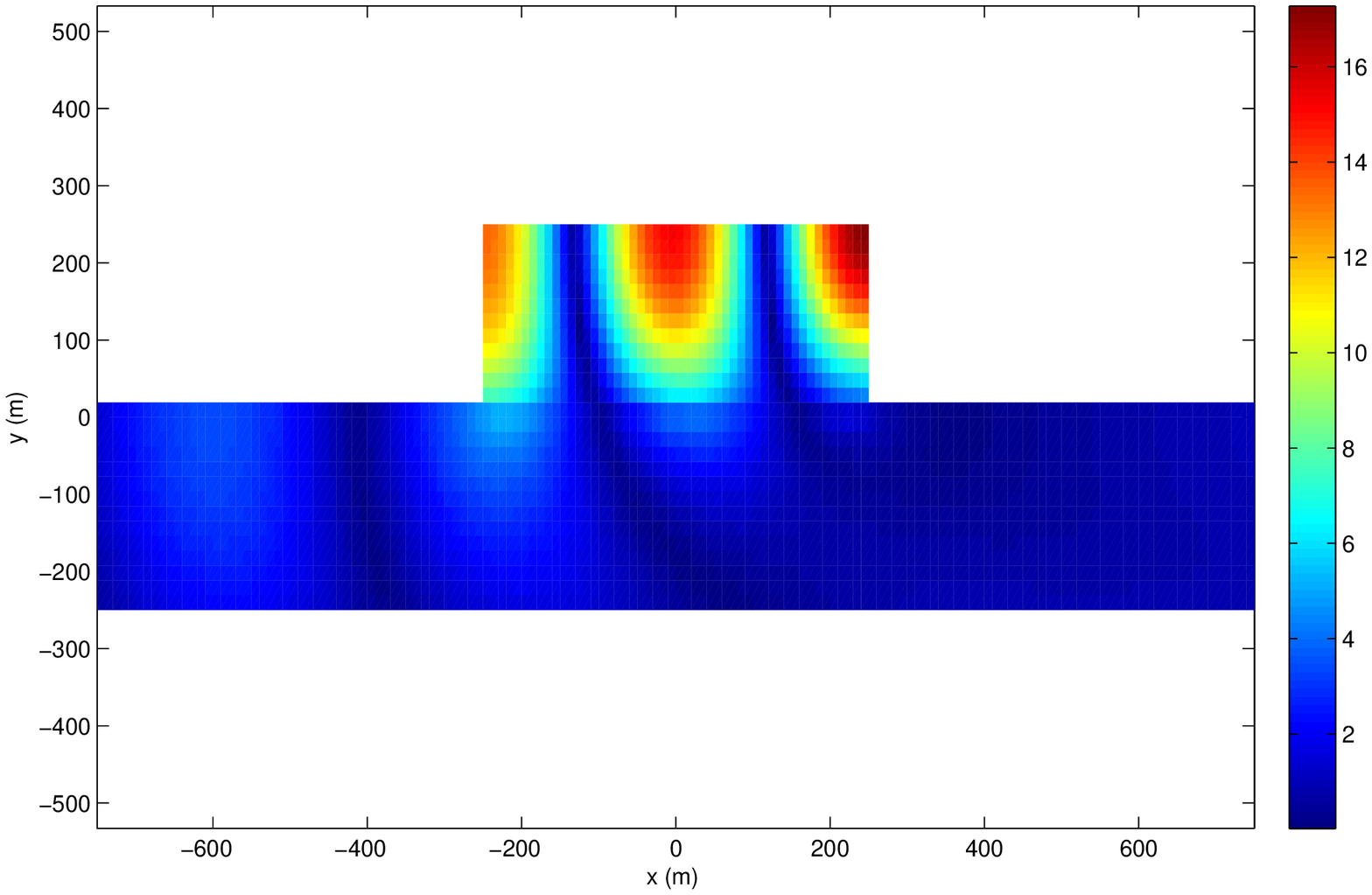}
      \label{b}
                         }
 \subfloat[]{
      \includegraphics[width=0.33\textwidth]{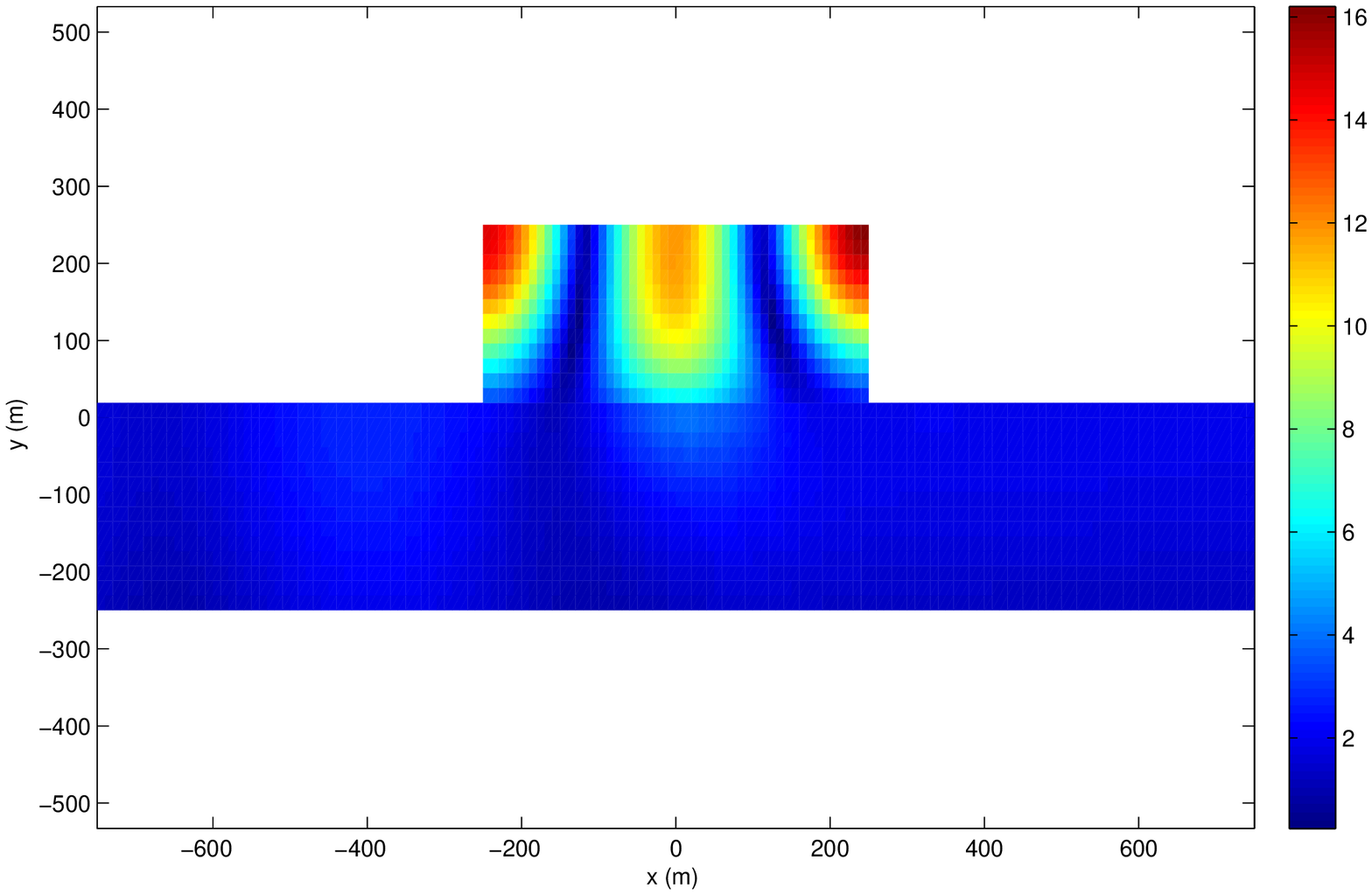}
      \label{c}
                        }
    \caption{
(a): $\|T(\mathbf{x},f_{3}^{R})\|$ for $\forall\mathbf{x}$ in the protuberance and in the upper portion of the underground. $\mathbf{x}=(0,h)$, $\mathbf{x}=(-w/2,h)$ and $\mathbf{x}=(w/2,h)$ in the (1,1), (2,1) and (2,2) panels respectively, and $1/\|D(f)\|$ in the (1,2) panel as a function of $f$. Case $\mu^{[0]}=6.85~GPa$ and $\beta^{[0]}=1629.4~ms^{-1}$ for which $f_{3}^{R}=3.482~Hz$ and Max=7.5.\\
(b): Same as (a) for case $\mu^{[0]}=12~GPa$ and $\beta^{[0]}=2500~ms^{-1}$  for which $f_{3}^{R}=3.530~Hz$ and Max=18.8.\\
(c): Same as (a) for case $\mu^{[0]}=25~GPa$ and $\beta^{[0]}=3500~ms^{-1}$  for which $f_{3}^{R}=3.563~Hz$ and Max=16.2.\\
}
    \label{ug-030}
  \end{center}
\end{figure}
\clearpage
\newpage
Figs \ref{ug-020}-\ref{ug-030} show that the transfer function maps  are {\it qualitatively} very similar, but quantitatively quite different, as a function of the rock undergrounds, this being true for the two resonant frequencies $f_{1}^{R}$ and $f_{3}^{R}$. Once again, the question of whether these changes are larger or smaller than those due to shape (i.e. aspect ratio) variations cannot be answered on the sole basis of these results.
\section{Concluding comments}
%
 \subsection{General comments}
 Even for a scattering configuration as simple as  a cylindrical rectangular-shaped protuberance solicited by a plane seismic wave, there exist twelve (incident angle, aspect ratio, size, and constitutive) parameters, aside from the frequency and spatial location, that condition the  seismic response. We found the angle of incidence, damping, size and shape, and composition of the protuberance and underground, to  exert the strongest influence (in this order) on the response.

This parametric study confirmed the empirically- and numerically-established facts that protuberances such as dikes, hills and mountains, emerging from flat ground give rise to a great variety of seismic responses, all the more so than the composition of the underground, the composition, shape and size of the protuberance is very diverse in the natural environment.

In spite of this, the seismic responses of all these configurations were shown theoretically in \cite{wi20b}, and confirmed numerically herein, to be largely-dominated by (surface shape) resonances, the manifestation of which are peaks of response at, or near, the resonance frequencies. In other words: large amplifications (with respect to that (=1) on flat ground) of  seismic response in a protuberance such as a mountain (like that in a dike or hill)  is possible only at, or in the neighborhood of, (shape-) resonant frequencies.

Actually, as shown herein, and explained in \cite{wi20b}, the spatial locations at which the response peaks occur are diverse, but the strongest displacement field amplifications usually occur  at hot spots (HS)  near the edges of, or along a horizontal strip (UTHS) near or on  the top segment of the protuberance. Moreover these amplifications turn out to be strongest at large incident angles so that  the oft-assumed normal incidence is a risky, if not bad, choice for predicting the maximal magnitude of plane seismic wave response of a configuration comprising a protuberance emerging from flat ground.

The 3-point transfer functions  are very different at the frequency locations of their maxima  and also quite different at the three points of the free-surface boundary of the protuberance.
Put in another way,   a resonance peak (i.e., a response peak at the location of a resonance frequency $f^{R}$, the latter being determined from the position of a local maximum of $1/|D(f)\|$) at a given $f^{R}$ is not always present at one or the other of the spatial locations of the 3-point transfer functions, but, as explained in \cite{wi20b}, this is simply the result of: a) how many, and the degree of coupling to, excited resonant modal coefficients occur at this frequency, as well as b) the geometrical factors that modulate the sum of these coefficients at that frequency;\\

At off-resonance frequencies, the displacement field in the underground, and on the free-surface flanks of the protuberance (i.e., $u^{[0]}(|x|>w/2,y\le 0)$, are often larger than the field at a point (usually the topmost or midpoint) of the summit  of the protuberance (e.g., $u^{[2]}(0,h)$ so that if, as is often the case in the seismic engineering community \cite{fa91,az93,bm14}, the ratio $\mathfrak{R}=\frac{\|u^{[2]}(0,h)\|}{\|u^{[0]}(|x|>w/2,0)\|}$ is adopted as the indication of amplified (for $\mathfrak{R}>1$ or de-amplified (for $\mathfrak{R}<1$) response, then one sees, at off-resonance frequencies,  that this ratio indicates almost systematically-large amplification although the field is almost systematically un-amplified within and on top of our rectangular protuberance.

Variations of the aspect ratio (for constant width $w$ or constant height $h$) of the protuberance produce little (for constant $w$) or fairly-large (for constant $h$) qualitative, and large quantitative, changes of resonant response within the protuberance. The question of whether these changes are larger or smaller than those due to compositional variations cannot be answered on the sole basis of these findings.

Our previous study \cite{wi20b} showed that  resonant seismic response is universal in that it occurs in an isolated protuberance (e.g., dike, hill or mountain) of arbitrary shape and aspect ratio. Herein, we confirmed this finding as concerns the aspect ratio, and  found, in addition, that the resonant response also occurs  for a variety of compositions of the protuberance. This means that the (theoretical/numerical) prediction of the seismic response of a protuberance  is meaningless unless a rather detailed knowledge of the subsurface  composition (i.e., the geological features of the protuberance as well as that of the medium underlying this feature)  is available and incorporated in the computation. At the best, parametric studies, based on a systematic variation of the convex feature geology, should be carried out to evaluate the expected variability of response of a feature with given geometry.

 A forthcoming contribution will be devoted to the seismic response of multiple protuberances spread out on the ground.
 \subsection{Specific comments}
 %
 \subsubsection{Variation of incident angle}                                                                                   %
  We found that the field generally tends to increase with $|\theta^{i}|$ beyond a certain angle of incidence (usually $\ge 60^{\circ}$).  For this reason, we chose the majority of the subsequent variations to apply to large angles of incidence.
 \subsubsection{Variation of the material damping within the protuberance}                                                                                   %
 We found that:\\
(i) the position of the resonant frequencies hardly changes with the (relatively-small) amount of material damping in the protuberance, which means that they can (and should) be inferred from the undamped response (for which the resonance peaks are the sharpest);\\
(ii) the heights of the resonant response peaks are diminished considerably by the introduction of material damping, all the more so than this damping is larger and the resonance order is larger (i.e., the first resonance peaks in the 3-point transfer functions are the ones that are the least affected by material damping, but these peaks are usually lower than those of the higher-order  resonance peaks).
 \subsubsection{Variation of wavespeed within a homogeneous protuberance}                                                                                   %
 We found that:\\
(i) if we concentrate our attention on the different response maxima for the lossy configurations (of greatest interest in practical applications) then   it seems that the maximal response (regardless of its frequency of occurrence) increases as the wavespeed $\beta^{[1]'}=\beta^{[2]'}$ increases;\\
(ii) the maximal response, for all the  choices of the wavespeed, is greater at the right-hand edge of the top segment of the protuberance than at the center and left-hand edge of this segment, which is related to the fact that the angle of incidence is $60^{\circ}$. Consequently, the center of the top segment is not the best location for predicting the maximal response  to obliquely-incident  seismic waves.
 \subsubsection{Variation of the shear modulus within a homogeneous protuberance}                                                                                   %
We found that:\\
(i) if we concentrate our attention on the different response maxima for the lossy configurations (of greatest interest in practical applications) then   it seems  that the maximal response (regardless of its frequency of occurrence) decreases as the shear modulus $\mu^{[1]}=\mu^{[2]}$ increases;\\
(ii) the maximal responses, for all the  choices of the shear modulus, is greater at the right-hand edge of the top segment of the protuberance than at the center and left-hand edge of this segment, which is related to the fact that the angle of incidence is $60^{\circ}$. This again underlines the fact that the center of the top segment is often not the best location for predicting the maximal response due to obliquely-incident  seismic waves.
 \subsubsection{Variation of the aspect ratio for constant protuberance width and composition}                                                                                 %
 We found that:
\\
(i) as one would expect, the resonance frequencies of order $j$, i.e., $f_{j}^{R}$, decrease as $h$ increases, more so for low $j$ than for high $j$;\\
(ii) the responses are {\it qualitatively} the same  for all three values of $h$ and  most incident angles and resonance frequencies;\\
(iii) however, there are significative {\it quantitative} differences in response as a function of $h$,  principally at the larger incident angles;\\
(iv) the introduction of material losses in the protuberance can result in important reductions of resonant response, particularly for the largest $h$.\\
 \subsubsection{Variation of the aspect ratio for constant protuberance height}                                                                                   %
We found that:\\
(i) the number and intensity of the HS (the latter appearing to be arranged in near-periodic, spatially-horizontal manner) seem to increase with $w$   and incident angle;\\
(ii) the intensity of the UTHS seems to change in less systematic fashion with increases of the  resonance frequency and incident angle;\\
(iii) $f_{j}^{R}$ increases as $w$ increases for low $j$, and then decreases with increasing $w$ for  for high $j$;\\
(iv) the responses are {\it qualitatively} different  for all three values of $w$ and  most incident angles and resonance frequencies;\\
(v) moreover, there are significative {\it quantitative} differences in response as a function of $w$.
 \subsubsection{Variation of the layering within the protuberance}                                                                                   %
We found that:\\
(i) a thin, soft, layer at the top of the  protuberance increases the maximal seismic response, at least  at the third resonance frequency,\\
(ii) as the amount of lower (upper) layer material increases (decreases), the (third) resonance frequency increases,\\
(iii) as the amount of lower upper) layer material increases (decreases), the resonant coupling for $0^{\circ}$ incidence first increases, to attain a maximum for equal heights of the two layers, and then decreases,\\
(iv) as the amount of lower (upper) layer material increases (decreases), the resonant coupling for  $40^{\circ}$ incidence systematically decreases, which seems to contradict the hypothesis that increased weathering increases the level of resonant seismic response\\
(v) as the amount of lower (upper) layer material increases (decreases), the resonant coupling for $80^{\circ}$ incidence systematically decreases, with the exception of the case of equal layer heights,\\
(vi) the resonant coupling is usually (but not always) larger for $40^{\circ}$ incidence than for $0^{\circ}$ incidence,\\
(vii) the resonant coupling is  always) larger for $80^{\circ}$ incidence than for $40^{\circ}$ incidence.
 \subsubsection{Variation of the rock properties of the homogeneous underground}                                                                                   %
We found that:\\
the transfer function maps  are {\it qualitatively} very similar, but quantitatively quite different, as a function of the rock undergrounds, this being true for the two resonant frequencies $f_{1}^{R}$ and $f_{3}^{R}$.

\end{document}